\documentclass[aps, pra, twocolumn, 10pt, amsmath, amssymb, superscriptaddress, floatfix, letterpaper, nofootinbib, nobibnotes]{revtex4-2}
\usepackage[utf8x]{inputenc}  % Tildar (input)
\usepackage[T1]{fontenc} % Tildar (output)
\usepackage[english]{babel}
\usepackage{graphicx}  % needed for figures
\usepackage{mathbbol} % all ten digits with \mathbb
\usepackage{physics} % For physics mathematical notation
\usepackage{kantlipsum}
\usepackage{float}
\usepackage{tabularx}
\usepackage{booktabs}
\usepackage{siunitx} % S column alignment
\usepackage[colorlinks,hyperindex]{hyperref}
\hypersetup{colorlinks, citecolor=blue, linkcolor=blue, urlcolor=blue}
\usepackage[caption=false]{subfig} % the subfigure package is deprecated and the subcaption package is incompatible with revtex4-2
\usepackage{mathtools}  % loads »amsmath«
\usepackage{amssymb}
\usepackage[noformat]{mattens} % use [noformat] for non-bold symbols % \bS for \widebar
\usepackage[dvipsnames,svgnames]{xcolor}
\DeclareMathOperator{\e}{\operatorname{e}}
\renewcommand{\r}{\mathrm{re}}
\renewcommand{\i}{\mathrm{im}}
\renewcommand{\c}{\mathrm{c}}
\newcommand{\m}{\mathrm{m}}
\newcommand{\h}{\mathrm{h}}
\newcommand{\p}{\mathrm{p}}
\newcommand{\zpf}{\mathrm{zpf}}
\renewcommand{\in}{\mathrm{in}}
\newcommand{\out}{\mathrm{out}}
\renewcommand{\th}{\mathrm{th}}
\newcommand{\B}{\mathrm{B}}
\newcommand{\G}{\mathcal{G}}
\newcommand{\D}{\mathcal{D}}
\newcommand{\E}{\mathcal{E}}
\newcommand{\K}{\mathcal{K}}
\renewcommand{\o}{\scriptscriptstyle \mathcal{O}}
\renewcommand{\O}{\mathcal{O}}
\renewcommand{\S}{\scriptscriptstyle \mathrm{S}}
\newcommand{\N}{\scriptscriptstyle \mathrm{N}}

\newcommand{\W}{\mathcal{W}}

\newcommand{\widebar}[1]{\bS{#1}}

\makeatletter
\def\NAT@def@citea{\def\@citea{\NAT@separator}}
\makeatother
\begin{document}

\title[Nonstationary force sensing]{Nonstationary force sensing under dissipative mechanical quantum squeezing}
\author{D. N. Bernal-García}
\email{dnbernalg@unal.edu.co}
\affiliation{Departamento de Física, Universidad Nacional de Colombia, Ciudad Universitaria, K. 45 No. 26-85, Bogotá D.C., Colombia}
\affiliation{School of Engineering and Information Technology, UNSW Canberra, ACT 2600, Australia}
\author{H. Vinck-Posada}
\affiliation{Departamento de Física, Universidad Nacional de Colombia, Ciudad Universitaria, K. 45 No. 26-85, Bogotá D.C., Colombia}
\author{M. J. Woolley}
\email{m.woolley@unsw.edu.au}
\affiliation{School of Engineering and Information Technology, UNSW Canberra, ACT 2600, Australia}
\date{\today}

%------------------------------------------------------------------------------
% Abstract
%
\begin{abstract}
We study the stationary and nonstationary measurement of a classical force driving a mechanical oscillator coupled to an electromagnetic cavity under two-tone driving.
For this purpose, we develop a theoretical framework based on the signal-to-noise ratio to quantify the sensitivity of linear spectral measurements.
Then, we consider stationary force sensing and study the necessary conditions to minimise the added force noise.
We find that imprecision noise and back-action noise can be arbitrarily suppressed by manipulating the amplitudes of the input coherent fields, however, the force noise power spectral density cannot be reduced below the level of thermal fluctuations.
Therefore, we consider a nonstationary protocol that involves non-thermal dissipative state preparation followed by a finite time measurement, which allows one to perform measurements with a signal-to-noise much greater than the maximum possible in a stationary measurement scenario.
We analyse two different measurement schemes in the nonstationary transient regime, a back-action evading measurement, which implies modifying the drive asymmetry configuration upon arrival of the force, and a nonstationary measurement that leaves the drive asymmetry configuration unchanged.
Conditions for optimal force noise sensitivity are determined, and the corresponding force noise power spectral densities are calculated.

\end{abstract}

%------------------------------------------------------------------------------
%
\maketitle 

%------------------------------------------------------------------------------
% Introduction
%
\section{Introduction}
\label{sec:intro}
The problem of measuring a force by monitoring the position of a mechanical quantum oscillator has served as a longstanding inspiration for the theory of quantum measurement ~\cite{Braginsky1980,Caves1980,Caves1987,Braginsky1992,Bocko1996,Chen2013}.
Further, the use of optomechanical systems as a sensitive platform for the measurement of very weak forces has been especially motivated by the endeavour to detect gravitational waves~\cite{Danilishin2012, Zeuthen2019}.
However, the development of ultra-sensitive force measurement technologies is also relevant in many other applications, such as atomic force microscopes~\cite{Milburn1994, Butt2005}, magnetic resonance force microscopy~\cite{Poggio2010}, absolute rotation detection~\cite{Davuluri2016a, Davuluri2016b}, proposals for the detection of dark matter~\cite{Carney2019b, Carney2019c}, and in studying the interplay of quantum mechanics and gravity on a tabletop scale~\cite{Bose2017, Marletto2017, Carney2019a, Carlesso2019a, Carlesso2019b}.
It was first described by Braginsky~\cite{Braginsky1967, *Braginsky1968}, that the maximum achievable sensitivity in the measurement of weak forces will be attained via an optimal trade-off between measurement imprecision and quantum back-action ~\cite{Caves1980a}, defining a lower limit which in the case of a simplified experimental scheme is known as the standard quantum limit (SQL) for force detection.
Nonetheless, this SQL may be beaten using more sophisticated measurement protocols; but even in that case, the sensitivity will ultimately be limited by thermal and quantum fluctuations of the mechanical oscillator and the electromagnetic field that make up the sensor~\cite{Clerk2010}.
\begin{figure}[b]
    \centering
    \includegraphics[width=0.85\linewidth]{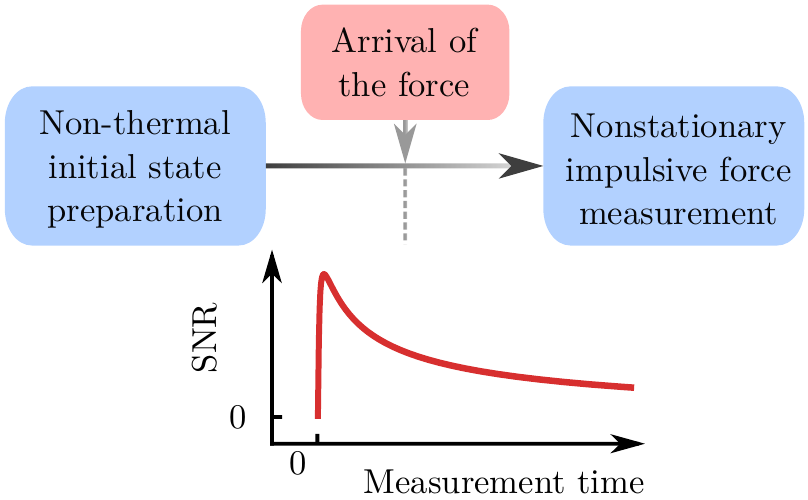}
    \caption{Nonstationary strategy for the measurement of impulsive forces.
    The strategy consists of two stages, non-thermal state preparation and nonstationary force measurement.
    Thus, first the mechanical oscillator is prepared in a dissipative squeezed state, and then upon arrival of the impulsive force, the force measurement is performed in the nonstationary transient regime before the re-thermalisation of the mechanical oscillator takes place.
    This strategy enables the measurement of impulsive forces with a signal-to-noise ratio (SNR) much greater than the maximum achievable with steady-state measurements (cf. Fig.~\ref{fig:snr}).
    \label{fig:nonstationary_strategy}}
\end{figure}
Since the 1970s there have been a number of proposals regarding an improvement of sensitivity beyond the SQL, and ultimately surpassing quantum and thermal fluctuations associated with the measurement~\cite{Braginsky1974,*Braginsky1975,Giffard1976,Caves1980}.
However, important experimental demonstrations in the last few years~\cite{Aspelmeyer2014,Aspelmeyer2014a}, and the first direct detection of gravitational waves~\cite{Abbott2016}, have motivated new ideas and sophisticated experiments feeding back to the problem of the detection of a weak classical force coupled to a quantum-mechanical oscillator.
Among the latest experimental breakthroughs, it is worth highlighting the demonstration of force and position measurement below the SQL~\cite{Mason2019}, and the achievement of quantum amplification of the displacement of a mechanical oscillator using a single-trapped ion~\cite{Burd2019}.
On the theoretical side, recent proposals in the modification of the sensor design have included, inserting a degenerate optical parametric amplifier in an optomechanical cavity~\cite{Huang2017,Zhao2019}, introducing an auxiliary mechanical oscillator~\cite{Wang2018,Zhang2019}, using hybrid atom-cavity optomechanical setups~\cite{Ivanov2015,Motazedifard2016,Motazedifard2019}, and taking advantage of the electromagnetically induced transparency in an ensemble of three-level atoms~\cite{Davuluri2018}.
Here, we propose an alternative route based on a  time-dependent protocol that does not require the inclusion of additional components to the optomechanical cavity.
We consider a mechanical oscillator parametrically coupled to an electromagnetic cavity which is driven at the two sidebands associated with the mechanical motion.
Such a system has been used before to perform back-action evading (BAE) measurements of a single mechanical quadrature in microwave electromechanical systems~\cite{Clerk2008,Woolley2008,Hertzberg2010,Suh2014}, and optical systems~\cite{Shomroni2019b}.
In addition, it has been used to achieve dissipative mechanical and electromagnetic squeezed states~\cite{Kronwald2013,Kronwald2014,Lecocq2015,Wollman2015,Pirkkalainen2015,Lei2016}.
Further, recently it has been used to demonstrate a two-tone optomechanical instability in backaction-evading measurements~\cite{Shomroni2019a}.
Moreover, considering two mechanical oscillators coupled to a common cavity mode, this scheme has been used to perform two-mode BAE measurements~\cite{Woolley2013,Ockeloen-Korppi2016} and to prepare entangled mechanical states~\cite{Woolley2014,Ockeloen-Korppi2018,Massel2019}.
We study two techniques for the measurement of a classical force using the aforementioned quantum optomechanical system, and determine the conditions for optimal signal-to-noise ratio (SNR) in the force measurement.
First, we consider force sensing in the steady-state under dissipative state preparation, for which we use the stationary force noise power spectral density (PSD) as a figure of merit to quantify the sensitivity of the force measurement.
Second, we consider a time-dependent sensing protocol, where the mechanical oscillator fluctuations are first reduced dissipatively, and then sensing is conducted in a finite measurement time before the re-thermalisation of the mechanical oscillator takes place (see Fig.~\ref{fig:nonstationary_strategy}).
Since nonstationary measurements depend on the initial state of the system, the careful manipulation of the initial conditions can lead to an improvement in the sensitivity of the force sensor.
We have identified regimes where such an approach is beneficial, and analysed this scenario quantitatively.
A nonstationary strategy similar to the one discussed here was presented in Refs.~\cite{Vitali2001, *Vitali2001Erratum} and~\cite{Vitali2002, *Vitali2002Erratum} in the context of feedback cooling.
This paper is organised as follows. 
In Sec.~\ref{sec:theoretical_model}, we introduce the model and obtain the Heisenberg-Langevin equations of motion describing the system dynamics. 
In Sec.~\ref{sec:signal-to-noise_ratio}, we consider the definition of SNR for a generic linear nonstationary force measurement, which gives a theoretical framework to the rest of this work.
In Sec.~\ref{sec:stationary_force_sensing}, we describe the stationary force sensing protocol under two-tone driving and determine the conditions for optimal force measurement based on the corresponding stationary force noise PSD.
In Sec.~\ref{sec:nonstationary_force_sensing}, we discuss a nonstationary strategy that significantly improves the SNR for force measurements, which considers a measurement in the nonstationary transient regime using a mechanical oscillator initially prepared in a dissipative squeezed state.
In Sec.~\ref{sec:conclusions}, we conclude.
% 
%------------------------------------------------------------------------------
% Theoretical model
%
\section{Theoretical model}
\label{sec:theoretical_model}
\begin{figure}[t]
    \centering
    \includegraphics[width=0.35\textwidth]{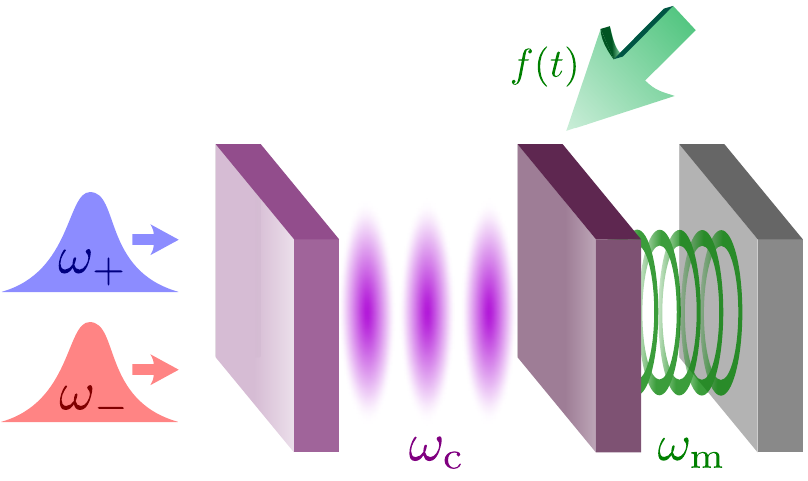}
    \caption{Sketch of the optomechanical/electromechanical system under consideration. A classical force $f(t)$ drives a mechanical oscillator with resonance frequency  $\omega_{\m}$, which in turn is coupled to an electromagnetic cavity with resonance frequency $\omega_{\c}$ under two-tone driving. The frequencies of the input coherent tones are $\omega_{\pm} = \omega_{\c} \pm \omega_{\m}$.  \label{fig:mechanical_squeezing}}
\end{figure}
In this section we present the theoretical model with which we will describe the dynamics of the physical system considered in this work.
As mentioned in the introduction, this system has been extensively studied in the past, and quite often a model very similar to the one we present here is used (e.g. Refs. \cite{Kronwald2013,Shomroni2019b}, and supplementary materials for Refs. \cite{Wollman2015,Pirkkalainen2015,Lei2016}).
However, different from what has been done before, here we describe the dynamics of the mechanical oscillator by a set of generalised quantum Langevin equations \cite{Ford1988, Giovannetti2001, Bowen2015} (see Appendix \ref{app:derivation_hamiltonian} for more details), which allow us to give a more accurate description of the system in the nonstationary transient regime.
Thus, the reader familiar with cavity quantum optomechanical (electromechanical) systems under two driving may skip this section, but should nevertheless be aware that the correlation functions for the mechanical noise operators are not as usual.
We consider a classical force acting on a mechanical oscillator, which is coupled to an electromagnetic cavity driven at the two sidebands detuned from the cavity resonance frequency by the mechanical resonance frequency, as represented in Fig. \ref{fig:mechanical_squeezing}.
The mechanical oscillator is described as a single quantum harmonic oscillator with mass $m$ and resonance frequency $\omega_{\m}$. 
This description is valid since the detuning of the input coherent drives is adjusted to select a particular mechanical mode.
Further, we consider a high quality factor electromagnetic cavity with free spectral range much greater than $\omega_{\m}$ and, hence, we focus on a single cavity mode with resonance frequency $\omega_{\c}$ selected by the external driving and neglect scattering into other electromagnetic modes.
The frequencies of the input coherent tones will be $\omega_{\pm} = \omega_{\c} \pm \omega_{\m}$, such that they drive the two sidebands corresponding to the chosen mechanical mode.
The weak classical force to be measured acts on the mechanical oscillator shifting its position, this accordingly modifies the effective length of the cavity whose change can be monitored through the output electromagnetic field.
The linearised Hamiltonian for this system, which is derived in Appendix \ref{app:derivation_hamiltonian}, is given by ($\hbar = 1$):
\begin{align}\label{eq:linearized_hamiltonian}
    H =\, &  \omega_{\c}\, a^{\dagger} a + \omega_{\m}\, b^{\dagger} b -  g\, \pqty{ \alpha\, a^{\dagger} +\, \alpha^* a}\, \pqty{b^{\dagger} +\, b} \nonumber \\
    &  - \frac{F}{\sqrt{2}} \pqty{b^{\dagger} +\, b},
\end{align}
where $a$ and $b$ are the electromagnetic and mechanical annihilation operators, respectively; $g$ is the single-photon optomechanical coupling strength and $F=F(t)$ corresponds to the classical force to be measured.
Further, $\alpha =\, \widebar{a}_+\e^{-i \omega_+ t} +\, \widebar{a}_- \e^{-i \omega_- t}$, where $\widebar{a}_{\pm}$ are real constants corresponding to the amplitudes of the coherent cavity field in the steady-state.
Now, with the intention of eliminating explicit time-dependence in the interaction terms due to $\alpha$, we move to an interaction picture with respect to $H_0 =\, \omega_{\c}\, a^{\dagger} a + \omega_{\m}\, b^{\dagger} b$, via $H_{\mathrm{I}}= U^{\dagger} H U - i\, U^{\dagger} \partial U/\partial t$ with $U = \e^{-i H_0 t}$.
Furthermore, if the external force is near-resonant with the mechanical oscillator, we may write
\begin{align}\label{eq:classical_force}
    F(t) =\, \widebar{F}(t)\, \e^{-i \omega_{\m} t} +\,  \widebar{F}^*(t)\, \e^{i \omega_{\m} t},
\end{align}
where $\widebar{F}(t)$ is a slowly-varying amplitude.
Thus, the Hamiltonian in the interaction picture will be given by
\begin{align}\label{eq:rotated_hamiltonian_2}
    H_{\mathrm{I}} =& - a^{\dagger} \bqty{ \pqty{G_+ + G_- \e^{2 i \omega_{\m} t}}\, b^{\dagger} + \pqty{G_+ \e^{-2 i \omega_{\m} t} + G_- }\, b }  \nonumber \\
    & -\frac{\widebar{F}}{\sqrt{2}} \pqty{ b^{\dagger} + b\, \e^{- 2 i \omega_{\m} t} }  + \mathrm{H.c.},
\end{align}
where $G_{\pm} = g\, \widebar{a}_{\pm}$ ($G_+, G_- \geq 0$) are the effective optomechanical coupling strengths.
Moreover, if $\omega_{\m} \gg G_{\pm}, \abs{\widebar{F}}$ [with $\abs{\widebar{F}}$ the magnitude of  $\widebar{F}(t)$], we can make a rotating-wave approximation (RWA) and neglect the fast-oscillating terms in Eq.~\eqref{eq:rotated_hamiltonian_2}, such that $H_{\mathrm{I}}$ reduces to the effective Hamiltonian
\begin{align}\label{eq:effective_hamiltonian:1}
    H_{\mathrm{eff}} =& - a^{\dagger}\, (G_+\, b^{\dagger} + G_- b) - \frac{\widebar{F}}{\sqrt{2}}\, b^{\dagger} + \mathrm{H.c.}
\end{align}
It is useful to write the effective Hamiltonian in Eq.~\eqref{eq:effective_hamiltonian:1} in terms of the dimensionless mechanical and electromagnetic quadratures.
Here, the mechanical quadratures are defined as $Q = ( b^{\dagger} + b)/\sqrt{2}$, $P = i\, ( b^{\dagger} - b)/\sqrt{2}$; while the electromagnetic quadratures are given by $X = ( a^{\dagger} + a)/\sqrt{2}$, $Y = i\, ( a^{\dagger} - a)/\sqrt{2}$.
Hence, the effective Hamiltonian takes the form
\begin{align}\label{eq:effective_hamiltonian:2}
    H_{\mathrm{eff}} =& -\big[ (G_- + G_+)\, Q\, X +   (G_- - G_+)\, P\, Y \big] \nonumber \\
    & -  \widebar{F}_{\r}\, Q  - \widebar{F}_{\i}\, P ,
\end{align}
where $\widebar{F}_{\r}$ and  $ \widebar{F}_{\i}$ are the real and imaginary parts of $\widebar{F}$, respectively.
Finally, from the effective Hamiltonian in Eq.~\eqref{eq:effective_hamiltonian:1}, the Heisenberg-Langevin equations in the interaction picture for the operators $a$ and $b$ will be given by,
% and considering a RWA in the interaction between system and reservoir
%
\begin{subequations}\label{eq:HLEs_rwa}
\begin{align}
    \dot{a} &= - \frac{\kappa}{2}\,   a + i\, \pqty{ G_+\,   b^{\dagger} +  G_-\,   b }  +  \sqrt{\kappa}\, a_{\in}, \\
    \dot{b} &= - \frac{\gamma}{2}\,   b + i\, \pqty{ G_+\,   a^{\dagger} +  G_-\,   a } + \frac{i}{\sqrt{2}} \pqty{ \widebar{F} + \widebar{\W} }.
\end{align}
\end{subequations}
The electromagnetic input noise $a_{\in} =  a_{\in}(t)$ satisfies the following the correlation functions,
\begin{align}
    \expval*{ a_{\in}(t)\, a_{\in}^{\dagger}(t') } &=  \delta(t-t'), \nonumber \\
    \expval*{ a_{\in}(t)\, a_{\in}(t') } =  \expval*{ a_{\in}^{\dagger}(t)\, a_{\in}^{\dagger}(t') } &= \expval*{ a_{\in}^{\dagger}(t)\, a_{\in}(t') } = 0,
\end{align}
as well as the input-output relation  
\begin{align}\label{eq:input-output}
     a_{\out}(t) + a_{\in}(t) = \sqrt{\kappa}\, a(t),
\end{align}
where $ a_{\out}(t)$ will be associated with the output electromagnetic field in the interaction picture.
The mechanical quantum Langevin force $\widebar{\W}$ is defined and treated in Appendix \ref{app:derivation_hamiltonian}.
In this work we focus on the dynamics of the mechanical and electromagnetic quadratures in the interaction picture, therefore, we stress that if the conditions considered to ensure the validity of Eqs.~\eqref{eq:HLEs_rwa} are satisfied (see Appendix \ref{app:derivation_hamiltonian} and previously in this section), then, the system quadratures will obey the following system of Heisenberg-Langevin equations~\cite{Kronwald2013},
\begin{align}\label{eq:system_dynamics}
    \dot{\vb*{v}}= \vb{M} \cdot \vb*{v} + \vb*{f} + \vb*{\xi},
\end{align}
where
\begin{align}\label{eq:vector_quadratures}
    \vb*{v} = \pqty{ Q,\, P,\, X,\, Y }^{\mathrm{T}}
\end{align}
is the vector of quadrature operators, $\vb{M}$ describes the system dynamics in the interaction picture
\small
\begin{align}
    &\vb{M} = \nonumber \\
    &\renewcommand\arraystretch{1.15}
    \setlength{\tabcolsep}{0.005cm}
    \begin{bmatrix}
        -\gamma/2 & 0 & 0 & -(G_- - G_+)\\
        0 & -\gamma/2 & G_- + G_+ & 0\\
        0 & -(G_- - G_+) & -\kappa/2 & 0\\
        G_- + G_+ & 0 & 0 & -\kappa/2\\
    \end{bmatrix},
\end{align}
\normalsize
and $\vb*{f}$ is the force vector, which contains the information about the force applied to the mechanical oscillator:
\begin{align}
    \vb*{f} = \pqty{ -\widebar{F}_{\i},\, \widebar{F}_{\r},\, 0,\, 0 }^{\mathrm{T}}.
\end{align}
Further,
\begin{align}
    \vb*{\xi} = \pqty{ -\widebar{\W}_{\i},\, \widebar{\W}_{\r},\, \sqrt{\kappa}\, X_{\in},\, \sqrt{\kappa}\, Y_{\in} }^{\mathrm{T}}
\end{align}
is the input noise vector describing the Langevin noise due to the mechanical and electromagnetic reservoirs, where $\widebar{\W}_{\r}$ and $\widebar{\W}_{\i}$ are the real and imaginary parts of $\widebar{\W}$, and $X_{\in} =  ( a_{\in}^{\dagger} +\, a_{\in} )/\sqrt{2}$ and $Y_{\in} = i\, ( a_{\in}^{\dagger} -\, a_{\in} )/\sqrt{2}$ are the input noises associated with the electromagnetic quadratures .
The correlation functions of the input electromagnetic noises are given by,
\begin{subequations}\label{eq:xy_corr_0}
    \begin{align}\label{eq:xy_corr}
        \expval{ X_{\in}(t) X_{\in}(t') } =& \expval{ Y_{\in}(t) Y_{\in}(t') } = \frac{1}{2}\,  \delta(t-t'), \\
        \expval{ X_{\in}(t) Y_{\in}(t') } =& \expval{ Y_{\in}(t) X_{\in}(t') }^* = \frac{i}{2}\, \delta(t-t').
    \end{align}
\end{subequations}
Moreover, from the input-output relation in Eq.~\eqref{eq:input-output}, we have,
\begin{subequations}
    \begin{align}
    &X_{\out}(t) + X_{\in}(t)= \sqrt{\kappa}\, X(t), \\
    &Y_{\out}(t) + Y_{\in}(t)= \sqrt{\kappa}\, Y(t); \label{eq:input-output_Y}
    \end{align}
\end{subequations}
where the output electromagnetic quadratures are given by, $X_{\out} =\, (a_{\out}^{\dagger} +\, a_{\out} )/\sqrt{2}$, $Y_{\out} =\, i\, (a_{\out}^{\dagger} -\, a_{\out} )/\sqrt{2}$.
On the other hand, the correlation functions involving the mechanical quantum Langevin forces $\widebar{\W}_{\r}(t)$  and $\widebar{\W}_{\i}(t)$ are as follows,
\begin{subequations}\label{eq:langevin_corr_1}
\begin{align}
    &\expval{\widebar{\W}_{\r}(t) \widebar{\W}_{\r}(t')} = \expval{\widebar{\W}_{\i}(t) \widebar{\W}_{\i}(t')} = \nonumber \\
    & \, \frac{\gamma}{4 \pi \omega_{\m}} \Bigg\{ \int_0^{\varpi} \dd{\omega} \omega \coth \pqty{\frac{\hbar \omega}{2 k_{\B} T} }  \cos{ \bqty{ (\omega - \omega_{\m}) (t-t') }} \nonumber \label{eq:langevin_corr_1:a} \\
    & \hspace{1.5cm} - i \int_0^{\varpi} \dd{\omega} \omega \sin{ \bqty{ (\omega - \omega_{\m}) (t-t') }} \Bigg\}, \\
    &\expval{\widebar{\W}_{\r}(t) \widebar{\W}_{\i}(t')} = - \expval{\widebar{\W}_{\i}(t) \widebar{\W}_{\r}(t')} = \nonumber \\
    & \, \frac{\gamma}{4 \pi \omega_{\m}} \Bigg\{ \int_0^{\varpi} \dd{\omega} \omega \coth \pqty{\frac{\hbar \omega}{2 k_{\B} T} }  \sin{ \bqty{ (\omega - \omega_{\m}) (t-t') }} \nonumber \\
    & \hspace{1.5cm} + i \int_0^{\varpi} \dd{\omega} \omega \cos{ \bqty{ (\omega - \omega_{\m}) (t-t') }} \Bigg\}, \label{eq:langevin_corr_1:b}
\end{align}
\end{subequations}
where $\varpi$ is a cutoff frequency for the continuous spectrum of reservoir quantum harmonic oscillators~\cite{Giovannetti2001, Bowen2015}.
From the Heisenberg-Langevin Eqs.~\eqref{eq:system_dynamics}, it is clear that if $G_+=G_-$, which can be achieved by tuning the input coherent drives, then, we can perform a BAE measurement of the mechanical $Q$ quadrature as described in Refs.~\cite{Woolley2008,Clerk2008}. 
Otherwise, if $G_+ \neq G_-$, the electromagnetic quadrature $Y$ will act as a force for the mechanical oscillator, introducing additional noise in the measurement process.
However, this unbalanced detection scheme allows one to obtain arbitrarily large dissipative squeezing of the mechanical quadratures, as demonstrated in Ref.~\cite{Kronwald2013}.
On the other hand, Eqs.~\eqref{eq:system_dynamics} show that each quadrature of the force affects a different mechanical quadrature, and each mechanical quadrature is coupled to only one quadrature of the electromagnetic field.
Thus, the continuous homodyne measurement of the output electromagnetic field quadrature $Y_{\out}$ ($X_{\out}$) corresponds to continuously monitoring the mechanical quadrature $Q$ ($P$), which in turn implies sensing the classical force quadrature $\widebar{F}_{\i}$ ($\widebar{F}_{\r}$).
% 

% 
%------------------------------------------------------------------------------
% Signal-to-noise ratio
%
\section{Signal-to-noise ratio in force measurements}
\label{sec:signal-to-noise_ratio}
The sensitivity of a measurement is commonly quantified by a signal-to-noise ratio (SNR), where a more sensitive measurement will have a greater associated SNR.
Thus, in this work we are interested in a definition of SNR that is able to describe the sensitivity of stationary and nonstationary force measurements.
A SNR with these characteristics has been presented by Vitali et al. in Refs. \cite{Vitali2001,*Vitali2001Erratum,Vitali2002,*Vitali2002Erratum}.
Here we provide a thorough justification for this metric.
In order to give a formal definition of SNR, first we shall consider how to obtain information about a classical force applied to a mechanical oscillator using an optomechanical scheme like the one we study here.
For this purpose, we shall focus on the sensing of the $\widebar{F}_{\i}(t)$ quadrature of the force, which can be done through the measurement of the electromagnetic output quadrature $Y_{\out}(t)$, as described in the following relationship:
\begin{align}\label{eq:yout_snr}
    Y_{\out}(t) = A(t)*\widebar{F}_{\i}(t) + N(t),
\end{align}
where $A(t)$ is the amplification of the force signal, and $N(t)$ is the zero-mean noise added due to the measurement.
% $*$ denotes convolution.
%
Eq.~\eqref{eq:yout_snr} can be evaluated from the Heisenberg-Langevin Eqs.~\eqref{eq:system_dynamics} and the input-output relation \eqref{eq:input-output}, as will be done in Sec. \ref{sec:nonstationary_force_sensing}.
To estimate $\widebar{F}_{\i}(t)$ from $Y_{\out}(t)$, we apply $Y_{\out}(t)$ to a linear filter with impulse response $h(t)$ and frequency response $H(\omega)$, such that
\begin{align}\label{eq:f_est}
    F_{\mathrm{est}}(t) \equiv h(t) * Y_{\out}(t),
\end{align}
where $F_{\mathrm{est}}(t)$ is the quantum estimator of the classical force quadrature $\widebar{F}_{\i}(t)$.
The estimated force may be broken down into a signal and a noise components,
\begin{align}
    F_{\mathrm{est}}(t) = F_{\mathrm{est}}^{\S}(t) + F_{\mathrm{est}}^{\N}(t), \label{eq:nonstationary_measurement}
\end{align}
where $ F_{\mathrm{est}}^{\S}(t)$ is the response due to the signal $\widebar{F}_{\i}(t)$ and $F_{\mathrm{est}}^{\N}(t)$ is the added force noise due to the measurement.
These are given by
\begin{subequations}\label{eq:f_signal_noise}
\begin{align}
    &F_{\mathrm{est}}^{\S}(t) = h(t) * A(t)*\widebar{F}_{\i}(t), \\
    &F_{\mathrm{est}}^{\N}(t) =  h(t) * N(t) \label{eq:f_noise},
\end{align}
\end{subequations}
respectively.

\subsection{Signal-to-noise ratio}
In the following, we shall consider the SNR of the linear force measurement described by Eqs.~\eqref{eq:yout_snr} -- \eqref{eq:f_signal_noise}.
The SNR is usually defined as the ratio of the mean to the standard deviation of a given filtered measured signal \cite{Papoulis2002}, where the signal is identified with the mean while the noise corresponds to the standard deviation.
Since we are interested in spectral measurements that are in general nonstationary, we will focus on making a description in frequency domain that accounts for the effects of a finite measurement time.
For this purpose, we shall consider the truncated Fourier transform, which is defined as
\begin{align}\label{eq:truncated_ft}
    \O(\omega, T_{\m}) = \int_{-\infty}^{+\infty} \dd{t} \e^{i \omega t} \Pi_{T_{\m}}(t)\, \O(t),
\end{align}
where $\O(t)$ is a generic operator, $\Pi_{T_{\m}}(t)$ is a rectangular window satisfying the normalisation condition
\begin{align}\label{eq:normalisation_condition}
    \frac{1}{T_{\m}} \int_{-\infty}^{+\infty} \dd{t} \abs{\Pi_{T_{\m}}(t)}^2 = 1,
\end{align}
and $T_{\m}$ is the measurement time.
Thus, we define the truncated SNR as 
\begin{align}\label{eq:snr-nonstationary_0}
    \mathrm{SNR}(\omega, T_{\m}) \equiv \frac{ \mathcal{S}(\omega, T_{\m}) }{ \mathcal{N}(\omega, T_{\m}) },
\end{align}
where the signal $\mathcal{S}(\omega, T_{\m})$ is given by  
\begin{align}\label{eq:signal}
    \mathcal{S}(\omega, T_{\m}) =\, \abs{ \big\langle\, F_{\mathrm{est}}(\omega, T_{\m})\, \big\rangle  },
\end{align}
while the noise $\mathcal{N}(\omega, T_{\m})$ is defined as
\begin{align}\label{eq:noise}
    \mathcal{N}(\omega, T_{\m}) = \sqrt{\, \mathrm{Var}\big[\, F_{\mathrm{est}}(\omega, T_{\m})\, \big]\, }\, .
\end{align}
Here, $F_{\mathrm{est}}(\omega, T_{\m})$ is the truncated Fourier transform of $F_{\mathrm{est}}(t)$ as defined in Eq.~\eqref{eq:truncated_ft}, while
\begin{align}
    \mathrm{Var}\big[\, F_{\mathrm{est}}(\omega, T_{\m})\, \big] =\,  &\big\langle\, F_{\mathrm{est}}^{\dagger}(\omega, T_{\m})\,  F_{\mathrm{est}}(\omega, T_{\m})\, \big\rangle \nonumber \\ 
    -\,  &\big\langle\, F_{\mathrm{est}}^{\dagger}(\omega, T_{\m})\, \big\rangle \big\langle\, F_{\mathrm{est}}(\omega, T_{\m})\, \big\rangle
\end{align}
is the variance of $F_{\mathrm{est}}(\omega, T_{\m})$ \cite{Anandan1990,Pati2015}.
Therefore, $\mathcal{S}(\omega, T_{\m})$ and $\mathcal{N}(\omega, T_{\m})$ correspond to the mean and standard deviation of the truncated quantum estimator $F_{\mathrm{est}}(\omega, T_{\m})$, respectively.
Further, the absolute value in the definition of the signal $\mathcal{S}(\omega, T_{\m})$ in Eq.~\eqref{eq:signal} was included to guarantee that $\mathrm{SNR}(\omega, T_{\m})$ is always positive and real-valued.
Taking the truncated Fourier transform of Eq.~\eqref{eq:nonstationary_measurement}, we have that $F_{\mathrm{est}}(\omega, T_{\m})$ is by 
\begin{align}\label{eq:truncated_fest}
    F_{\mathrm{est}}(\omega, T_{\m}) = F_{\mathrm{est}}^{\S}(\omega, T_{\m}) + F_{\mathrm{est}}^{\N}(\omega, T_{\m}), 
\end{align}
with $F_{\mathrm{est}}^{\S}(\omega, T_{\m})$ and $F_{\mathrm{est}}^{\N}(\omega, T_{\m})$ the truncated Fourier transforms of $F_{\mathrm{est}}^{\S}(t)$ and $F_{\mathrm{est}}^{\N}(t)$, respectively.
Now, from Eq.~\eqref{eq:truncated_fest} and taking into account that  $\langle F_{\mathrm{est}}^{\N}(\omega, T_{\m}) \rangle=0$ given that $\langle F_{\mathrm{est}}^{\N}(t) \rangle=0$, we may write signal and noise in Eqs.~\eqref{eq:signal} and \eqref{eq:noise} as,
\begin{subequations}\label{eq:signal_noise_def}
\begin{align}
    \mathcal{S}(\omega, T_{\m}) =& \, \abs{  F_{\mathrm{est}}^{\S}(\omega, T_{\m}) }, \label{eq:signal_def}
    \\
    \mathcal{N}(\omega, T_{\m}) =& \,\sqrt{\, \big\langle\, F_{\mathrm{est}}^{\N \dagger}(\omega, T_{\m})\, F_{\mathrm{est}}^{\N}(\omega, T_{\m})\, \big\rangle\, } \nonumber \\
    =& \, \sqrt{\, T_{\m}\, S_{F_{\mathrm{est}}}(\omega, T_{\m})\, }\,. \label{eq:noise_def}
\end{align}
\end{subequations}
Here,
\begin{align}\label{eq:truncated_psd}
    S_{F_{\mathrm{est}}}(\omega, T_{\m}) &= \frac{1}{T_{\m}} \big\langle\, F_{\mathrm{est}}^{\N \dagger}(\omega, T_{\m})\, F_{\mathrm{est}}^{\N}(\omega, T_{\m})\, \big\rangle
\end{align}
is the truncated force noise PSD, which mimics the classical definition of \emph{periodogram PSD estimator} \cite{Cooper1998, GroverBrown2012, Prabhu2014} (see Appendix \ref{app:wiener-khinchin} for details).
Therefore, the truncated SNR in Eq.~\eqref{eq:snr-nonstationary_0} takes the form
\begin{align}\label{eq:snr-nonstationary}
    \mathrm{SNR}(\omega, T_{\m}) =\, \frac{ \abs{  F_{\mathrm{est}}^{\S}(\omega, T_{\m}) } }{\sqrt{\, T_{\m}\, S_{F_{\mathrm{est}}}(\omega, T_{\m})\, } },
\end{align}
which we identify as the most suitable figure of merit for quantifying the sensitivity of any nonstationary quantum measurement of a classical force.
Thus, our goal is to find the optimal conditions that maximise $\mathrm{SNR}(\omega, T_{\m})$ and, accordingly, the sensitivity of the measurement.
\subsection{Signal-to-noise ratio: stationary case}
It is common for measurements to be made in the stationary regime, where the measurement time is much larger than the relaxation times of the system.
In this regime, the definition of SNR given in Eq.~\eqref{eq:snr-nonstationary} will describe the sensitivity of the force measurement, thus, we take into account that in this limit  $\Pi_{T_\m}(t) \simeq 1$  \cite{Vitali2001, *Vitali2001Erratum, Vitali2002, *Vitali2002Erratum}, which is consistent with the normalisation condition in Eq. \eqref{eq:normalisation_condition} if we consider $\Pi_{T_\m}(t) = \theta(t + T_\m/2) - \theta(t - T_\m/2)$ and then we assume that $T_\m \to \infty$.
Therefore, we obtain the following expressions for  signal and noise in the stationary regime,
\begin{subequations}
\begin{align}
    &\mathcal{S}_{\mathrm{st}}(\omega) = \abs{ F_{\mathrm{est}}^{\S}(\omega) }, \\
    &\mathcal{N}_{\mathrm{st}}(\omega, T_\m) = \sqrt{\, T_\m\, S_{F_{\mathrm{est}}}(\omega)\, };
\end{align}
\end{subequations}
where $F_{\mathrm{est}}^{\S}(\omega)$ is the Fourier transform of $F_{\mathrm{est}}^{\S}(t)$, while $S_{F}(\omega)$ is the stationary force noise PSD given by (see Appendix \ref{app:wiener-khinchin}),
\begin{align}\label{eq:force_noise_spectrum}
    S_{F_{\mathrm{est}}}(\omega) =& \int_{-\infty}^{+\infty} \frac{\dd{\omega'}}{2\pi} \big\langle  F_{\mathrm{est}}^{\N}(\omega')\,  F_{\mathrm{est}}^{\N}(\omega)\, \big\rangle,
\end{align}
with $F_{\mathrm{est}}^{\N}(\omega)$ the Fourier transform of $F_{\mathrm{est}}^{\N}(t)$, and it was assumed that the force signal $F_{\mathrm{est}}^{\S}(t)$ is different than zero from $t=0$.
Hence, the SNR in the stationary regime will be given by
\begin{align}\label{eq:snr_stationary_0}
    \mathrm{SNR}_{\mathrm{st}}(\omega, T_\m) = \frac{\mathcal{S}_{\mathrm{st}}(\omega)}{\mathcal{N}_{\mathrm{st}}(\omega,  T_\m)} = \frac{\abs{ F_{\mathrm{est}}^{\S}(\omega) }}{\sqrt{\, T_\m\, S_{F_{\mathrm{est}}}(\omega)\, }}.
\end{align}
If an inverse filter is considered (see below in this section), the expression in Eq. \eqref{eq:snr_stationary_0} is equivalent to the one used in Ref. \cite{Vitali2002, *Vitali2002Erratum} to calculate the SNR for force measurements in the stationary regime.
However, since in the stationary regime the measurement time is an arbitrary parameter, the SNR may be rescaled with respect to $1/\sqrt{T_\m}$ such that we have
\begin{align}\label{eq:snr_stationary_1}
    \mathrm{SNR}(\omega) &=\,  \sqrt{T_\m}\,\, \mathrm{SNR}_{\mathrm{st}}(\omega, T_\m) \nonumber \\
    &= \frac{\abs{ F_{\mathrm{est}}^{\S}(\omega) }}{\sqrt{ S_{F_{\mathrm{est}}}(\omega)\, }}\, ,
\end{align}
which corresponds to the standard definition of stationary SNR for a steady-state force measurement \cite{Motazedifard2019}.
It is important to note that this rescaling will modify the units of the SNR, making it no longer dimensionless but now with units of $\mathrm{Hz}^{-1/2}$.
A stationary SNR equal to one, $\mathrm{SNR}(\omega) = 1$, is often associated with the minimum force that can be measured using a given sensing protocol \cite{Lucamarini2006, Motazedifard2019}; therefore, from Eq.~\eqref{eq:snr_stationary_1} we can see that $\sqrt{S_{F_{\mathrm{est}}}(\omega)}$ will correspond to the minimum magnitude of the frequency component of the force $\abs{F_{\mathrm{est}}^{\S}(\omega)}$ that can be measured in a given bandwidth.
Hence, the stationary force noise PSD $S_{F_{\mathrm{est}}}(\omega)$ as given in Eq.~\eqref{eq:force_noise_spectrum} will be a good figure of merit for the sensitivity of force measurements in the stationary regime, such that the smaller $S_{F_{\mathrm{est}}}(\omega)$ is, the more sensitive the measurement will be \cite{Wimmer2014,Motazedifard2016}.
As a final step in describing the quantification of force sensitivity for a stationary force measurement, we shall consider the filtering of the force signal in real frequency domain.
To this end, first we must take the Fourier transform of Eqs.~\eqref{eq:f_est} -- \eqref{eq:f_signal_noise}.
Thus, from Eq.~\eqref{eq:f_est}, we have
\begin{align}\label{eq:f_est_omega}
    F_{\mathrm{est}}(\omega) = H(\omega)\, Y_{\out}(\omega),
\end{align}
where $F_{\mathrm{est}}(\omega)$ and $Y_{\out}(\omega)$ are the Fourier transforms of $F_{\mathrm{est}}(\omega)$ and $Y_{\out}(t)$, respectively, while $H(\omega)$ is the frequency response of the linear filter.
It is worth noting that $F_{\mathrm{est}}(\omega)$ corresponds to the estimated frequency component of the force quadrature $\widebar{F}_{\i}(t)$.
Further, following Eqs.~\eqref{eq:nonstationary_measurement} and \eqref{eq:f_signal_noise}, $F_{\mathrm{est}}(\omega)$ may be broken down as
\begin{align}
    F_{\mathrm{est}}(\omega) = F_{\mathrm{est}}^{\S}(\omega) + F_{\mathrm{est}}^{\N}(\omega), \label{eq:stationary_spectral_measurement}
\end{align}
where 
\begin{subequations}
\begin{align}
    &F_{\mathrm{est}}^{\S}(\omega) = H(\omega)\, A(\omega)\,  \widebar{F}_{\i}(\omega), \\
    &F_{\mathrm{est}}^{\N}(\omega) = H(\omega) N(\omega);
\end{align}
\end{subequations}
being $F_{\mathrm{est}}^{\S}(\omega)$, $F_{\mathrm{est}}^{\N}(\omega)$, $A(\omega)$, and $N(\omega)$, the Fourier transforms of $F_{\mathrm{est}}^{\S}(t)$, $F_{\mathrm{est}}^{\N}(t)$, $A(t)$, and $N(t)$, respectively.
Here, as it is standard for stationary linear measurements, we will consider an inverse filter, which frequency response is given by
\begin{align}\label{eq:filter_frequency_response}
    H(\omega) = \frac{1}{A(\omega)}.
\end{align}
Hence, the inverse filter will rescale $Y_{\out}(\omega)$ in such a way that $F_{\mathrm{est}}(\omega)$ will have the same units of $\widebar{F}_{\i}(\omega)$, and $F_{\mathrm{est}}^{\S}(\omega)$ and $F_{\mathrm{est}}^{\N}(\omega)$ will be given by
\begin{subequations}
\begin{align}
    &F_{\mathrm{est}}^{\S}(\omega) = \widebar{F}_{\i}(\omega), \label{eq:f-signal_0} \\
    &F_{\mathrm{est}}^{\N}(\omega) = \frac{N(\omega)}{A(\omega)}. \label{eq:f-noise_0}
\end{align}
\end{subequations}
Therefore, using Eqs.~\eqref{eq:f-signal_0} and \eqref{eq:f-noise_0}, we are able to calculate the stationary force noise PSD as well as as the stationary SNR using Eqs.~\eqref{eq:force_noise_spectrum} and \eqref{eq:snr_stationary_1}, respectively.
\subsection{Signal-to-noise ratio: exponential window}
\label{subsec:snr_output}
The definition of truncated SNR given in Eq.~\eqref{eq:snr-nonstationary_0}, which yields to Eqs.~\eqref{eq:snr-nonstationary} and \eqref{eq:snr_stationary_1}, assumes that the measurement record is truncated using the rectangular window function $\Pi_{T_{\m}}(t)$; however, this is not the best option when analysing experimental data, since the use of rectangular windows reduces the frequency resolution of PSD estimates \cite{Cooper1998}.
In addition, a rectangular window function is not convenient either to obtain simple analytical results in the nonstationary regime, which is possible with other window functions as we will see below.
Therefore, in order to consider a more convenient window function in the definition of SNR we introduce the windowed Fourier transform
\begin{align}
    \mathcal{F}_{w} \{ \O(t) \} = \int_{-\infty}^{+\infty} \dd{t} \e^{i \omega t} w(t)\, \O(t),
\end{align}
where $w(t)$ is a window function.
This windowed Fourier transform will replace the truncated Fourier transform in the definition of signal and noise in Eqs.~\eqref{eq:signal_def} and \eqref{eq:noise_def}, respectively, allowing us to define what we will call below the \emph{nonstationary SNR}.
Using the windowed Fourier transform it is possible to obtain a windowed version of the PSD in Eq.~\eqref{eq:truncated_psd}, which will correspond to the classical definition of \emph{modified periodogram PSD estimator} \cite{Prabhu2014}. 
Here, $w(t)$ must satisfy the condition
\begin{align}
    \frac{1}{T_{\m}} \int_{-\infty}^{+\infty} \dd{t} \abs{w(t)}^2 = 1,
\end{align}
which guarantees that in the stationary regime the estimated average power is the same as that in the signal.
It is important to emphasise that this normalisation condition is already satisfied by $\Pi_{T_{\m}}(t)$.
Further, the chosen window must ensure that the PSD estimate is asymptotically unbiased, i.e., that in the limit of infinite measurement time ($T_{\m} \to \infty$) it reduces to the stationary PSD satisfying the Wiener-Khinchin theorem (see Appendix \ref{app:wiener-khinchin}).
On the other hand, the definition of truncated SNR in Eq.~\eqref{eq:snr-nonstationary_0} relies on the existence of a filter that allows us to represent the quantum  estimator as described in Eq.~\eqref{eq:nonstationary_measurement}.
This assumption implies that we must explicitly consider some filter in order to calculate the SNR.
A standard approach to filter the output signal of a linear measurement is to apply an inverse filter in time or frequency domain to the complete measurement record.
Unfortunately, this procedure is not suitable for the estimation of nonstationary signals in the transient regime (see Appendix \ref{app:inverse_filter}), ability that we identify as the nonstationary operation of the transducer.
However, it is possible to follow a similar procedure but in complex frequency domain, such that one is able to describe the nonstationary measurement of impulsive forces.
Thus, for the purpose of filtering the signal in the nonstationary transient regime, we will use a one-sided decaying exponential window function
\begin{align}\label{eq:exponential_window}
    w_{T_{\m}}(t) = \e^{-t/2 T_{\m}} \theta(t), 
\end{align}
so that the windowed Fourier transform may be written as a Laplace transform:
\begin{align}
    \mathcal{F}_{w}\{ \O(t) \} \equiv \mathcal{L}\{ \O(t) \} = \int_{0}^{+\infty} \dd{t} \e^{- s t} \O(t),
\end{align}
where the complex variable $s$ is given by
\begin{align}
    s=-i\omega + 1/2T_{\m}.
\end{align}
Further, if we take the Laplace transform of Eq.~\eqref{eq:yout_snr} and we take into account the convolution theorem, we will have
\begin{align}
    Y_{\out}(s) =\,  &A(s)\, \widebar{F}_{\i}(s) + N(s),
\end{align}
where $Y_{\out}(s)$, $A(s)$, $\widebar{F}_{\i}(s)$ and $N(s)$
are the Laplace transforms of $Y_{\out}(t)$, $A(t)$, $\widebar{F}_{\i}(t)$ and $N(t)$; respectively. 
Hence, we can use an inverse filter with transfer function
\begin{align}
    H(s) = \frac{1}{A(s)},
\end{align}
to obtain
\begin{align}
    F_{\mathrm{est}}(s) = F_{\mathrm{est}}^{\S}(s) + F_{\mathrm{est}}^{\N}(s);
\end{align}
where estimator, signal, and noise, are given by
\begin{subequations}
\begin{align}
    &F_{\mathrm{est}}(s) = \frac{Y_{\out}(s)}{A(s)}, \\
    &F_{\mathrm{est}}^{\S}(s) = \widebar{F}_{\i}(s), \\
    &F_{\mathrm{est}}^{\N}(s) = \frac{N(s)}{A(s)}. \label{eq:f_noise-nonstationary}
\end{align}
\end{subequations}
Now, we will use $F_{\mathrm{est}}^{\S}(s)$ and $F_{\mathrm{est}}^{\N}(s)$ to replace the truncated Fourier transforms of $F_{\mathrm{est}}^{\S}(t)$ and $F_{\mathrm{est}}^{\N}(t)$ in the definitions of signal and noise in Eqs.~\eqref{eq:signal_def} and \eqref{eq:noise_def} and, accordingly, in the SNR in Eq.~\eqref{eq:snr-nonstationary}.
Further, we will drop the $s$ notation and we will refer to $\omega$ and $T_{\m}$ explicitly.
Hence, we define the nonstationary SNR as
\begin{align}\label{eq:snr-nonstationary-laplace_0}
    \mathrm{SNR}(\omega, T_{\m}) =\, \frac{ \mathcal{S}(\omega, T_{\m}) }{ \mathcal{N}(\omega, T_{\m}) };
\end{align}
where $\mathcal{S}(\omega, T_{\m})$ is the nonstationary signal while $\mathcal{N}(\omega, T_{\m})$ is the nonstationary noise, which are given by, 
\begin{subequations}\label{eq:signal-noise-laplace}
\begin{align}
    \mathcal{S}(\omega, T_{\m}) &=\, \abs{  \widebar{F}_{\i}(-i \omega + 1/2 T_{\m}) }, \label{eq:signal-laplace} \\
    \mathcal{N}(\omega, T_{\m}) &=\, \sqrt{\, T_{\m}\, S_{F_{\mathrm{est}}}(\omega, T_{\m})  }\,. \label{eq:noise-laplace}
\end{align}
\end{subequations}
Here, $S_{F_{\mathrm{est}}}(\omega, T_{\m})$ is the nonstationary force noise PSD given by
\begin{align}\label{eq:imp:nonstationary_psd_definition}
    &S_{F_{\mathrm{est}}}(\omega, T_{\m}) = \nonumber \\
    & \hspace{0.75cm} \frac{1}{T_{\m}} \big\langle F_{\mathrm{est}}^{\N \dagger}(-i \omega + 1/2 T_{\m})\, F_{\mathrm{est}}^{\N}(-i \omega + 1/2 T_{\m}) \big\rangle,
\end{align}
which, as is shown in Appendix \ref{app:wiener-khinchin}, satisfies the Wiener-Khinchin theorem in the stationary regime.
We included the explicit dependence on $T_{\m}$ in the nonstationary force noise PSD in Eq.~\eqref{eq:imp:nonstationary_psd_definition} in order to emphasise its nonstationary nature.
Therefore, explicitly, the nonstationary SNR will be given by
\begin{align}\label{eq:snr-nonstationary-laplace}
    \mathrm{SNR}(\omega, T_{\m}) =\, \frac{ \abs{  \widebar{F}_{\i}(-i \omega + 1/2 T_{\m}) } }{ \sqrt{\, T_{\m}\, S_{F_{\mathrm{est}}}(\omega, T_{\m})  }\, }\,.
\end{align}
This expression will allow us calculate analytically the nonstationary SNR in a relatively simple and straightforward manner.
It is important to note that Eq.~\eqref{eq:snr-nonstationary-laplace} is equivalent to the expression that is presented without justification in Refs.~\cite{Vitali2001, *Vitali2001Erratum} and~\cite{Vitali2002, *Vitali2002Erratum}.
The deduction that we presented here corresponds to one of the main results of this work.
In the deduction of Eq.~\eqref{eq:snr-nonstationary-laplace}, the inclusion of the exponential window function $w_{T_{\m}}(t)$  may be seen as an approximation that is made solely for analytical convenience.
However, more than an approximation it is included as part of a technique to estimate the SNR for finite measurement times.
A technique that may be implemented in an experimental scenario as well.
In order to explore this possibility, it is useful to consider the performance of the exponential window function $w_{T_{\m}}(t)$, as defined in Eq.~\eqref{eq:exponential_window}, when calculating PSD estimates in the presence of broadband white noise.
To quantify this performance it is standard to use the equivalent noise bandwidth (ENBW), which is defined as the bandwidth of an ideal filter (with rectangular frequency response) that would pass the same average power as the window of interest when each is driven by stationary random classical noise \cite{Prabhu2014}.
Thus, the ENBW will be given by (normalised to $1/T_{\m}$),
\begin{align}
    \mathrm{ENBW} = \frac{ \frac{1}{T_{\m}} \int_{-\infty}^{+\infty} \dd{t} \abs{w(t)}^2 }{ \abs{ \frac{1}{T_{\m}} \int_{-\infty}^{+\infty} \dd{t} w(t) }^2 };
\end{align}
where the smaller the ENBW, the better the performance of the window in the presence of broadband noise.
Therefore, it is easy to see that for a rectangular window function $\Pi_{T_{\m}}(t)$, $\mathrm{ENBW}=1$; while for the exponential window function $w_{T_{\m}}(t)$, $\mathrm{ENBW}=1/4$.
This proves that the smoothing effect of $w_{T_{\m}}(t)$ on the measurement record increases its performance in comparison to $\Pi_{T_{\m}}(t)$, which is a desirable effect when analysing experimental data.
% 
%------------------------------------------------------------------------------
% Stationary force sensing
%
\section{Stationary force sensing}
\label{sec:stationary_force_sensing}
\begin{figure}[t]
    %\hspace{0.7cm}
    \centering
    \includegraphics[width=0.48\textwidth]{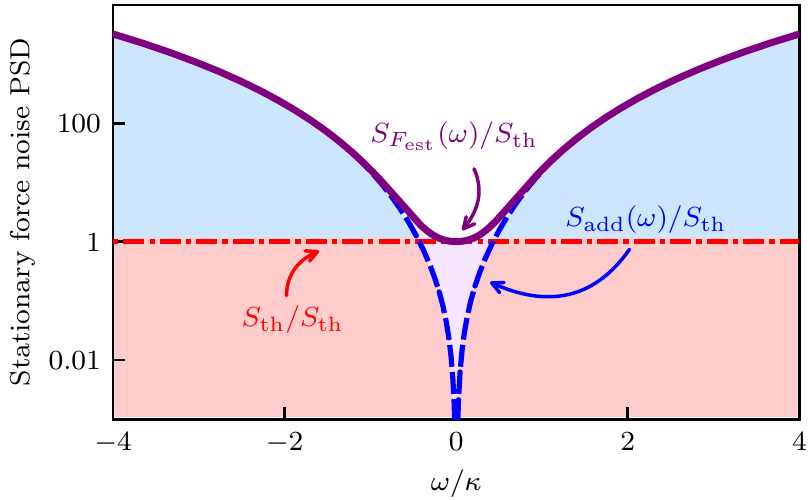}
    \vspace{-0.5cm}
    \caption{Contributions to the stationary force noise PSD, as described in Eqs. \eqref{eq:total_noise}--\eqref{eq:thermal_noise},  scaled by the thermal noise floor. The blue dashed line represents the added force noise PSD $S_{\mathrm{add}}(\omega) = S_{\mathrm{imp}}(\omega) + S_{\mathrm{rp}} + S_{\mathrm{imp-rp}}(\omega)$, which is the joint contribution of imprecision and back-action noise; the red dot-dashed line corresponds to the thermal noise floor $S_{\mathrm{th}}$; and the continuous purple line is the stationary force noise PSD $S_{F_{\mathrm{est}}}(\omega)$. As mentioned in the main text, $S_{\mathrm{add}}(0)$ can reach zero for appropriate sets of parameters, leaving thermal noise as the ultimate limit for stationary force sensing. The parameters used here were, $G_-/\kappa = G_+/\kappa =  1$, $\gamma/\kappa = 10^{-4}$. \label{fig:force_noise_1}}
\end{figure}
\begin{figure}[t]
    \centering
    \includegraphics[width=0.46\textwidth]{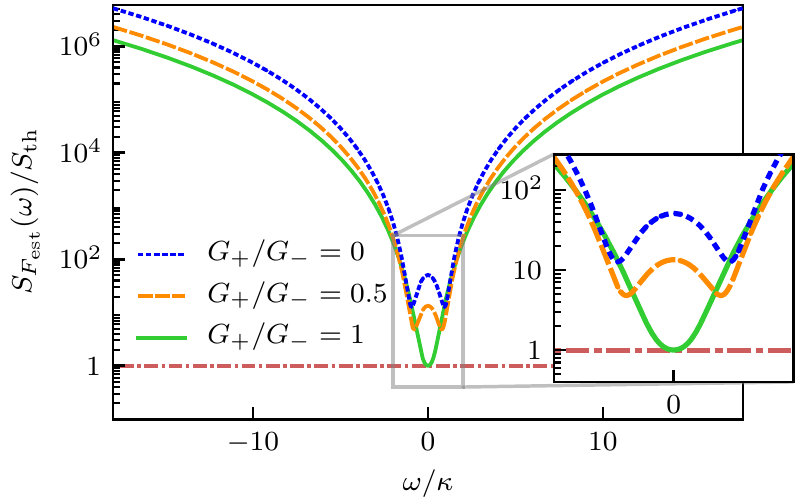}
    \vspace{-0.3cm}
    \caption{Stationary force noise PSD as given by Eq.~\eqref{eq:force_spectrum_2}, scaled by the thermal noise floor, for different drive asymmetries $G_+/G_-$. Here, $G_+$ was tuned to obtain each curve, whilst $G_-/\kappa = 1$ and $\gamma/\kappa = 10^{-4}$. For the chosen parameters, $C_- = 4 \times 10^4 $, where the BAE measurement ($G_+/G_-=1$) gives lowest added force noise in the stationary regime.}
    \label{fig:force_noise_2}
\end{figure}
\begin{figure}[t]
    \centering
    \includegraphics[width=0.45\textwidth]{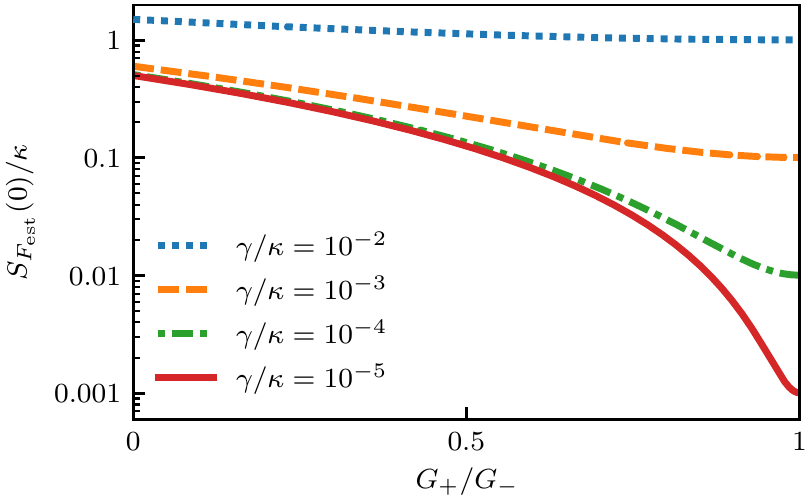}
    \vspace{-0.3cm}
    \caption{Resonant stationary force noise PSD $S_{F_{\mathrm{est}}}(0)$ in units of $\kappa$, as a function of drive asymmetry $G_+/G_-$ for different ratios of the dissipation rates $\gamma/\kappa$.  The curves were obtained making $\omega=0$ in Eq.~\eqref{eq:force_spectrum_2}. Here $G_-$ remained fixed at $G_-/\kappa = 1$ and $\bar{n}_{\mathrm{th}} = 100$. For each $\gamma$, the resonant stationary force noise PSD reaches its minimum when $G_+/G_-=1$. As expected, the sensitivity of the measurement on resonance increases when the mechanical dissipation decreases.}
    \label{fig:force_noise_3}
\end{figure}
\begin{figure}[t]
    \bigskip
    \centering
    \includegraphics[width=0.48\textwidth]{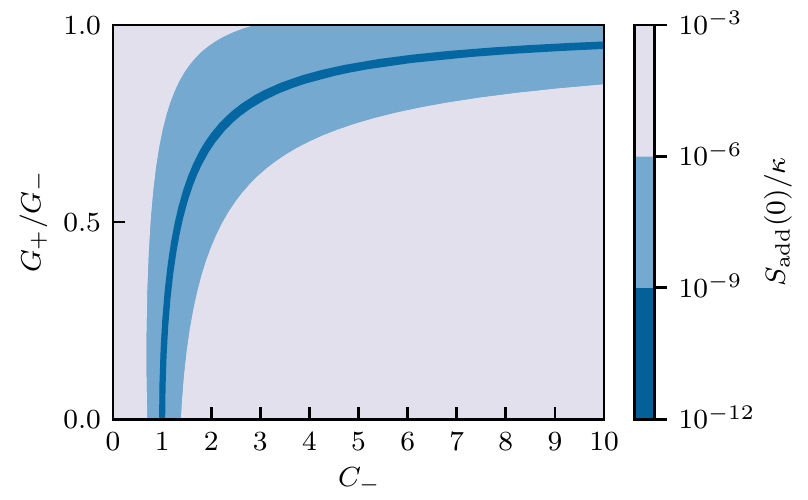}
    \vspace{-0.6cm}
    \caption{Added force noise PSD at resonance $S_{\mathrm{add}}(0)$ in units of $\kappa$, as a function of the cooperativity $C_-$ and drive asymmetry $G_+/G_-$ as given in Eq.~\eqref{eq:added_psd}. The added force noise PSD is given by $S_{\mathrm{add}}(\omega) = S_{F_{\mathrm{est}}}(\omega) - S_{\mathrm{th}}$. In the limit $C_- \to \infty$, the minimum added force noise PSD corresponds to $G_+/G_-=1$; however, for $C_- \sim 1$, the minimum occurs for $0 \leq G_+/G_- < 1$, corresponding to $G_+/G_-=0$ for $C_-=1$. Here we considered $\gamma/\kappa = 10^{-4}$.}
    \label{fig:optimal_drive_asymmetry}
\end{figure}

In this section we shall focus on force measurements in the stationary regime, where the system dynamics is time-invariant in the rotating frame under consideration and a description in real frequency domain is sufficient.
The proposal to use two-tone driving of an electromagnetic cavity coupled to a mechanical oscillator in order to perform an optimal measurement of a single-quadrature of the mechanical motion and, consequently, perform an optimal measurement of a single-quadrature of an external classical force, was first suggested in 1980 by Braginsky et al. \cite{Braginsky1980}.
This early proposal relied on the idea of making a BAE measurement of the mechanical quadrature of interest, which implies that the back-action due to the measurement is redirected to the unmeasured canonical conjugate quadrature.
Almost 30 years later, this idea was brought into the context of cavity quantum optomechanics in Ref. \cite{Clerk2008}, where a fully quantum description of the steady-state BAE measurement of a single-quadrature of the mechanical motion was made.
This BAE scheme becomes evident in Eqs. \eqref{eq:system_dynamics} if we consider $G_+=G_-$, which can be achieved through the appropriate manipulation of the powers of the input drives.
However, despite it seems to be the most obvious approach for the ultrasensitive sensing of weak forces in the stationary regime, there are limits for which a BAE measurement is not the best option for the enhancement of the sensitivity of force measurements, as we will see below.
Thus, here we study a more general scenario, where in general there is an asymmetry between the coupling constants $G_+$ and $G_-$, and a BAE measurement is just a particular case.  
In Sec. \ref{sec:signal-to-noise_ratio} we saw that the sensitivity of a stationary force measurement is well quantified by the stationary force noise PSD $S_{F_{\mathrm{est}}}(\omega)$ as given by Eq.~\eqref{eq:force_noise_spectrum}.
Therefore, considering the measurement of the $\widebar{F}_{\i}(\omega)$ force quadrature through the output electromagnetic quadrature $Y_{\out}(\omega)$, we shall determine the quantum noise process $F_{\mathrm{est}}^{\N}(\omega)$ using Eq.~\eqref{eq:f-noise_0} with the ultimate goal of calculating the figure of merit $S_{F_{\mathrm{est}}}(\omega)$.
We will do this using the frequency domain representations of the input-output relation in Eq.~\eqref{eq:input-output_Y} and the Heisenberg Langevin Eqs.~\eqref{eq:system_dynamics}, in such a way that we arrive at an expression that explicitly shows the estimation of the force quadrature $\widebar{F}_{\i}(\omega)$.
A description of the measurement of $\widebar{F}_{\r}(\omega)$ through the output electromagnetic quadrature $X_{\out}(\omega)$ would require following a procedure completely analogous to the one presented here.
Thus, from the input-output relation in Eq.~\eqref{eq:input-output_Y}, the output signal in frequency domain $Y_{\out}(\omega)$ will be given by
\begin{align}\label{eq:yout_omega}
    Y_{\out}(\omega) = \sqrt{\kappa}\, Y(\omega) - Y_{\in}(\omega),
\end{align}
where $Y(\omega)$ can be determined from the Fourier transform of the Heisenberg-Langevin Eqs.~\eqref{eq:system_dynamics}.
This yields the following coupled equations,
\begin{subequations}\label{eq:fourier}
  \begin{align}
        Y(\omega) &=\, \chi_{\c}(\omega)\, \Big[\pqty{G_- + G_+}\, Q(\omega) + \sqrt{\kappa}\, Y_{\in}(\omega) \Big], \label{eq:fourier:a} \\
        Q(\omega) &= -\chi_{\m}(\omega)\,  \Big[ (G_- - G_+)\, Y(\omega) +  \widebar{F}_{\i}(\omega) \nonumber \\ 
        & \hspace{4.52cm} + \widebar{\W}_{\i}(\omega)  \Big], \label{eq:fourier:b}
    \end{align} 
\end{subequations}
where $\chi_{\c}(\omega) = (-i \omega + \kappa/2 )^{-1}$ is the susceptibility of the electromagnetic mode, and $\chi_{\m}(\omega) =  (-i \omega + \gamma/2)^{-1}$ is the mechanical susceptibility.
Combining Eqs. \eqref{eq:fourier:a} and \eqref{eq:fourier:b}, we obtain
\begin{align}\label{eq:y_omega}
    Y(\omega) = -\D(\omega) \Big\{ &\pqty{G_- + G_+}\, \bqty{ \widebar{F}_{\i}(\omega) + \widebar{\W}_{\i}(\omega) } \nonumber \\
    & \hspace{1cm} + \sqrt{\kappa}\, (i \omega - \gamma/2)\, Y_{\in}(\omega) \Big\},
\end{align}
where
\begin{align}\label{eq:d_omega}
    \D(\omega) = \big[G_-^2 - G_+^2 + (i \omega - \gamma/2 )\, (i \omega - \kappa/2 )\big]^{-1}.
\end{align}
Therefore, substituting Eq.~\eqref{eq:y_omega} into \eqref{eq:yout_omega} leads to
\begin{align}\label{eq:out_phase_2}
    Y_{\out}(\omega) =&\,  A(\omega)\, \widebar{F}_{\i}(\omega) + N(\omega),
\end{align}
where the signal amplification $A(\omega)$ is given by
\begin{align}\label{eq:signal_amplification}
    A(\omega) = - \sqrt{\kappa}\, (G_- + G_+)\, \D(\omega),
\end{align}
and the measurement noise $N(\omega)$ is
\begin{align}\label{eq:measurement_noise}
    N(\omega) =&\,  A(\omega)\, \widebar{\W}_{\i}(\omega) -  Y_{\in}(\omega) \nonumber \\
    & +  \frac{\sqrt{\kappa}}{(G_- + G_+)}\, A(\omega)\, \pqty{i \omega - \gamma/2} \, Y_{\in}(\omega).
\end{align}
Now, to estimate $\widebar{F}_{\i}(\omega)$ we apply the output signal $Y_{\out}(\omega)$ to an inverse filter as described in Sec. \ref{sec:signal-to-noise_ratio}, and we use Eq.~\eqref{eq:f-noise_0} to determine $F_{\mathrm{est}}^{\N}(\omega)$.
Hence, we have
\begin{align}\label{eq:force_noise}
    &F_{\mathrm{est}}^{\N}(\omega) =\, \frac{ (i \omega - \gamma/2)\, (i \omega + \kappa/2) }{ \sqrt{\kappa}\, (G_- + G_+)}\, Y_{\in}(\omega) \nonumber \\
    & \hspace{2.75cm} + \frac{G_- - G_+}{\sqrt{\kappa}}\, Y_{\in}(\omega) + \widebar{\W}_{\i}(\omega).
\end{align}
This expression shows three different contributions to $F_{\mathrm{est}}^{\N}(\omega)$, each with different scalings with respect to the effective optomechanical coupling rates $G_{\pm}$, which in turn relate to the power of the coherent drives $\wp_{\pm}$ as discussed in Appendix \ref{app:derivation_hamiltonian}.
Thus, the first term in Eq.~\eqref{eq:force_noise} represents the measurement imprecision noise which is inversely proportional to $\sqrt{\wp_{\pm}}$, the second term describes radiation pressure noise which is proportional to $\sqrt{\wp_{\pm}}$, and the third term corresponds to mechanical thermal and quantum fluctuations which are independent of the input powers $\wp_{\pm}$.
Furthermore, in order to evaluate $S_{F_{\mathrm{est}}}(\omega)$ using $F_{\mathrm{est}}^{\N}(\omega)$ in Eq.~\eqref{eq:force_noise}, we need the correlation functions associated with $Y_{\in}(\omega)$ and $\widebar{\W}_{\i}(\omega)$, which can be obtained by taking the Fourier transform of the correlation functions involving $Y_{\in}(t)$ and $\widebar{\W}_{\i}(t)$ in Eqs. \eqref{eq:xy_corr} and \eqref{eq:langevin_corr_1:a}, respectively [see Appendix \ref{app:derivation_hamiltonian} for details on the calculation of the frequency correlation function involving $\widebar{\W}_{\i}(t)$].
Therefore, we have
\begin{subequations}
\begin{align}
    & \expval{ Y_{\in}(\omega') Y_{\in}(\omega) } = \pi\, \delta(\omega' + \omega), \\
    &\expval{ \widebar{\W}_{\i}(\omega') \widebar{\W}_{\i}(\omega) }  =  2 \pi \gamma\, \pqty{\widebar{n}_{\mathrm{th}} + 1/2 }\, \delta(\omega' + \omega); \label{eq:fourier_langevin_force}
\end{align}
\end{subequations}
where  $ \widebar{n}_{\mathrm{th}} = ( \e^{\hbar \omega_\m/k_{\B} T } -1  )^{-1}$ corresponds to the mean number of thermal phonons in the reservoir.
Finally, in the following we present the stationary force noise PSD, which is the main result of this section.
As before, we distinguish the different contributions according to their dependence on the powers of the input coherent drives.
Thus, we may write $S_{F_{\mathrm{est}}}(\omega)$ as
\begin{align}\label{eq:total_noise}
    S_{F_{\mathrm{est}}}(\omega) = S_{\mathrm{imp}}(\omega) + S_{\mathrm{rp}} +
    S_{\mathrm{imp-rp}}(\omega) + S_{\mathrm{th}},
\end{align}
where the imprecision noise corresponds to
\begin{equation}
    S_{\mathrm{imp}}(\omega) = \frac{ \pqty{\omega^2 + \gamma^2/4} \pqty{\omega^2 + \kappa^2/4} }{2 \kappa\, \pqty{G_- + G_+}^2},
\end{equation}
the radiation-pressure noise contribution is
\begin{equation}
    S_{\mathrm{rp}} = \frac{\pqty{G_- - G_+}^2}{2 \kappa},
\end{equation}
the cross-correlation between imprecision and back-action noises is given by
\begin{equation}
     S_{\mathrm{imp-rp}}(\omega) =- \frac{\pqty{ \omega^2+ \gamma \kappa/4 }\, \pqty{G_- - G_+}}{\kappa\, \pqty{G_- + G_+}},
\end{equation}
and
\begin{equation}\label{eq:thermal_noise}
    S_{\mathrm{th}} = \gamma\, ( \widebar{n}_{\mathrm{th}} + 1/2)
\end{equation}
is the thermal noise floor associated with the thermal fluctuations of the oscillator.
Explicitly, putting all contributions together, we have
\begin{align}\label{eq:force_spectrum_2}
    S_{F_{\mathrm{est}}}(\omega) &= \frac{ \pqty{\omega^2 + \gamma^2/4}\, \pqty{\omega^2 + \kappa^2/4} }{ 2 \kappa\, \pqty{G_- + G_+}^2}  + \frac{\pqty{G_- - G_+}^2}{2 \kappa}  \nonumber \\
    & - \frac{\pqty{ \omega^2+ \gamma \kappa/4 }\, \pqty{G_- - G_+}}{\kappa\, \pqty{G_- + G_+}} + \gamma\, ( \widebar{n}_{\mathrm{th}} + 1/2).
\end{align}
This stationary force noise PSD and its fundamental components, scaled by the thermal noise floor, are shown in Fig. \ref{fig:force_noise_1}.
Note that since $F_{\mathrm{est}}^{\N}(\omega)$ is dimensionless,  $S_{F_{\mathrm{est}}}(\omega)$ will have units of $\mathrm{Hz}$, then, in order to describe the force sensitivity in $\mathrm{N^2 Hz^{-1}}$, as it is commonly done, we have to multiply the force noise spectrum by $(\sqrt{2}\, p_{\zpf})^2 = \hbar m \omega_{\m}$, such that $S_{f_{\mathrm{est}}}(\omega) = \hbar m \omega_{\m} S_{F_{\mathrm{est}}}(\omega)$.
In Fig. \ref{fig:force_noise_2}, we represent graphically the stationary force noise PSD, as given by Eq.~\eqref{eq:force_spectrum_2}, scaled by the thermal noise floor [$S_{F_{\mathrm{est}}}(\omega)/S_{\mathrm{th}}$] as a function of the dimensionless frequency $\omega/\kappa$ for different values of the drive asymmetry $G_+/G_-$.
We found that as a consequence of the mutual cancellation of the noise contributions due to imprecision and radiation-pressure,  the proposed two-tone driving scheme allows one to reduce the force noise PSD to thermal noise at resonance, i.e., $S_{F_{\mathrm{est}}}(0) = S_{\mathrm{th}}$ and $S_{\mathrm{add}}(0) = 0$. 
Further, in Fig. \ref{fig:force_noise_3}, we analyse the  behaviour of the resonant force noise PSD $S_{F_{\mathrm{est}}}(0)$ as a function of the drive asymmetry $G_+/G_-$ for different values of the mechanical dissipation rate $\gamma$. As expected, a lower dissipation rate will reduce the noise present in the force sensing and, therefore, will improve the sensitivity of the measurement.
The results shown in Figs. \ref{fig:force_noise_2} and \ref{fig:force_noise_3} correspond to a regime of parameters for which a BAE measurement ($G_+/G_-=1$) is the best approach to enhance the sensitivity of a stationary force measurement, however, this is not always the case.
To prove this assertion, we shall use the added force noise PSD $S_{\mathrm{add}}(\omega)$, which we define as the sum of the contributions due to the measurement
\begin{align}\label{eq:added_noise}
    S_{\mathrm{add}}(\omega) = S_{\mathrm{imp}}(\omega)+ S_{\mathrm{rp}} + S_{\mathrm{imp-rp}}(\omega),
\end{align}
such that the stationary force noise PSD will be given by
\begin{align}
    S_{F_{\mathrm{est}}}(\omega) = S_{\mathrm{add}}(\omega) + S_{\mathrm{th}}.
\end{align}
Thus, we shall consider the added force noise PSD at resonance $S_{\mathrm{add}}(0)$, which written in terms of the cooperativity of the red sideband drive $C_- = 4 G_-^2/\gamma \kappa$ and the drive asymmetry $G_+/G_-$, will be given by
\begin{align}\label{eq:added_psd}
    S_{\mathrm{add}}(0) &= \frac{\gamma}{8}\, \Bigg\{ \frac{1}{C_-\, \pqty{1+G_+/G_-}^2} + C_-\, \pqty{1-G_+/G_-}^2  \nonumber \\
    & \hspace{3.2cm} - \frac{ 2\, \pqty{1-G_+/G_-}}{\pqty{1+G_+/G_-}} \Bigg\}.
\end{align}
This expression will allow us to find the optimal conditions for the reduction of the force noise PSD to the thermal noise floor, as is shown in Fig. \ref{fig:optimal_drive_asymmetry}.
From Eq.~\eqref{eq:added_psd}, we can see that given $G_+/G_-=1$, $S_{\mathrm{add}}(0) = 0$ for $C_- \to \infty$; however, from $C_- = 10$ the optimal drive asymmetry is close enough to $G_+/G_-=1$.
Thus, we can say that the optimal drive asymmetry configuration for $C_- \geq 10$ corresponds to $G_+/G_- \approx 1$.
This result will again be relevant in the next section to establish the optimal configuration for nonstationary force measurements.
In Appendix \ref{app:stationary_psd} we do a more thorough analysis of the different regimes defined by the drive asymmetry $G_+/G_-$, and we consider the conditions under which the stationary force noise PSD $S_{F_{\mathrm{est}}}(\omega)$ reduces to the thermal noise floor at resonance. 
This conditions define the optimal configuration for the realisation of stationary force measurements in each regime.
%

% 
%------------------------------------------------------------------------------
% Nonstationary force sensing
%
\section{Nonstationary force sensing}
\label{sec:nonstationary_force_sensing}
It is often the case in force sensing experiments that the applied force is impulsive, and the measurement is necessarily nonstationary.
Further, since nonstationary measurements depend on the initial state of the system, the careful manipulation of the initial conditions can lead to an improvement in the sensitivity of the force sensor.
Therefore, we now consider a nonstationary protocol that involves non-thermal state preparation followed by a finite time measurement (see Fig. \ref{fig:nonstationary_strategy}).
Thus, first the mechanical oscillator is prepared in a dissipative squeezed state, then, upon arrival of the impulsive force, the measurement is performed before the re-thermalisation of the mechanical oscillator takes place.
This protocol allows one to use different drive asymmetry configurations at the two stages of the measurement process, one for state preparation and another for force measurement.
However, as we will show below, it is not necessary to change the drive asymmetry in order to improve sensitivity beyond what can be achieved with a stationary force measurement.
A nonstationary strategy similar to the one discussed here was presented in Refs.~\cite{Vitali2001, *Vitali2001Erratum} and~\cite{Vitali2002, *Vitali2002Erratum}, where it was introduced as a technique to improve the sensitivity of force measurements in the presence of feedback cooling schemes.
Furthermore, in this context of feedback cooling, it was initially shown in Ref.~\cite{Harris2013} and then experimentally demonstrated for the nonstationary strategy in Ref.~\cite{Hosseini2014}, that the effect of state preparation can be reproduced through an estimation procedure.
However, estimation methods require a precise knowledge of the parameters of the system and the system dynamics, and can be computationally expensive~\cite{Hosseini2014}.
In the following, first, we consider the time-dependent dynamics of the system in order to determine the explicit relationship between classical force and output electromagnetic field.
Second, we quantify the noise present in the measurement using the nonstationary force noise PSD.
Next, we study the preparation of the initial state of the mechanical oscillator in a dissipative squeezed state.
Finally, considering an impulsive Dirac delta force, we calculate and analyse the nonstationary SNR and we establish its relationship with the initial squeezed state of the mechanical oscillator.
Since we are interested in studying the sensitivity of nonstationary force measurements,  we shall use the expressions for nonstationary signal and noise in Eqs.~\eqref{eq:signal-laplace} and \eqref{eq:noise-laplace}, respectively, together with the nonstationary SNR defined in Eq.~\eqref{eq:snr-nonstationary-laplace}.
\subsection{Time-dependent dynamics of the force sensor}
In this subsection we solve the dynamics of the output electromagnetic field, the result obtained does not differ significantly from the well-known dynamics of the electromagnetic field in a canonical optomechanical system and therefore this calculation can be skipped by the experienced reader.
As stated before, to estimate the force signal $\widebar{F}_{\i}(t)$  we need to evaluate the quadrature of the output electromagnetic field $Y_{\out}(t)$.
To this end, we recall the input-output relation in Eq \eqref{eq:input-output_Y}, from which we get
\begin{align}\label{eq:input-output_time}
    Y_{\out}(t) = \sqrt{\kappa}\, Y(t) - Y_{\in}(t).
\end{align}
Therefore, we need to determine the dynamics of $Y(t)$, which can be obtained from the Heisenberg-Langevin Eqs.~\eqref{eq:system_dynamics}.
Thus, decoupling Eqs.~\eqref{eq:system_dynamics} we have that the equation of motion for the $Y(t)$ quadrature is given by
\begin{align}\label{eq:ddho}
    \ddot{Y} + 2 \Gamma\, \dot{Y}  + \Omega^2\, Y = \xi_{_Y},
\end{align}
where,
\begin{subequations}
\begin{align}
    &2 \Gamma = \gamma/2 + \kappa/2, \label{eq:gamma}\\
    &\Omega^2 =  G_-^2 - G_+^2 + \gamma \kappa/4.
\end{align}
\end{subequations}
Further, the inhomogeneity $\xi_{_Y} = \xi_{_Y}(t)$ is
\begin{align}\label{eq:inhomogeneity_y}
    \xi_{_Y}(t) =& - ( G_- + G_+ )\, \Big[ \widebar{F}_{\i}(t) + \widebar{\W}_{\i}(t) \Big]  \nonumber \\
    & \hspace{1.75cm} +\, \sqrt{\kappa}\, \Big[ \dot{Y}_{\in}(t) + (\gamma /2)\, Y_{\in}(t)  \Big].
\end{align}
Eq.~\eqref{eq:ddho} corresponds to the dynamical equation of a driven damped harmonic oscillator, and since it is a linear differential equation,  the complete solution for the dynamics of $Y(t)$ may be written as
\begin{align}\label{eq:y_solution}
    Y(t) =&\, Y_{\p}(t) +  Y_{\h}(t),
\end{align}
where $Y_{\p}(t)$ is the particular solution while $Y_{\h}(t)$ is the solution to the corresponding homogeneous problem.
The particular solution $Y_{\p}(t)$ will be given by
\begin{align}\label{eq:y_p}
    Y_{\p}(t) = \D(t) * \xi_{_Y}(t) = \int_{0}^{t} \dd{t'} \D(t')\, \xi_{_Y}(t-t'), 
\end{align}
where $\D(t)$ is the classical Green's function of a driven damped harmonic oscillator, defined  as the solution to
\begin{align}\label{eq:greens-function-equation}
    \ddot{\D} + 2 \Gamma\, \dot{\D} + \Omega^2 \D = \delta(t).
\end{align}
Note that by calculating the Fourier transform of the latter equation, we can realise that $\D(t)$ corresponds to the inverse Fourier transform of $\D(\omega)$, which was previously defined in Eq.~\eqref{eq:d_omega} and may be written as $\D(\omega) = ( -\omega^2 - 2 i \omega\, \Gamma + \Omega^2 )^{-1}$.
To solve Eq.~\eqref{eq:greens-function-equation} and determine $\D(t)$ we use the discriminant of the associated homogeneous differential equation, $\Delta = \Omega^2 - \Gamma^2 = G_-^2 - G_+^2 - [(\gamma - \kappa)/4]^2$, which allows us to distinguish among three different responses of the oscillator: $\Delta > 0$ (under-damping), $\Delta < 0$ (over-damping), and $\Delta = 0$ (critical damping).
Thus, the (retarded) Green's function will be classified in three cases:
\begin{align}\label{eq:green_function}
    \D(t) = \theta(t)\, \e^{-\Gamma t} \times
    \begin{dcases}
        \sin{\pqty{ \sqrt{\Delta\,}\, t\,}} / \sqrt{\Delta}, & \mathrm{if}\,\,\, \Delta > 0 \\
        \sinh{\pqty{ \sqrt{-\Delta\,}\, t\,}} /\sqrt{-\Delta\,}, & \mathrm{if}\,\,\, \Delta < 0 \\
        \, \, t, & \mathrm{if} \,\,\, \Delta = 0.
    \end{dcases}\,\,\, 
\end{align}
On the other hand, the homogeneous solution to Eq.~\eqref{eq:ddho} is given by
\begin{align}\label{eq:y_h}
    Y_{\h}(t) = (G_- + G_+) \D(t)\, Q(0) + \K(t)\, Y(0),
\end{align}
where we used the relationship $\dot{Y}(0)=(-\kappa/2)\, Y(0) + (G_-+G_+)\, Q(0)$, which was obtained from the Heisenberg-Langevin Eqs.~\eqref{eq:system_dynamics}.
Further, $\K(t)$ is given by
\begin{align}
    \K(t) = \theta(t)\, \e^{-\Gamma t} \times
    \begin{dcases}
        \frac{\gamma - \kappa}{4 \, \sqrt{\Delta}}\, \sin{\pqty{ \sqrt{\Delta}\, t}}  & \\
        \hspace{0.65cm}  + \cos{\pqty{\sqrt{\Delta}\, t}}, & \mathrm{if}\,\,\, \Delta > 0 \\
        \frac{\gamma - \kappa}{4 \, \sqrt{-\Delta}}\, \sinh{\pqty{ \sqrt{-\Delta}\, t}} & \\
        \hspace{0.95cm} + \cosh{\pqty{\sqrt{-\Delta}\, t}}, & \mathrm{if}\,\,\, \Delta < 0 \\
        \,\, \frac{\gamma - \kappa}{4 }\,t  + 1 , & \mathrm{if} \,\,\, \Delta = 0;
    \end{dcases}\,\,\,
\end{align}
which for $t>0$ may be expressed compactly as
\begin{align}\label{eq:k_t}
    \K(t) = \dot{\D}(t) + \frac{\gamma}{2} \D(t),
\end{align}
relationship that will be useful below.
Combining Eqs.~\eqref{eq:input-output_time} and \eqref{eq:y_solution}, $Y_{\out}(t)$ will be given by
\begin{align}\label{eq:yout_1}
    Y_{\out}(t) =&\, \sqrt{\kappa}\, \big[  Y_{\p}(t) + Y_{\h}(t) \big] - Y_{\in}(t),
    %=& \, \sqrt{\kappa}\, \Big[ \D(t)*\Sigma_{_Y}(t) + Y_{\h}(t) \Big] - Y_{\in}(t),
    %& \sqrt{\kappa}\, \Big[ (G_- + G_+) \D(t)\, Q(0) + \K(t)\, Y(0)  \nonumber \\
    %& \hspace{2.25cm} + \D(t)*\Sigma_{_Y}(t) \Big] - Y_{\in}(t).
\end{align}
where $Y_{\p}(t)$ and $Y_{\h}(t)$ are given by Eqs.~\eqref{eq:y_p} and \eqref{eq:y_h}, respectively.
Thus, using Eqs.~\eqref{eq:inhomogeneity_y} and \eqref{eq:y_p}, we can rewrite Eq.~\eqref{eq:yout_1} in a form that makes explicit the amplification of the force signal and the noise added due to the measurement, as per Eq.~\eqref{eq:yout_snr},
\begin{align}\label{eq:yout_t}
    Y_{\out}(t) = A(t)*\widebar{F}_{\i}(t) + N(t),
\end{align}
where the time-dependent signal amplification $A(t)$ is given by
\begin{align}\label{eq:amplitude-t}
    A(t) = - \sqrt{\kappa}\, (G_- + G_+)\, \D(t),
\end{align}
and the time-dependent added noise due to the measurement is
\begin{align}\label{eq:noise-t}
    N(t) &=\, A(t)*\widebar{\W}_{\i}(t) + \sqrt{\kappa}\, Y_{\h}(t) - Y_{\in}(t) \nonumber \\
    & - \frac{\sqrt{\kappa}}{(G_- + G_+)}\, A(t) * \Big[  \dot{Y}_{\in}(t) + (\gamma /2)\, Y_{\in}(t) \Big] .
\end{align}
In the following, we will study the sensitivity of the nonstationary force measurement using the nonstationary SNR defined in Eq.~\eqref{eq:snr-nonstationary-laplace}.
\subsection{Nonstationary force noise PSD}
The nonstationary force noise PSD $S_{F_{\mathrm{est}}}(\omega, T_{\m})$, as defined in Eq.~\eqref{eq:imp:nonstationary_psd_definition}, will allow us to describe the behaviour of the added noise as a function of the parameters involved in the problem and, consequently, the parametric influence on the SNR.
Thus, in order to calculate $S_{F_{\mathrm{est}}}(\omega, T_{\m})$ we must first determine $F_{\mathrm{est}}^{\N}(s)$, which can be obtained using  Eq.~\eqref{eq:f_noise-nonstationary}.
Therefore, taking the Laplace transform of Eqs.~\eqref{eq:y_h}, \eqref{eq:k_t}, \eqref{eq:amplitude-t}, and \eqref{eq:noise-t}; we can calculate $N(s)$ and $A(s)$,  which are the Laplace transforms of  $N(t)$ and $A(t)$, respectively.
Then, replacing these quantities into Eq.~\eqref{eq:f_noise-nonstationary}, we have
\begin{align}\label{eq:nonstationary_f_noise}
    &F_{\mathrm{est}}^{\N}(s) =\, \frac{(s+\gamma/2) (s-\kappa/2)}{\sqrt{\kappa}\, (G_- + G_+)}\, Y_{\in}(s) + \frac{(G_- - G_+)}{\sqrt{\kappa}}\, Y_{\in}(s) \nonumber \\
    & \hspace{1.8cm} + \widebar{\W}_{\i}(s) - Q(0) - \frac{(s+\gamma/2)}{(G_- + G_+)}\, Y(0),
\end{align}
where $\widebar{\W}_{\i}(s)$ and $Y_{\in}(s)$ are the Laplace transforms of $\widebar{\W}_{\i}(t)$ and $Y_{\in}(t)$, respectively.
It is worth noting that the first three terms on the right-hand side of Eq.~\eqref{eq:nonstationary_f_noise} correspond to the nonstationary version of the quantum noise process $F_{\mathrm{est}}^{\N}(\omega)$ in Eq.~\eqref{eq:force_noise}, which was used in Sec.~\ref{sec:stationary_force_sensing} to study the sensitivity of stationary force measurements.
The expression for $F_{\mathrm{est}}^{\N}(s)$ in Eq.~\eqref{eq:nonstationary_f_noise} shows three different types of contributions to the added noise according to their scalings with respect to the effective optomechanical coupling rates $G_{\pm}$, which in turn are related to the input powers $\wp_{\pm}$ as described in Appendix \ref{app:derivation_hamiltonian}.
Thus, as per Eq.~\eqref{eq:force_noise}, the first three terms on the right-hand side of Eq.~\eqref{eq:nonstationary_f_noise} correspond to imprecision noise, radiation-pressure noise, and mechanical thermal and quantum fluctuations, respectively.
On the other hand, the fifth term contains information on fluctuations in the position of the mechanical oscillator, while the sixth is related to imprecision noise.
Now, we can calculate the nonstationary force noise  PSD in Eq.~\eqref{eq:imp:nonstationary_psd_definition} using $F_{\mathrm{est}}^{\N}(s)$ in Eq.~\eqref{eq:nonstationary_f_noise} together with the correlation functions associated with $Y_{\in}(s)$ and $\widebar{\W}_{\i}(s)$, which are given by,
\begin{subequations}\label{eq:imp:correlations_laplace}
\begin{align}
    &\big\langle Y_{\in}^{\dagger}(s) Y_{\in}(s) \big\rangle =\, T_{\m}/2, \label{eq:correlation_y_s} \\
    &\big\langle \widebar{\W}_{\i}^{\dagger}(s) \widebar{\W}_{\i}(s)  \big\rangle =\, \gamma\, T_{\m}\, \pqty{\bar{n}_{\th} + 1/2} \nonumber \\
    & \hspace{2.65cm} \times\bqty{ \frac{1}{2} + \frac{\arctan{ \pqty{2 \omega_{\m} T_{\m}} }}{\pi} }. \label{eq:correlation_w_s}
\end{align}
\end{subequations}
These correlation functions in complex frequency domain were obtained taking the Laplace transform of Eqs.~\eqref{eq:xy_corr} and \eqref{eq:langevin_corr_1:a}, respectively.
Therefore, if we substitute Eq.~\eqref{eq:nonstationary_f_noise} into Eq.~\eqref{eq:imp:nonstationary_psd_definition}, and we take into account the correlation functions in Eqs.~\eqref{eq:imp:correlations_laplace}, we obtain that the nonstationary force noise PSD $S_{F_{\mathrm{est}}}(\omega, T_{\m})$ may be expressed as
\begin{align}\label{eq:nonstationary_psd_0}
    S_{F_{\mathrm{est}}}(\omega,T_{\m}) =  S_{\mathrm{ss}}(\omega,T_{\m}) + S_{\mathrm{tr}}(\omega,T_{\m}) + S_{\mathrm{th}}(T_{\m}),
\end{align}
where,
\begin{align}\label{eq:steady-state_psd}
    &S_{\mathrm{ss}}(\omega,T_{\m}) =\, \,  \frac{\pqty{G_- - G_+}^2}{2 \kappa} \nonumber \\
    & + \frac{ \abs{-i \omega + 1/2 T_{\m} + \gamma/2}^2\,  \abs{-i \omega + 1/2 T_{\m} - \kappa/2}^2 }{ 2 \kappa\, \pqty{G_- + G_+}^2}  \nonumber \\
    & + \mathrm{Re}\big[ \pqty{-i \omega + 1/2 T_{\m} + \gamma/2} \pqty{-i \omega + 1/2 T_{\m} - \kappa/2} \big]  \nonumber \\
    & \hspace{5.1cm} \times \frac{\pqty{G_- - G_+}}{\kappa\, \pqty{G_- + G_+}}
\end{align}
is the steady-state contribution due to input noise $Y_{\in}(t)$ which does not depend on the system initial conditions; while
\begin{align}\label{eq:transient_psd}
    &S_{\mathrm{tr}}(\omega,T_{\m}) =\, \nonumber \\
    & \hspace{0.25cm} \frac{1}{T_{\m}}\, \bigg[\, \expval{Q^2}_{0} + \frac{ \abs{ -i \omega + 1/2 T_{\m} + \gamma/2 }^2 }{(G_- + G_+)^2}\, \expval{Y^2}_{0} \nonumber \\
    & \hspace{1.1cm} + \frac{ \mathrm{Re} (-i \omega + 1/2 T_{\m} + \gamma/2 ) }{(G_- + G_+)}\, \big\langle Q Y + Y Q \big\rangle_{0} \bigg]\,
\end{align}
is the transient contribution due to homogeneous solution $Y_{\h}(t)$, which carries the information about the initial state of the system and vanishes in the limit of an infinite measurement time.
The subscript 0 in the second moments in Eq.~\eqref{eq:transient_psd} stands for its value just before the arrival of the force at $t=0$.
Further, 
\begin{align}\label{eq:nonstationary_thermal_psd}
    S_{\th}(T_{\m}) = \gamma\, (\bar{n}_{\th} + 1/2)\, \bqty{ \frac{1}{2} + \frac{\arctan{ \pqty{2 \omega_{\m} T_{\m}} }}{\pi} }
\end{align}
is the thermal noise floor, whose explicit dependence on the measurement time accounts for the re-thermalisation of the transducer.
Thus, for $T_{\m} \to \infty$ the thermal noise floor reduces to its value in the steady-state, while for $T_{\m} \ll 1/\omega_{\m}$ it takes half of this value.
As expected, when $T_{\m} \to \infty$ the nonstationary force noise PSD $S_{F_{\mathrm{est}}}(\omega,T_{\m})$ reduces to the stationary force noise PSD $S_{F_{\mathrm{est}}}(\omega)$ in Eq.~\eqref{eq:force_spectrum_2}.
It is important to emphasise that an optimal nonstationary force measurement will not only depend on the ratio between the powers of the coherent drives but also on how the measurement time is related to the system parameters.
\subsection{Initial state preparation}
Since the transient component of the nonstationary force PSD $S_{\mathrm{tr}}(\omega,T_{\m})$ in Eq.~ \eqref{eq:transient_psd} depends on the initial conditions for the second moments, we now consider the preparation of the initial state of the system such that nonstationary signal can be sensed with an optimal SNR.
We shall study the situation in which the system is in a steady-state prior to the arrival of the force.
In particular, we are interested in the steady-state solution of the second moments associated with the quadratures $Q$ and $Y$ which are the ones involved in $S_{\mathrm{tr}}(\omega,T_{\m})$ as shown in Eq.~\eqref{eq:transient_psd}.
Decoupling the Heisenberg-Langevin Eqs.~\eqref{eq:system_dynamics}, in the absence of external force (signal) we have
\begin{align}\label{eq:driven_damped_quadratures}
    \ddot{\vb*{v}} + 2\Gamma\, \dot{\vb*{v}} + \Omega_0^2\, \vb*{v}  =  \vb*{\xi}_{_{0}} ,
\end{align}
where $\Gamma$ was defined in Eq.~\eqref{eq:gamma}, and $\Omega_0^2=G_{-_0}^2-G_{+0}^2 +\gamma \kappa/4$.
Furthermore, $\vb*{v}$ is the vector of quadrature operators defined in Eq.~\eqref{eq:vector_quadratures}, while the noise vector $\vb*{\xi}_{_{0}} = \vb*{\xi}_{_{0}}(t)$ is given by
\begin{align}
    \vb*{\xi}_{_{0}} = \pqty{ \xi_{_{Q 0}},\, \xi_{_{P 0}},\, \xi_{_{X 0}},\, \xi_{_{Y 0}} }^{\mathrm{T}},
\end{align}
where the driving terms are,
\begin{subequations}\label{eq:noise_elements}
    \begin{align}
        &\xi_{_{Q 0}} = -  \bS[\Dot]{\W}_{\i} - ( \kappa/2 )\, \widebar{\W}_{\i}  - \sqrt{\kappa}\, ( G_{-0} - G_{+0} )\, Y_{\in}, \\
        &\xi_{_{P 0}} = \bS[\Dot]{\W}_{\r} + ( \kappa/2 )\, \widebar{\W}_{\r} +  \sqrt{\kappa}\, ( G_{-0} + G_{+0} )\, X_{\in}, \nonumber \\ \\
        &\xi_{_{X 0}} = - ( G_{-0} - G_{+0} )\, \widebar{\W}_{\r} + \sqrt{\kappa}\, \big[ \dot{X}_{\in} + (\gamma /2)\, X_{\in} \big], \\
        &\xi_{_{Y 0}} = - ( G_{-0} + G_{+0} )\, \widebar{\W}_{\i} + \sqrt{\kappa}\, \big[ \dot{Y}_{\in} + (\gamma /2)\, Y_{\in} \big].
    \end{align}
\end{subequations}
Here the subscript $0$ notation indicates the scenario for initial state preparation before the arrival of the force for those quantities that can easily take different values for different instants of time.
Since the dynamics of the quadratures is given by the driven damped harmonic oscillator Eq.~\eqref{eq:driven_damped_quadratures}, the evolution of $\vb*{v}(t)$ will be given by 
\begin{align}
    \vb*{v}(t) = \D_{_{0}}(t)*\vb*{\xi}_{_{0}}(t) + \vb*{v}_{\mathrm{h}}(t),
\end{align}
where convolution is defined element-wise and $\D_{_{0}}(t)$ is the classical Green's function described in Eq.~\eqref{eq:green_function} depending now on the initial state parameter $\Omega_0$.
Further, $\vb*{v}_{\mathrm{h}}(t)$ is the vector of homogeneous solutions to Eq.~\eqref{eq:driven_damped_quadratures}.
In the steady-state the evolution $\vb*{v}(t)$ reduces to
\begin{align}
    \vb*{v}_{\mathrm{ss}}(t) = \D_{_{0}}(t)*\vb*{\xi}_{_{0}}(t) = \mathcal{F}^{-1} \big\{ \D_{_{0}}(\omega)\ \vb*{\xi}_{_{0}}(\omega) \big\},
\end{align}
where $\vb*{\xi}_{_{0}}(\omega)$ and $\D_{_{0}}(\omega)$ are the Fourier transforms of $\vb*{\xi}_{_{0}}(t)$ and $\D_{_{0}}(t)$, respectively, with $\D_{_{0}}(\omega)$ given by $\D_{_{0}}(\omega) = ( -\omega^2 - 2 i \omega\, \Gamma + \Omega_0^2 )^{-1}$.
Therefore, the initial state non-symmetrically ordered covariance matrix 
\begin{align}
    \renewcommand\arraystretch{1.5}
    \setlength{\tabcolsep}{2pt}
    \vb*{\Theta} \equiv \big\langle \vb*{v}_{\mathrm{ss}}\, \vb*{v}_{\mathrm{ss}}^{\mathrm{T}} \big\rangle_0 =
    \begin{bmatrix}
        \expval{Q^2}_{0} & \expval{Q P}_{0} & \expval{Q X}_{0} & \expval{Q Y}_{0}\\
        \expval{P Q}_{0} & \expval{P^2}_{0} & \expval{P X^2}_{0} & \expval{P Y}_{0}\\
        \expval{X Q}_{0} & \expval{X P}_{0} & \expval{X^2}_{0} & \expval{X Y}_{0}\\
        \expval{Y Q}_{0} & \expval{Y P}_{0} & \expval{Y X}_{0} & \expval{Y^2}_{0}\\
    \end{bmatrix}
\end{align}
will be given by 
\begin{align}
    &\vb*{\Theta} = \int_{-\infty}^{+\infty} \frac{\dd{\omega}}{2 \pi}\, \int_{-\infty}^{+\infty} \frac{\dd{\omega'}}{2 \pi}\, \e^{-i (\omega+\omega') t} \D_{_{0}}(\omega) \D_{_{0}}(\omega') \nonumber \\
    & \hspace{4.7cm} \times \big\langle \vb*{\xi}_{_{0}}(\omega)\, \vb*{\xi}_{_{0}}^{\mathrm{T}}(\omega') \big\rangle.
\end{align}
The elements of  $\big\langle \vb*{\xi}_{_{0}}(\omega) \vb*{\xi}_{_{0}}^{\mathrm{T}}(\omega') \big\rangle$ will depend on the correlation functions involving $\widebar{\W}_{\r}(\omega)$, $\widebar{\W}_{\i}(\omega)$, $X_{\in}(\omega)$, and $Y_{\in}(\omega)$; which can be found from the time dependent correlation functions in Eqs.~\eqref{eq:xy_corr_0} and \eqref{eq:langevin_corr_1} [see Appendix \ref{app:derivation_hamiltonian} for details on the calculation of the frequency correlation functions of the Langevin force quadratures $\widebar{\W}_{\r}(\omega)$ and $\widebar{\W}_{\i}(\omega)$].
The required correlation functions are given by
\begin{subequations}\label{eq:frequency_correlations}
\begin{align}
    & \expval{ X_{\in}(\omega) Y_{\in}(\omega') } = \expval{ Y_{\in}(\omega) Y_{\in}(\omega') } = \pi\, \delta(\omega + \omega'), \\
    & \expval{ X_{\in}(\omega) Y_{\in}(\omega') } = \expval{ Y_{\in}(\omega) X_{\in}(\omega') }^* = i \pi\, \delta(\omega + \omega'); \\
    &\expval{ \widebar{\W}_{\r}(\omega) \widebar{\W}_{\r}(\omega') } = \expval{ \widebar{\W}_{\i}(\omega) \widebar{\W}_{\i}(\omega') }  \nonumber \\
    & \hspace{2.9cm}=  2 \pi \gamma\, \pqty{\widebar{n}_{\mathrm{th}} + 1/2 }\, \delta(\omega + \omega'),\\
    &\expval{ \widebar{\W}_{\r}(\omega) \widebar{\W}_{\i}(\omega') } = \expval{ \widebar{\W}_{\i}(\omega) \widebar{\W}_{\r}(\omega') }^*  \nonumber \\
    & \hspace{4.7cm} =  i \pi \gamma\, \delta(\omega + \omega').
\end{align}
\end{subequations}
Hence, $\big\langle \vb*{\xi}_{_{0}}(\omega) \vb*{\xi}_{_{0}}^{\mathrm{T}}(\omega') \big\rangle$ will be proportional to $\delta(\omega+\omega')$ and, accordingly, $\vb*{\Theta}$ reduces to
\begin{align}\label{eq:covariance_matrix}
    \vb*{\Theta} = \int_{-\infty}^{+\infty} \frac{\dd{\omega}}{2 \pi}\, \abs{\D_{_{0}}(\omega)}^2 \int_{-\infty}^{+\infty} \frac{\dd{\omega'}}{2 \pi}\, \big\langle \vb*{\xi}_{_{0}}(\omega) \vb*{\xi}_{_{0}}^{\mathrm{T}}(\omega') \big\rangle.
\end{align}
Finally, taking the Fourier transform of the driving terms in Eqs.~\eqref{eq:noise_elements} to get $\vb*{\xi}_{_{0}}(\omega)$ and then using the correlation functions in Eqs.~\eqref{eq:frequency_correlations} to obtain $\big\langle \vb*{\xi}_{_{0}}(\omega) \vb*{\xi}_{_{0}}^{\mathrm{T}}(\omega') \big\rangle$, we can use Eq.~\eqref{eq:covariance_matrix} to find the second moments in $\vb*{\Theta}$ as functions of the system parameters.
Among these, we want to emphasise the following,
\begin{subequations}
\begin{align}
    &\expval{Q^2}_{0} =\,  \gamma\, (\widebar{n}_{\mathrm{th}} + 1/2)\, \Big[ (\kappa^2/4)\, \mathcal{I}_0 + \mathcal{I}_2 \Big] \nonumber \\
    & \hspace{3.3cm} + (\kappa/2)\, (G_{-0} - G_{+0})^2\, \mathcal{I}_0, \label{eq:q2} \\
    &\expval{P^2}_{0} =\,  \gamma\, (\widebar{n}_{\mathrm{th}} + 1/2)\, \Big[ (\kappa^2/4)\, \mathcal{I}_0 + \mathcal{I}_2 \Big] \nonumber \\
    & \hspace{3.3cm} + (\kappa/2)\, (G_{-0} + G_{+0})^2\, \mathcal{I}_0, \label{eq:p2} \\
    &\expval{X^2}_{0} =\, \gamma\, (\widebar{n}_{\mathrm{th}} + 1/2)\, (G_{-0} - G_{+0})^2\, \mathcal{I}_0 \nonumber \\
    & \hspace{3.3cm} + (\kappa/2)\, \Big[ (\gamma^2/4)\, \mathcal{I}_0 + \mathcal{I}_2 \Big], \label{eq:x2} \\
    &\expval{Y^2}_{0} =\, \gamma\, (\widebar{n}_{\mathrm{th}} + 1/2)\, (G_{-0} + G_{+0})^2\, \mathcal{I}_0 \nonumber \\
    & \hspace{3.3cm} + (\kappa/2)\, \Big[ (\gamma^2/4)\, \mathcal{I}_0 + \mathcal{I}_2 \Big], \label{eq:y2} \\
    & \big\langle Q Y \big\rangle_{0} = \big\langle Y Q \big\rangle_{0} \nonumber \\
    & \hspace{1.05cm} = (\gamma \kappa/2)\,  \big[\, \widebar{n}_{\mathrm{th}}\, (G_{-0} + G_{+0}) +  G_{+0} \big]\, \mathcal{I}_0. \label{eq:qy}
\end{align}
\end{subequations}
To calculate the elements of $\vb*{\Theta}$ it was necessary to take into account the solution to the following integrals,
\begin{subequations}\label{eq:integrals}
\begin{align}
    &\mathcal{I}_0 = \int_{-\infty}^{+\infty} \frac{\dd{\omega}}{2 \pi}\, \abs{\D_{_{0}}(\omega)}^2, \\
    &\mathcal{I}_1 = \int_{-\infty}^{+\infty} \frac{\dd{\omega}}{2 \pi}\, \abs{\D_{_{0}}(\omega)}^2\, \omega, \\
    &\mathcal{I}_2 = \int_{-\infty}^{+\infty} \frac{\dd{\omega}}{2 \pi}\, \abs{\D_{_{0}}(\omega)}^2\, \omega^2;
\end{align}
\end{subequations}
where $\mathcal{I}_1 = 0$, while the solutions to $\mathcal{I}_0$ and $\mathcal{I}_2$ are too lengthy
to be reported here.
From Eqs.~\eqref{eq:q2} and \eqref{eq:p2}, we can see that for $G_+ \neq G_-$ the two-tone driving scheme under consideration lead to dissipative mechanical squeezing of the $Q$ quadrature and anti-squeezing of the $P$ quadrature~\cite{Kronwald2013}.
At the same time, Eqs.~\eqref{eq:x2} and \eqref{eq:y2}  make evident that this scheme is also producing dissipative squeezing of the electromagnetic quadrature $X$ and anti-squeezing of the quadrature $Y$, as reported in Ref.~\cite{Kronwald2014}.
Furthermore, from Eq.~\eqref{eq:qy} it is clear that this scheme also allows us to obtain entangled steady-states between light and matter.
However, in accordance with Eq.~\eqref{eq:transient_psd}, a higher cross-correlation between electromagnetic and mechanical operators will increase the added noise due to the measurement and, accordingly, reduce the sensitivity of the force measurement.
\begin{figure}[t]
    %\hspace{0.7cm}
    \centering
    \includegraphics[width=\linewidth]{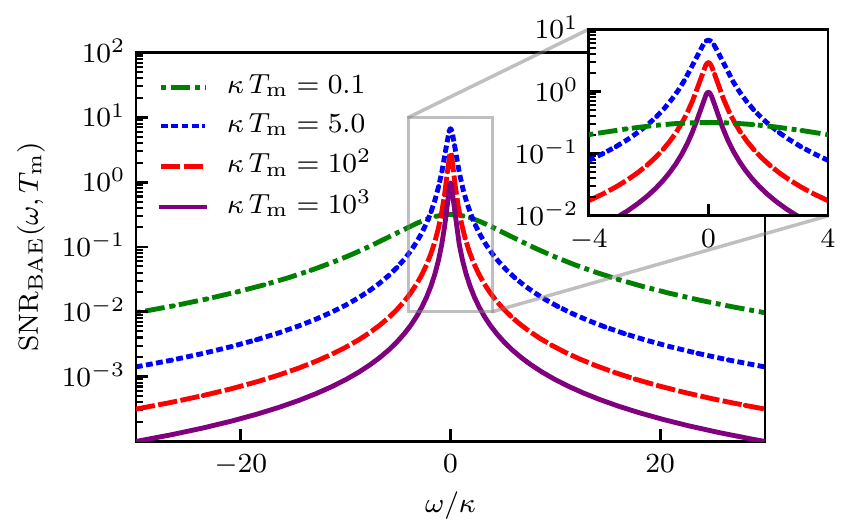}
    \vspace{-0.65cm}
    \caption{%
    Signal-to-noise ratio for an impulsive Dirac delta force under a nonstationary BAE measurement  ($G_+/G_- = 1$, $G_{+0}/G_{-0} \neq 1$) as a function of frequency for different measurement times.
    Here, the maximum SNR at resonance occurs for $T_{\m}=5/\kappa$ (blue dotted line).
    The parameters used were: $\omega_{\m}/\kappa = 10$, $\gamma/\kappa = 10^{-4}$, $\widebar{n}_{\mathrm{th}} = 10$,  $G_+/\kappa = G_-/\kappa = G_{-0}/\kappa = 1$.
    The initial state corresponds to $G_{-0}/G_{+0}=0.97$,  which for the considered parameters maximises the mechanical squeezing before the arrival of the force. 
    For the signal, $f_0= 1$ and $t_0=0^+$ were used. \label{fig:snr_spectral} %
    }
\end{figure}
\begin{figure}[t]
    %\hspace{0.7cm}
    \centering
    \includegraphics[width=\linewidth]{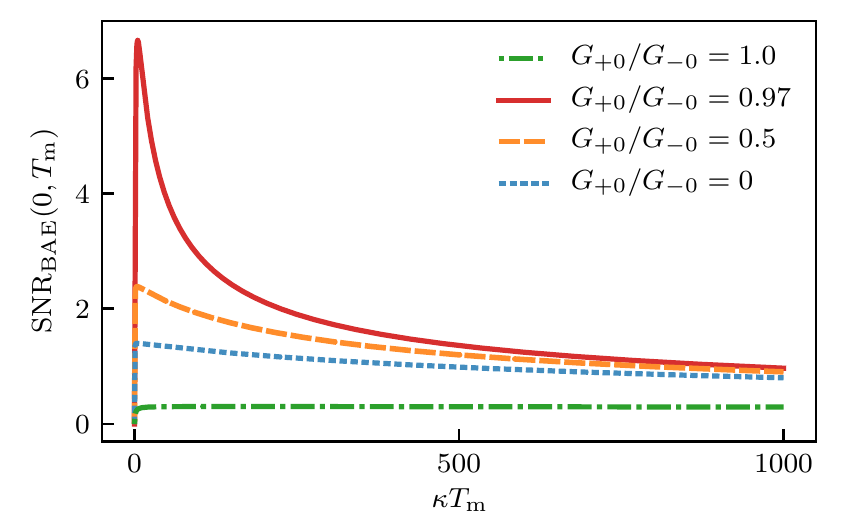}
    \vspace{-0.65cm}
    \caption{Signal-to-noise ratio at resonance $\mathrm{SNR}(0,T_{\m})$ for an impulsive Dirac delta force under a nonstationary BAE measurement  ($G_+/G_- = 1$, $G_{+0}/G_{-0} \neq 1$) as a function of the measurement time for different state preparation drive asymmetries.
    For each curve where $G_{+ 0}/G_{- 0}<1$, the maximum SNR occurs when $T_{\m} \sim 1/\kappa$; while for  $G_{+ 0}/G_{- 0}=1$, a finite measurement time does not improve the sensitivity of the force measurement.
    Here, the optimal SNR takes place for $G_{+ 0}/G_{- 0}=0.97$ (red full line), which for the parameters under consideration corresponds to a system prepared in an optimal mechanical squeezed initial state.
    The parameters used were: $G_+/\kappa = G_-/\kappa = G_{- 0}/\kappa = 1$, $\omega_{\m}/\kappa = 10$, $\gamma/\kappa = 10^{-4}$, and $\widebar{n}_{\mathrm{th}} = 10$. For the signal, $f_0= 1$ and $t_0=0^+$ were used. \label{fig:snr}}
\end{figure}
\begin{figure*}[t]
    %\hspace{0.7cm}
    \centering
    \includegraphics[width=\linewidth]{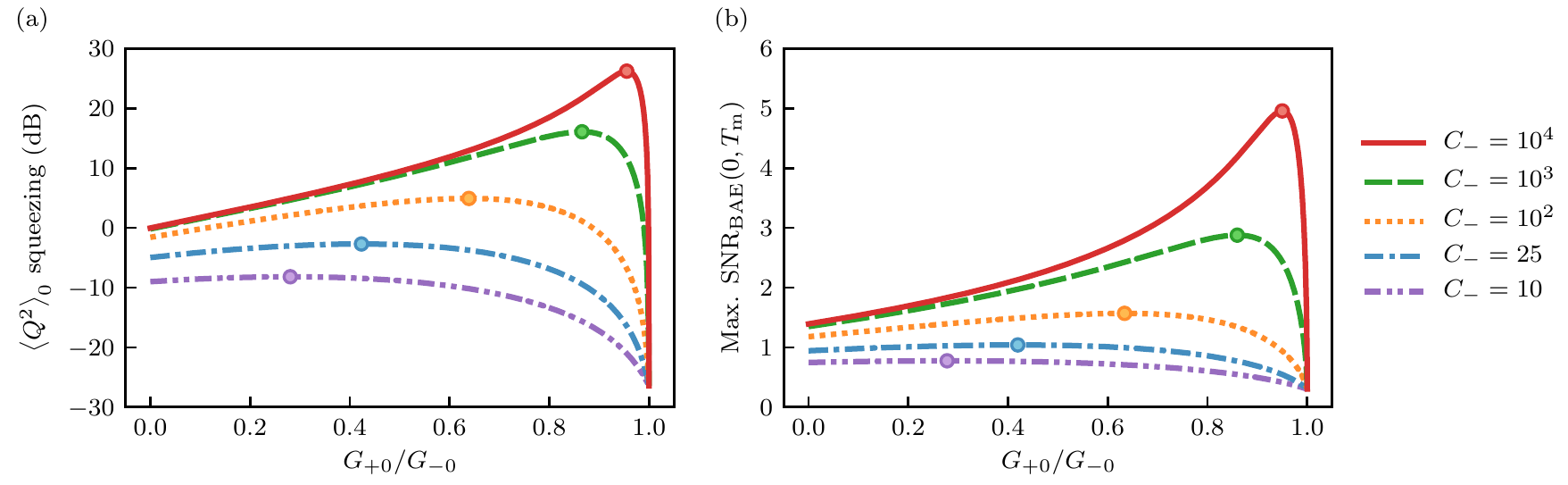}
    \vspace{-0.65cm}
    \caption{Initial state mechanical squeezing (a) and maximum signal-to-noise ratio for an impulsive Dirac delta force under a nonstationary BAE measurement  ($G_+/G_- = 1$, $G_{+0}/G_{-0} \neq 1$) (b) as a function of the state preparation drive asymmetry ($G_{+0}/G_{-0}$) for different cooperativities of the red sideband drive ($C_-=4 G_{-0}^2/\gamma \kappa$). (a) $\langle Q^2 \rangle_0$ steady-state squeezing [$-10 \log_{10}{(\langle Q^2 \rangle_0/\langle Q^2 \rangle_{\zpf})}\, \mathrm{dB}$], here  $\langle Q^2 \rangle_0$ was calculated using Eq.~\eqref{eq:q2} and $\langle Q^2 \rangle_{\zpf}=1/2$. (b) Maximum signal-to-noise ratio at resonance calculated using Eq.~\eqref{eq:snr-nonstationary-laplace}, with $\widebar{F}_{\i}(-i \omega + 1/2 T_{\m})$ as given by Eq.~\eqref{eq:dirac_force} and $S_{F_{\mathrm{est}}}(\omega, T_{\m})$ as described in Eqs.~\eqref{eq:nonstationary_psd_0}--\eqref{eq:nonstationary_thermal_psd}. The dot in each curve marks its maximum value, which in both plots corresponds almost exactly to the same value of $G_{+0}/G_{-0}$. The parameters used were:  $\omega_{\m}/\kappa = 10$, $\gamma/\kappa = 10^{-4}$, and $\widebar{n}_{\mathrm{th}} = 10$. For the signal, $f_0= 1$ and $t_0=0^+$ were used. Further, for the measurement configuration it was considered $G_{-}=G_{-0}$ and $G_{+}/G_{-}=1$. \label{fig:snr_squeezing}}
\end{figure*}
Since all three $\expval{Q^2}_{0}$, $\expval{Y^2}_{0}$, and $\expval{QY+YQ}_{0}$, appear in the transient contribution to the nonstationary force noise PSD $S_{\mathrm{tr}}(\omega,T_{\m})$ defined in Eq.~\eqref{eq:transient_psd}, it will be necessary to consider a parameter regime for which the anti-squeezing of $\expval{Y^2}_{0}$ and the cross-correlation $\expval{QY+YQ}_{0}$ do not counteract the noise reduction due to the squeezing of $\expval{Q^2}_{0}$.
Thus, we may note from Eq.~\eqref{eq:transient_psd} that at resonance the coefficients associated with $\expval{Y^2}_{0}$ and $\expval{QY+YQ}_{0}$ depend on the relationship among the effective coupling constants $G_{\pm}$, the mechanical dissipation rate $\gamma$, and the measurement time $T_{\m}$.
However, since in general $G_{\pm} \gg \gamma$, the aforementioned coefficients will depend only on the relationship between $G_{\pm}$ and $T_{\m}$, such that within the nonstationary transient regime if $T_{\m} \gg 1/G_{\pm}$, the only non-negligible contribution to $S_{\mathrm{tr}}(0,T_{\m})$ will be the term associated with $\expval{Q^2}_{0}$.
Therefore,  regardless of the values of $\expval{Y^2}_{0}$ and $\expval{QY+YQ}_{0}$, it is to be expected that the mere preparation of the system in a dissipative mechanical squeezed state will allow us to significantly reduce the added noise due to the measurement and increase the sensitivity of the force measurement.
\subsection{Signal-to-noise ratio for impulsive forces}
\begin{figure}[t]
    %\hspace{0.7cm}
    \centering
    \includegraphics[width=\linewidth]{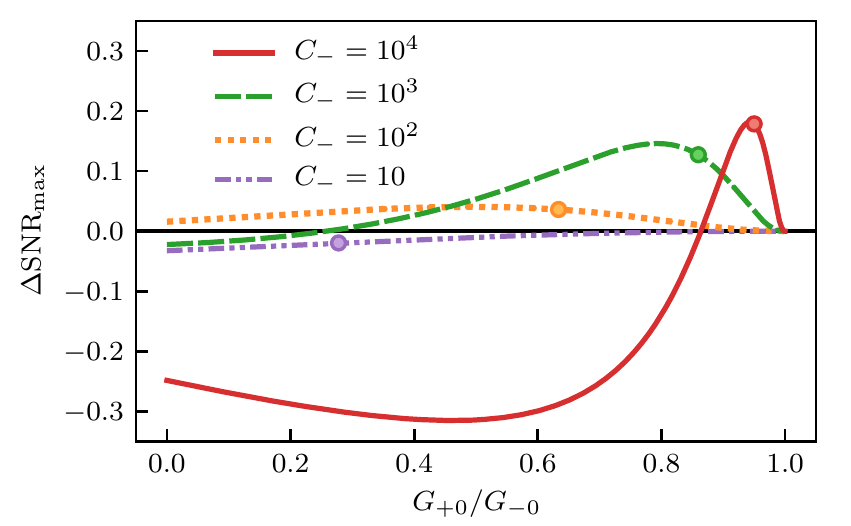}
    \vspace{-0.65cm}
    \caption{Difference between the maximum signal-to-noise ratio (SNR) at resonance achieved by leaving the drive asymmetry configuration unchanged upon arrival of the force ($G_+ = G_{+0}$, $G_- = G_{-0}$) and the nonstationary BAE measurement  ($G_+/G_- = 1$, $G_{+0}/G_{-0} \neq 1$, $G_{-}=G_{-0}$) described in Fig.~\ref{fig:snr_squeezing} [ $\Delta \mathrm{SNR}_{\mathrm{max}} = \max{ \{\, \mathrm{SNR}_{\mathrm{SMS}}(0, T_{\m})\, \}} - \max{\{\, \mathrm{SNR}_{\mathrm{BAE}}(0, T_\m)}\, \}$ ], as a function of $G_{+0}/G_{-0}$ for different cooperativities $C_-$. 
    The dot in each curve marks the value of $G_{+0}/G_{-0}$ for which $\mathrm{SNR}_{\mathrm{BAE}}(0,T_\m)$ reaches its maximum (see Fig.~\ref{fig:snr_squeezing}).
    The parameters here, except for the coupling $G_+$ that was used to calculate $\mathrm{SNR}_{\mathrm{SMS}}(0, T_{\m})$, are the same as those that were used to obtain Fig.~\ref{fig:snr_squeezing}.
    Since the difference $\Delta \mathrm{SNR}_{\mathrm{max}}$ is in general very small (cf. Fig. \ref{fig:snr}), the SNR achieved by leaving the drive asymmetry configuration unchanged is still much better than what can be obtained by means of stationary force measurements.
    This result yields the considerable advantage that we do not need to know the arrival time of the force in order to improve the sensitivity of force measurements using the proposed nonstationary protocol.
    \label{fig:snr_difference} }
\end{figure}

The results of the previous subsections give us the ingredients to analyse the sensitivity of nonstationary measurements under the proposed dissipative mechanical squeezing state preparation.
Here, we shall consider a Dirac delta force in order to analyse the SNR in the nonstationary measurement of impulsive forces.
It is important to note that although a Gaussian force would correspond more exactly to an experimental scenario, in the impulsive limit for a Gaussian envelope ($\sigma \ll T_{\m}$, with  $\sigma$ the standard deviation) the SNR results are not significantly different from those obtained for a Dirac delta force.
Thus, we consider an impulsive Dirac delta force given by
\begin{align}
    \widebar{F}_{\i}(t) = f_0\,\, \delta(t-t_0),
\end{align}
with $f_0$ the amplitude of the force and $t_0>0$ the arrival time.
Taking the Laplace transform of $\widebar{F}_{\i}(t)$, we get
\begin{align}\label{eq:dirac_force}
    \widebar{F}_{\i}(-i \omega + 1/2 T_{\m}) = f_0\, \e^{-(-i \omega + 1/2 T_{\m})\, t_0}
\end{align}
and, therefore, from Eq.~\eqref{eq:signal-laplace} we have that the signal will be given by
\begin{align}\label{eq:dirac_signal}
    \mathcal{S}(\omega, T_{\m}) = \abs{\widebar{F}_{\i}(-i \omega + 1/2 T_{\m})} = f_0\, \e^{-t_0/2 T_{\m}}.
\end{align}
Finally, we may replace the signal $|\widebar{F}_{\i}(-i \omega + 1/2 T_{\m})|$ defined in Eq.~\eqref{eq:dirac_signal} and the nonstationary force noise PSD $S_{F_{\mathrm{est}}}(\omega,T_{\m})$ given by Eqs.~\eqref{eq:nonstationary_psd_0} -- \eqref{eq:nonstationary_thermal_psd}, into Eq.~\eqref{eq:snr-nonstationary-laplace} to obtain the nonstationary signal-to-noise ratio $\mathrm{SNR}(\omega,T_{\m})$ for the force measurement under consideration.
The explicit form of $\mathrm{SNR}(\omega,T_{\m})$ is too cumbersome to be shown here, however, in Figs.~\ref{fig:snr_spectral} -- \ref{fig:snr_difference} we use the resulting expression to study the nonstationary SNR as a function of the parameters involved in the problem.
Now, it is important to emphasise that the drive asymmetry used for the preparation of the initial state in general can be different from that used for the nonstationary force measurement.
Thus, as mentioned before, for the preparation of the initial state we will consider a drive asymmetry that optimises the squeezing of $\expval{Q^2}_{0}$ as per Eq.~\eqref{eq:q2}, while for the force measurement we will use a configuration that reduces the added noise in accordance with the nonstationary force noise PSD $S_{F_{\mathrm{est}}}(\omega,T_{\m})$ in Eqs.~\eqref{eq:nonstationary_psd_0} -- \eqref{eq:nonstationary_thermal_psd}.
Therefore, in the following we shall consider two different schemes for the drive asymmetry configuration to be used during the nonstationary measurement. The first is a nonstationary BAE measurement ($G_+/G_- = 1$, $G_{+0}/G_{-0} \neq 1$), while the second assumes that the drive asymmetry is left unchanged upon arrival of the impulsive force  ($G_+ = G_{+0}$, $G_- = G_{-0}$).

\subsection{Nonstationary back-action evading measurement (\texorpdfstring{$G_+/G_- = 1,\, G_{+0}/G_{-0} \neq 1$}{})}
Since the thermal noise floor $S_{\mathrm{th}}(T_{\m})$ in Eq.~\eqref{eq:nonstationary_thermal_psd} does not depend on the drive asymmetry $G_+/G_-$, and the transient contribution to the nonstationary force noise PSD $S_{\mathrm{tr}}(\omega, T_{\m})$ in Eq.~\eqref{eq:transient_psd} fundamentally depends on the preparation of the initial state only, then, we can choose the configuration for the nonstationary measurement according to the steady-state contribution to the nonstationary PSD $S_{\mathrm{ss}}(\omega, T_{\m})$ given by Eq.~\eqref{eq:steady-state_psd}.
It is worth noting that $S_{\mathrm{ss}}(\omega, T_{\m})$ is the nonstationary version of the stationary added force noise PSD $S_{\mathrm{add}}(\omega)$ defined in Eq.~\eqref{eq:added_noise} and, therefore, the analysis made in Sec.~\ref{sec:stationary_force_sensing} for the reduction of $S_{\mathrm{add}}(\omega)$ at resonance can be used here in order to minimise $S_{\mathrm{ss}}(\omega, T_{\m})$.
Thus, in accordance with Eq.~\eqref{eq:added_psd} and Fig. \ref{fig:optimal_drive_asymmetry}, first we shall consider $G_+/G_- = 1$ (BAE measurement), since it corresponds to the optimal configuration for the reduction of the stationary force noise PSD for a red sideband cooperativity $C_- \geq 10$.
We are going to refer to the nonstationary SNR associated with a finite time BAE measurement as $\mathrm{SNR}_{\mathrm{BAE}}(\omega, T_\m)$.
Then, in Fig.~\ref{fig:snr_spectral} we plot the SNR as a function of frequency for different measurement times, finding that the SNR is increased beyond the steady-state limit when the measurement time belongs to the nonstationary transient regime.
It is worth noting that enhancement of the SNR occurs even far from resonance.
Further, in Fig.~\ref{fig:snr} we plot the SNR at resonance as a function of the measurement time for different initial states of the transducer.
The initial state is characterised by the drive asymmetry $G_{+0}/G_{-0}$ used before the arrival of the force to prepare the system in a given steady-state. 
From Figs.~\ref{fig:snr_spectral} and \ref{fig:snr} we can see that the SNR at resonance for $G_{+0}/G_{-0} \neq 1$ reaches its maximum for $T_{\m} \sim 1/\kappa $; while from Fig.~\ref{fig:snr} we may note that for $T_{\m} \gg 1/\kappa$ ($T_{\m} \sim 1/\gamma $) the SNR is gradually reduced until it reaches a point were the effect of the initial conditions is not noticeable.
On the other hand,  in Fig. \ref{fig:snr_squeezing}, we show the relationship between the initial state mechanical squeezing and the maximum nonstationary SNR at resonance as a function of state preparation drive asymmetry for different red sideband cooperativities.
Thus, it was confirmed that there is a correlation between the dissipative mechanical squeezing before the arrival of the force and the enhancement of the SNR for the measurement of impulsive forces, such that the drive asymmetry $G_{+0}/G_{-0}$ required to maximise the dissipative mechanical squeezing corresponds almost exactly with the drive asymmetry necessary to maximise the nonstationary SNR. 
\subsection{Unchanged drive asymmetry upon arrival of the force  \texorpdfstring{$(G_+= G_{+0},\, G_- = G_{-0})$}{}}
Next, we shall consider the effect on the SNR of not changing the drive asymmetry configuration upon arrival of the impulsive force.
This scenario is of special relevance when thinking of implementations of the nonstationary strategy we propose here, since in many of the possible applications the arrival time of the force is unknown.
Thus, as described before, we prepare the system in an optimal dissipative squeezed initial state such that when the force kicks the mechanical oscillator the fluctuations in the position quadrature had been reduced, then, without changing the drive asymmetry configuration, the nonstationary measurement is performed. 
Since the drive asymmetry configuration used during the nonstationary measurement corresponds to the one used for the preparation of the mechanical squeezed initial state, we will refer to the nonstationary SNR obtained using this scheme as $ \mathrm{SNR}_{\mathrm{SMS}} (\omega, T_\m)$, where SMS stands for steady-state mechanical squeezing.
In Fig.~\ref{fig:snr_difference} we consider the difference between the maximum SNR achievable at resonance if we leave the drive asymmetry configuration unchanged [$\mathrm{SNR}_{\mathrm{SMS}}(0, T_{\m})$] and the maximum SNR that can be obtained by performing a nonstationary BAE measurement [$\mathrm{SNR}_{\mathrm{BAE}}(0, T_\m)$].
Thus, we plot
\begin{align}\label{eq:snr_difference}
    \Delta \mathrm{SNR}_{\mathrm{max}} =\, & \max \{\, \mathrm{SNR}_{\mathrm{SMS}}(0, T_{\m})\, \} \nonumber \\
    & \hspace{1.2cm} - \max \{\, \mathrm{SNR}_{\mathrm{BAE}}(0, T_\m)\, \}.
\end{align}
It is very interesting to note that there are regions of the space of parameters where leaving the drive asymmetry configuration unchanged allows one to obtain a maximum SNR greater than that achievable with a nonstationary BAE measurement.
This behaviour is possible because in the nonstationary transient regime the noise reduction due to the initial conditions is greater than that due to the drive asymmetry used during the measurement.
Moreover, even in those regions where a nonstationary BAE measurement provides a better sensitivity, the difference $\Delta \mathrm{SNR}_{\mathrm{max}}$ is still very small (cf. Fig.~\ref{fig:snr}) and, therefore, the SNR achieved by leaving the drive asymmetry configuration unchanged is still much better than what can be obtained by means of stationary force measurements.
This result makes the proposed ``squeeze and measure'' strategy highly relevant to improve the sensitivity in classical force measurements where the arrival time of the force is unknown.
% 

%------------------------------------------------------------------------------
% Conclusions
%
\section{Conclusions}
\label{sec:conclusions}
In this paper, we have analysed the measurement of a classical force driving a mechanical oscillator coupled to an electromagnetic cavity under two-tone driving.
The applied force shifts the position of the mechanical oscillator, whose change can be monitored through the output electromagnetic field.
Thus, we studied stationary and nonstationary protocols for the sensing of a classical force through the output electromagnetic field, and determined the conditions for optimal sensitivity in the force measurement.
For the purpose of analysing the force sensitivity quantitatively, first, we developed a theoretical framework based on the signal-to-noise ratio of linear spectral measurements, stationary or nonstationary.
Further, for the case of nonstationary force sensing, we used a one-sided decaying exponential window function to construct an inverse filter in complex frequency domain that allow us to describe the nonstationary measurement of impulsive forces preserving information on the initial conditions of the transducer upon arrival of the force.
Then, we considered force sensing in the steady-state under dissipative state preparation, for which we used the stationary force noise PSD as a figure of merit to quantify the sensitivity of the force measurement.
We found that as a consequence of the mutual cancellation of the noise contributions due to imprecision and radiation-pressure, the proposed two-tone driving scheme allows one to reduce the stationary force noise PSD to the thermal noise floor at resonance.
Furthermore, we found that a BAE measurement is not always the best approach to enhance the sensitivity of a stationary force measurement, but there are regimes of parameters for which the dissipative preparation of the mechanical oscillator in a squeezed state is optimal.
Finally, we considered a nonstationary protocol that involves non-thermal state preparation followed by a finite time measurement. This protocol allows one to use different drive asymmetry configurations at the two stages of the measurement process, one for state preparation and another for force measurement.
Thus, first the fluctuations are reduced dissipatively and, then, sensing is conducted in a finite measurement time.
We analysed this scenario quantitatively using a nonstationary SNR and identified regimes where such an approach is beneficial.
Hence, it was confirmed that there exists a correlation between the dissipative mechanical squeezing before the arrival of the force and the enhancement of the SNR for the measurement of impulsive forces, such that the state preparation drive asymmetry required to maximise the dissipative mechanical squeezing corresponds almost exactly to that necessary to maximise the SNR when a nonstationary BAE measurement is performed.
Furthermore, we found that leaving the drive asymmetry configuration unchanged upon arrival of the force is not particularly detrimental to the sensitivity of the force measurement, but on the contrary, in certain regions of the parameter space it allows us to obtain a SNR greater than that achievable with a nonstationary BAE measurement.
This result is of particular relevance in applications where the arrival time of the force is unknown, since it is no longer necessary to resort to cyclic repetitions of the squeezing and measurement steps.
% 

% 
%------------------------------------------------------------------------------

%------------------------------------------------------------------------------
% Acknowledgments
%
\section*{Acknowledgments}
We thank Glen I. Harris and Aashish A. Clerk for valuable discussions.
D.N.B.-G. acknowledges support from the Australian Government Department of Education and Training and the Colombian Administrative Department of Science, Technology and Innovation (Colciencias, now Minciencias) through the Australia--Americas PhD Internship Program 2018.
D.N.B.-G. also gratefully acknowledges support from UNSW Canberra as a Visiting Research Fellow.
D.N.B.-G. acknowledges funding from Colciencias through ``Convocatoria No. 727 de 2015 -- Doctorados Nacionales 2015''.
D.N.B.-G. and H.V.-P. acknowledge partial financial support from Colciencias under the projects,  ``Emisión en sistemas de qubits superconductores acoplados a la radiación'', code 110171249692, CT 293-2016, HERMES 31361; ``Control dinámico de la emisión en sistemas de qubits acoplados con cavidades no-estacionarias'', HERMES 41611; and ``Electrodinámica cuántica de cavidades no estacionarias'', HERMES 43351.
M.J.W. acknowledges support from ARC CE 110001013 and AFOSR FA 2386-18-1-4026.

%------------------------------------------------------------------------------

%------------------------------------------------------------------------------
% Appendices
%
\bigskip
\appendix
% 
%------------------------------------------------------------------------------
% Derivation of Hamiltonian
%
\section{Derivation of Hamiltonian}
\label{app:derivation_hamiltonian}
The total Hamiltonian describing the dynamics of this system is given by
\begin{align}\label{eq:app:hamiltonian}
    H_{\mathrm{tot}} = H_{\c} + H_{\m} + H_{\c\m} + H_{\mathrm{d}} + H_{\mathrm{f}},
\end{align}
where $H_{\c}$ represents the cavity mode, $H_{\m}$ describes the mechanical oscillator, $H_{\c\m}$ accounts for the optomechanical (electromechanical) coupling, $ H_{\mathrm{d}}$ describes the coherent driving, and $H_{\mathrm{f}}$ describes the contribution to the dynamics due to the external classical force.
The terms in Eq.~\eqref{eq:app:hamiltonian} are given by
\begin{subequations}\label{eq:app:hamiltonian_terms}
    \begin{align}
        &H_{\c} = \hbar\, \omega_{\c}\, a^{\dagger} a, \\
        &H_{\m} = \frac{\hbar\, \omega_{\m}}{2} \pqty{ Q^2+P^2 }, \\
        &H_{\c\m} = - \hbar\, g\, \sqrt{2} Q a^{\dagger} a, \\
        &H_{\mathrm{d}} = i \hbar\, \pqty{\E_+  \e^{-i \omega_+ t} +\, \E_-  \e^{-i \omega_- t} }\, a^{\dagger} + \mathrm{H.c.}, \label{eq:app:two-tone_driving} \\
        &H_{\mathrm{f}} = - \hbar\, F(t)\, Q,
    \end{align}
\end{subequations}
where $a^{\dagger}$ and $a$ are the electromagnetic creation and annihilation operators, respectively, obeying the bosonic commutation relation $\comm{a}{a^{\dagger}}=1$.
$Q$ and $P$ are the dimensionless mechanical position and momentum defined as, $q = \sqrt{2}\, q_{\zpf}\, Q$ and $p = \sqrt{2}\, p_{\zpf}\, P$, where $q_{\zpf} = \sqrt{\hbar/2 m \omega_{\m}}$ and $p_{\zpf} = \sqrt{\hbar m \omega_{\m}/2 }$ are the zero-point fluctuations of the oscillator position and momentum operators, respectively, such that $\comm{Q}{P}=i$.
Further, $g = -q_{\zpf} [ \partial\omega_{\mathrm{cav}}(q) / \partial q]_{q=0}$ is the single-photon optomechanical coupling strength, where $\omega_{\mathrm{cav}}(q)$ is the position-dependent frequency of the cavity mode with $\omega_{\mathrm{cav}}(0) = \omega_{\c}$;
while $f(t) = \sqrt{2}\, p_{\zpf} F(t)$ is the classical force to be measured, which is defined such that the force $F(t)$ has units of $\mathrm{Hz}$.
Moreover, $\E_{\pm}$ are the coherent driving strengths, which are in general complex numbers such that
\begin{align}\label{eq:app:amplitudes}
    \E_{\pm} = \widebar{\E}_{\pm} \e^{i \theta_{\pm}}, 
\end{align}
where $\widebar{\E}_{\pm}$ are real constants related to the input powers $\wp_{\pm}$ by 
\begin{align}\label{eq:app:amplitude_magnitudes}
    \widebar{\E}_{\pm} = \sqrt{ \frac{\kappa\, \wp_{\pm}}{\hbar\, \omega_{\pm}}\, },
\end{align}
with $\kappa$ the decay rate of the cavity mode, and the phases $\theta_{\pm}$ can be chosen at convenience as discussed below.
Since we are considering two-tone driving, the driving Hamiltonian in Eq.~\eqref{eq:app:two-tone_driving} is different from that considered in a canonical optomechanical scenario in which force sensing has previously been carefully considered.
We shall describe the dynamics of the mechanical oscillator by a set of generalised quantum Langevin equations~\cite{Ford1988, Giovannetti2001, Bowen2015}, while the electromagnetic field dynamics will be described by the input-output theory of quantum optics~\cite{Gardiner1985, Gardiner2004, Walls2007}.
The input-output formalism corresponds to a generalised quantum Langevin equation under a RWA on the system-reservoir interaction Hamiltonian.
This approximation is generally valid when the frequency of the subsystem is typically much greater than the system-reservoir coupling strength and any other relevant rate in the system.
Therefore, is suitable for the description of the electromagnetic field, but not always for the mechanical oscillator.
In fact, the RWA turns out to be a good approximation for the mechanical oscillator dynamics only when the mechanical quality factor $\mathcal{Q}_{\m}=\omega_{\m}/\gamma$ is such that $\mathcal{Q}_{\m} \gg 1$, and $\omega_{\m}^{-1}$ is faster than the time-scales associated with the phenomena of interest~\cite{Bowen2015}, restrictions that in principle we are not considering.
Thus, the Heisenberg-Langevin equations representing the system dynamics read
\begin{subequations}\label{eq:app:HLEs1}
    \begin{align}
        &\dot{Q} = \omega_{\m} P, \\
        &\dot{P} = -\omega_{\m} Q - \gamma P +\,  g\, \sqrt{2}\, a^{\dagger} a + F + \W, \\
        &\dot{a} = - \pqty{ i \omega_{\c} + \frac{\kappa}{2} } a + i  g\, \sqrt{2}\, Q\, a  \nonumber \\
        & \hspace{0.6cm} +  \pqty{\E_+  \e^{-i \omega_+ t} +\, \E_-  \e^{-i \omega_- t} } + \sqrt{\kappa}\, a_{\in},
    \end{align}
\end{subequations}
where $\W=\W(t)$ is the stochastic Langevin force due to the thermal mechanical reservoir, with the correlation function
\begin{equation}\label{eq:app:mechanical_correlation}
    \expval{\W(t) \W(t')} = \frac{\gamma}{\pi \omega_{\m}} \Bqty{ \mathcal{F}_{\mathrm{r}}(t-t') + i \mathcal{F}_{\mathrm{i}}(t-t') },
\end{equation}
where
\begin{subequations}
\begin{align}
    \mathcal{F}_{\mathrm{r}}(t) =& \int_0^{\varpi} \dd{\omega} \omega \cos{(\omega t)} \coth{(\hbar \omega/ 2 k_{\B} T)}, 
    \\ \mathcal{F}_{\mathrm{i}}(t) =& -\int_0^{\varpi} \dd{\omega} \omega \sin{(\omega t)},
\end{align}
\end{subequations}
such that the reservoir is assumed to be in thermal equilibrium at temperature $T$ and $\varpi$ is a cutoff frequency for the continuous spectrum of reservoir quantum harmonic oscillators~\cite{Giovannetti2001, Bowen2015}. 
In the high-temperature limit ($k_{\B} T \gg \hbar \omega$, $\varpi \to \infty$), $\coth{(\hbar \omega / 2 k_{\B} T)} \approx  2 k_{\B} T/\hbar \omega$, and the symmetric part of the correlation function $\mathcal{F}_{\mathrm{r}}$ becomes proportional to a Dirac delta function, $\mathcal{F}_{\mathrm{r}}(t) = (2 \pi k_{\B} T / \hbar)\,  \delta(t)$, whilst the antisymmetric part $\mathcal{F}_{\mathrm{i}}$ reduces to $\mathcal{F}_{\mathrm{i}}(t) = \pi \dot{ \delta }(t)$.
On the other hand, $a_{\in}=a_{\in}(t)$ is the input white noise of the electromagnetic quantum vacuum, which satisfies the correlation functions,
\begin{align}\label{eq:app:optical_correlation}
    \expval*{ a_{\in}(t)\, a_{\in}^{\dagger}(t') } &=  \delta(t-t'), \nonumber \\
    \expval*{ a_{\in}(t)\, a_{\in}(t') } =  \expval*{ a_{\in}^{\dagger}(t)\, a_{\in}^{\dagger}(t') } &= \expval*{ a_{\in}^{\dagger}(t)\, a_{\in}(t') } = 0,
\end{align}
where we assumed zero thermal photons in the electromagnetic field reservoir.
Further, the electromagnetic operators satisfy the input-output relation
\begin{align}\label{eq:app:input-output}
    a_{\out}(t)+a_{\in}(t) = \sqrt{\kappa}\, a(t),
\end{align}
where $a_{\out}(t)$ is associated with the output electromagnetic field~\cite{Gardiner1985,Walls2007}.
\subsection{Linearisation of the Heisenberg-Langevin equations}
In the regime where the coherent drive is strong enough to  efficiently extract information about the mechanical oscillator motion, the dynamics of the physical system is well described by linearising the Heisenberg-Langevin equations around the semiclassical steady-state, such that the operators correspond to semiclassical evolution plus quantum noise fluctuations.
The semiclassical  steady-state solutions to Eqs.~\eqref{eq:app:HLEs1}  are obtained from the expectation value of the Heisenberg-Langevin equations in the absence of external force and under a mean-field approximation,
\begin{subequations}
    \begin{align}
        \dot{\expval{Q}} =\, & \omega_{\m} \expval{P}, \label{eq:app:semiclassical:1}\\
        \dot{\expval{P}} =& -\omega_{\m} \expval{Q} - \gamma \expval{P} +\,  g\, \sqrt{2}\, |\alpha|^2, \label{eq:app:semiclassical:2} \\
        \dot{\alpha}\, \, \, =& - \pqty{ i \omega_{\c} + \frac{\kappa}{2} } \alpha + i g\, \sqrt{2} \expval{Q} \alpha  \nonumber \\
        &  \hspace{0.4cm} +  \pqty{\E_+  \e^{-i \omega_+ t} +\, \E_-  \e^{-i \omega_- t} },  \label{eq:app:semiclassical:3}
    \end{align}
\end{subequations}
where $\expval{Q}$, $\expval{P}$, and $\alpha = \expval{a}$ are all time-dependent.
Due to the coherent driving, $\expval{Q}$ and  $\expval{P}$ will oscillate around constant values and, therefore, they can be written as $\expval{Q} =  \widebar{Q} + \widetilde{Q}(t)$, $\expval{P} = \widebar{P} + \widetilde{P}(t)$, where $\widebar{Q}$ and $\widebar{P}$ are constant time averages whilst $\widetilde{Q}(t)$ and $\widetilde{P}(t)$ are oscillations around these averages.
Taking the time averages of Eqs.~\eqref{eq:app:semiclassical:1} and \eqref{eq:app:semiclassical:2} over a very long period of time after the system reached the steady-state, we have
\begin{subequations}
\begin{align}
    \widebar{Q} =&\, \frac{1}{t_1-t_0} \int_{t_0}^{t_1} \dd{t} \expval{Q} =\, \frac{g\, \sqrt{2}}{\omega_{\m}}\, \pqty{\widebar{a}_+^2 + \widebar{a}_-^2}, \label{eq:app:constant_q}\\
    \widebar{P} =&\, \frac{1}{t_1-t_0} \int_{t_0}^{t_1} \dd{t} \expval{P} =\,  0,
\end{align}
\end{subequations}
with $t_0$ a time in the steady-state and $t_1>t0$ much greater that all the time-scales involved in the problem.
Besides, if $|\widetilde{Q}(t)| \ll \kappa/(2\sqrt{2}\, g) $ which is the case for stable optomechanical systems~\cite{Bowen2015}, then, the contribution of $\widetilde{Q}(t)$ to Eq.~\eqref{eq:app:semiclassical:3} is negligible and the cavity field will be in a time-dependent coherent state $\ket{\alpha}$, having amplitude
\begin{align} 
    \alpha &= \widebar{a}_+\e^{-i \omega_+ t} +\, \widebar{a}_- \e^{-i \omega_- t};
\end{align}
where 
\begin{align}
    \widebar{a}_{\pm} = \frac{\E_{\pm}}{\kappa/2 - i \pqty{ g\, \sqrt{2}\, \widebar{Q} \pm  \omega_{\m} } }.
\end{align}
To simplify calculations, and without loss of generality, we may assume the steady-state amplitudes $\widebar{a}_{\pm}$ to be real.
This corresponds to adjusting the phase reference for the input coherent drives, such that in Eq.~\eqref{eq:app:amplitudes}, $\theta_{\pm} =  -\arctan{[ 2\, (g \sqrt{2}\, \widebar{Q} \pm  \omega_{\m} )/ \kappa]}$.
Thus, the steady-state electromagnetic amplitudes will be given by
\begin{align}\label{eq:app:steady-state_amplitudes}
    \widebar{a}_{\pm} = \frac{\widebar{\E}_{\pm}}{\sqrt{ (\kappa/2)^2 +  \pqty{ g \sqrt{2}\, \widebar{Q} \pm  \omega_{\m} }^2\, } },
\end{align}
which is a nonlinear equation for the amplitudes $\widebar{a}_{\pm}$,  given the definition of $\widebar{Q}$ in Eq.~\eqref{eq:app:constant_q}.
To linearise the Heisenberg-Langevin equations we make the replacements $Q \to \expval{Q} + Q$, $P \to \expval{P} + P$, and $a \to \expval{a} + a$. Since the coherent drive is assumed to be strong, then $\widebar{a}_{\pm}$ will be large in comparison to the other parameters involved in the problem and, in consequence, any interaction term (proportional to $g$) that is not enhanced by $\alpha$ will be neglected in the resulting linearised equations.
Thus, the linearised Heisenberg-Langevin equations take the form,
\begin{subequations}\label{eq:app:HLEs2}
    \begin{align}
        &\dot{  Q} = \omega_{\m}   P, \\
        &\dot{  P} = -\omega_{\m}   Q - \gamma   P + g\, \sqrt{2}\, (\alpha\,   a^{\dagger} + \alpha^*\,   a) + F + \W, \\
        &\dot{ a} = - \pqty{i \omega_{\c} + \frac{\kappa}{2}} a + i\,  g\, \sqrt{2}\, \alpha\,   Q\, +  \sqrt{\kappa}\, a_{\in}; \label{eq:app:linear_field}
    \end{align}
\end{subequations}
which can be obtained from the linearised Hamiltonian
\begin{align}\label{eq:app:linearized_hamiltonian}
    H =\, & \frac{\hbar \omega_{\m}}{2} (Q^2 + P^2) + \hbar \omega_{\c} a^{\dagger} a - \hbar \sqrt{2}\, g\, (\alpha\, a^{\dagger} + \alpha^* a)\, Q \nonumber \\
    &  - \hbar F\, Q,
\end{align}
It is useful to write the Heisenberg-Langevin equations corresponding to the Hamiltonian in Eq.~\eqref{eq:app:linearized_hamiltonian} in terms of creation and annihilation operators only, where $b = (Q + i P)/\sqrt{2}$ and $ b^{\dagger} = (Q - i P)/\sqrt{2}$  are introduced as the mechanical annihilation and creation operators, respectively. 
Thus,
\begin{subequations}
\begin{align}
    & \dot{a} = - \pqty{i \omega_{\c} + \frac{\kappa}{2}}   a + i\, g\, \alpha\, \pqty{b^{\dagger} + b} +  \sqrt{\kappa}\, a_{\in}, \\
    &\dot{b} = - i \omega_{\m} b  + \frac{\gamma}{2} \pqty{ b^{\dagger} - b } + i\, g\, \sqrt{2}\, \pqty{ \alpha\, a^{\dagger} + \alpha^*\,   a } \nonumber \\
    & \hspace{4.5cm} + \frac{i}{\sqrt{2}} \pqty{ F + \W },
\end{align}
\end{subequations}
and the linearised Hamiltonian takes the form
\begin{align}
    H =\, &  \hbar \omega_{\c} a^{\dagger} a + \hbar \omega_{\m} b^{\dagger} b - \hbar g\, (\alpha\, a^{\dagger} + \alpha^* a)\, \pqty{b^{\dagger} + b} \nonumber \\
    &  - \frac{\hbar F}{\sqrt{2}} \pqty{b^{\dagger} + b} .
\end{align}
This Hamiltonian is the starting point in the main text.

\subsection{Langevin force in the interaction picture}
Moving to the interaction picture (rotating frame), as explicitly shown in the main text, not only modifies the system Hamiltonian but also the interaction between system and reservoir contained in the open quantum system model used to describe the system dynamics. 
Hence, to avoid unwanted time dependencies in the Heisenberg-Langevin equations,  we consider a relatively narrow band around the frequency of the  mechanical oscillator, such that we may write the stochastic Langevin force $\W(t)$ as
\begin{align}\label{eq:app:stochastic_force}
    \W(t) = \widebar{\W}(t) \e^{-i \omega_{\m} t} +  \widebar{\W}^*(t) \e^{i \omega_{\m} t},
\end{align}
where $\widebar{\W}(t)$ is a slowly-varying stochastic amplitude which preserves the statistical properties of $\W(t)$.
Thus, if $\omega_{\m} \gg \gamma, \abs{\widebar{\W}}$, with $\abs{\widebar{\W}}$ the magnitude of $\widebar{\W}(t)$, then, we will be able to perform a further RWA, but now in the interaction of the mechanical oscillator with its reservoir.
It is important to note that the fulfillment of these conditions will require a weak coupling between system and reservoir.
The correlation functions involving $\widebar{\W}(t)$ and $\widebar{\W}^*(t)$ are the following,
\begin{subequations}\label{eq:app:langevin_corr_0}
\begin{align}
    &\langle \widebar{\W}(t)\, \widebar{\W}(t') \rangle = \langle \widebar{\W}^*(t)\, \widebar{\W}^*(t') \rangle = 0, \\
    &\langle \widebar{\W}(t)\, \widebar{\W}^*(t') \rangle = \nonumber \\
    &\frac{\gamma}{2 \pi \omega_{\m}} \Bigg\{ \int_0^{\varpi} \dd{\omega} \omega \bqty{\coth \pqty{\frac{\hbar \omega}{2 k_{\B} T} }  + 1 }\e^{ -i (\omega - \omega_{\m}) (t-t') }  \Bigg\},  \label{eq:langevin_corr_0:1} \\
    &\langle \widebar{\W}^*(t)\, \widebar{\W}(t') \rangle = \nonumber \\
    &\frac{\gamma}{2 \pi \omega_{\m}} \Bigg\{ \int_0^{\varpi} \dd{\omega} \omega \bqty{\coth \pqty{\frac{\hbar \omega}{2 k_{\B} T} }  -1 }\e^{ i (\omega - \omega_{\m}) (t-t') }  \Bigg\}. \label{eq:app:langevin_corr_0:2}
\end{align}
\end{subequations}
From this expressions the correlation functions in Eqs.~\eqref{eq:langevin_corr_1} may be calculated.

\subsection{Frequency correlation functions of the Langevin force quadratures}
In Sec. \ref{sec:theoretical_model} we showed that the Heisenberg-Langevin equations describing the dynamics of the mechanical and electromagnetic quadratures depend on the real and imaginary parts of $\widebar{\W}(t)$, i.e., $\widebar{\W}_{\r}(t)$ and $\widebar{\W}_{\i}(t)$, respectively.
Moreover, in order to evaluate the stationary force noise PSD $S_{F_{\mathrm{est}}}(\omega)$ in Sec. \ref{sec:stationary_force_sensing} and the initial conditions for the nonstationary measurement in Sec. \ref{sec:nonstationary_force_sensing}, we need the Fourier transform of the correlation functions involving $\widebar{\W}_{\r}(t)$ and $\widebar{\W}_{\i}(t)$ in Eqs.~\eqref{eq:langevin_corr_1}.
Thus, considering $\varpi \to \infty$, for $\abs{\omega} < \omega_{\m}$ we have,
\begin{subequations}\label{eq:app:fourier_langevin_force}
\begin{align}
    &\expval{ \widebar{\W}_{\r}(\omega) \widebar{\W}_{\r}(\omega') } = \expval{ \widebar{\W}_{\i}(\omega) \widebar{\W}_{\i}(\omega') } = \nonumber \\
    & \frac{\pi \gamma}{2 \omega_{\m}} \bigg\{ (\omega + \omega_{\m}) \bqty{ \coth{ \pqty{ \frac{\hbar\, (\omega + \omega_{\m})}{2 k_{\B} T} } } + 1} \nonumber \\
    & \hspace{0.59cm} +  (\omega - \omega_{\m}) \bqty{ \coth{ \pqty{ \frac{\hbar\, (\omega - \omega_{\m})}{2 k_{\B} T} } } + 1 } \bigg\}\, \delta(\omega + \omega'), \\
    & \expval{\widebar{\W}_{\r}(\omega) \widebar{\W}_{\i}(\omega')} =
    \expval{\widebar{\W}_{\i}(\omega) \widebar{\W}_{\r}(\omega')}^* = \nonumber \\
    & \frac{i \pi \gamma}{2 \omega_{\m}} \bigg\{ (\omega + \omega_{\m}) \bqty{ \coth{ \pqty{ \frac{\hbar\, (\omega + \omega_{\m})}{2 k_{\B} T} } } + 1} \nonumber \\
    & \hspace{0.59cm} -  (\omega - \omega_{\m}) \bqty{ \coth{ \pqty{ \frac{\hbar\, (\omega - \omega_{\m})}{2 k_{\B} T} } } + 1 } \bigg\}\, \delta(\omega + \omega').
\end{align}
\end{subequations}
Here,  it is important to take into account that $\coth{(\hbar \omega / 2 k_{\B} T)} = 2\,  \widebar{n}_{\mathrm{th}}(\omega) + 1 $, where $\widebar{n}_{\mathrm{th}}(\omega)$ is the Bose-Einstein occupation factor given by $ \widebar{n}_{\mathrm{th}}(\omega) = ( \e^{\hbar \omega/k_{\B} T } -1  )^{-1}$.
Now, for simplicity, we shall consider the high-temperature limit ($ k_{\B} T \gg \hbar \omega $), where the approximation $\coth{(\hbar \omega / 2 k_{\B} T)} \approx 2 k_{\B} T / \hbar \omega $ holds, and the correlation functions in Eqs.~\eqref{eq:app:fourier_langevin_force} reduce to 
\begin{subequations}
\begin{align}
    &\expval{ \widebar{\W}_{\r}(\omega) \widebar{\W}_{\r}(\omega') } = \expval{ \widebar{\W}_{\i}(\omega) \widebar{\W}_{\i}(\omega') } = \nonumber \\
    & \hspace{4.65cm} 2 \pi \gamma\, (\widebar{n}_{\mathrm{th}} + 1/2)\, \delta(\omega + \omega'), \label{eq:app:qp_corr_frequency}\\
    &\expval{\widebar{\W}_{\r}(\omega) \widebar{\W}_{\i}(\omega')} = \expval{\widebar{\W}_{\i}(\omega) \widebar{\W}_{\r}(\omega')}^* = \nonumber \\
    &\hspace{5.05cm}  i \pi \gamma\, \delta(\omega + \omega'); 
\end{align}
\end{subequations}
where  $ \widebar{n}_{\mathrm{th}} = \widebar{n}_{\mathrm{th}}(\omega_{\m}) $ corresponds to the mean number of thermal phonons in the reservoir.
%

% 
%------------------------------------------------------------------------------
% Wiener-Khinchin theorem
%
\section{Power spectral density and the Wiener-Khinchin theorem}
\label{app:wiener-khinchin}
In this appendix our goal is to define the PSD for a generic quantum noise process and to prove the stationary Wiener-Khinchin theorem, which relates the first-order correlation function of a given noise process to its stationary PSD as Fourier transform pairs.
For this purpose, it is necessary to apply a window function to the operator that represents the quantum noise process in order to guarantee the convergence of the integrals involved in the calculations.
Thus, we will first consider a rectangular window as it is commonly done, and then we will consider an exponential window which we use in this work for convenience in the analytical calculation of the PSD in the nonstationary regime.
\subsection{Power spectral density}
First, we consider the PSD of a quantum noise process represented by a generic operator $\mathcal{O}(t)$, which is defined as
\begin{align}\label{eq:app:periodogram_psd}
    S_{\o}(\omega, T_{\m}) =
    %& \big\langle | \O(\omega, T_{\m})|^2  \big\rangle \nonumber \\
    \frac{1}{T_{\m}}\, \big\langle \O^{\dagger}(\omega, T_{\m}) \O(\omega, T_{\m})\, \big\rangle,
\end{align}
where
\begin{align}\label{eq:app:truncated_ft}
    \O(\omega, T_{\m}) = \int_{-\infty}^{+\infty} \dd{t} \e^{i \omega t} \Pi_{T_{\m}}(t)\, \O(t)
\end{align}
is the truncated Fourier transform of $\mathcal{O}(t)$~\cite{Gea-Banacloche1990, Clerk2010}.
Here $T_{\m}$ is the measurement time, while $\Pi_{T_{\m}}(t)$ is a rectangular window function equal to one in the interval $(0,T_{\m})$ and zero elsewhere.
Therefore, from Eqs.~\eqref{eq:app:periodogram_psd} and \eqref{eq:app:truncated_ft}, it follows that the PSD takes the form
\begin{align}\label{eq:app:finite-time_psd_0}
    S_{\o}(\omega,T_{\m}) =
    \frac{1}{T_{\m}} \int_0^{T_{\m}} &\dd{t} \int_0^{T_{\m}} \dd{t'} \nonumber \\
    & \times \e^{-i \omega\, (t-t')}  \big\langle \O^{\dagger}(t) \O(t')\,  \big\rangle.
\end{align}
The definition of PSD given in Eq.~\eqref{eq:app:periodogram_psd} mimics the classical definition of the \emph{periodogram PSD estimator}~\cite{GroverBrown2012, Prabhu2014}, where the quantity $\O^{\dagger}(\omega, T_{\m}) \O(\omega, T_{\m})/T_{\m}$ will correspond here to the the \emph{periodogram} of the quantum signal $\O(t)$.
Now, we shall prove the nonstationary Wiener-Khinchin theorem,
which is finite measurement time generalisation of the stationary result we will see below.
For this purpose, we will follow an idea analogous to the classical result presented in Ref.~\cite{Dechant2015}.
Here, it is important to note that we call nonstationary Wiener-Khinchin theorem to a rule valid for nonstationary quantum noise processes that relates the two-time correlation function to the PSD. 
Therefore, reorganising the integration domain in the Eq.~\eqref{eq:app:finite-time_psd_0}, we have
\begin{align}
    S_{\o}(\omega,T_{\m}) =\, & \frac{1}{T_{\m}} \Bqty{ \int_0^{T_{\m}} \dd{t} \int_0^t \dd{t'} +  \int_0^{T_{\m}} \dd{t'} \int_0^{t'} \dd{t} } \nonumber \\
    & \hspace{1.45cm} \times \e^{-i \omega (t-t')}  \big\langle \O^{\dagger}(t) \O(t') \, \big\rangle,
\end{align}
and making a change of variable in each pair of integrals,
\begin{align}\label{eq:app:nonstationary_psd_00}
    & S_{\o}(\omega,T_{\m}) = \frac{1}{T_{\m}} \int_0^{T_{\m}} \dd{t} \int_0^t \dd{\tau} \e^{-i \omega \tau} \big\langle \O^{\dagger}(t) \O(t-\tau)\, \big\rangle  \nonumber \\
    & \hspace{0.5cm} \, +  \frac{1}{T_{\m}} \int_0^{T_{\m}} \dd{t'} \int_0^{t'} \dd{\tau}  \e^{i \omega \tau}  \big\langle \O^{\dagger}(t'-\tau) \O(t') \, \big\rangle,
\end{align}
where in each term $\tau$ was chosen in such a way that it is always positive, $\tau = t-t'$ in the first term and $\tau = t'-t$ in the second term. 
Since $t'$ is a dummy variable in the second term on the right-hand side of Eq.~\eqref{eq:app:nonstationary_psd_00}, and $\big\langle\O^{\dagger}(t-\tau) \O(t)\, \big\rangle^* = \big\langle \O^{\dagger}(t) \O(t-\tau)\, \big\rangle$, hence,
\begin{align}\label{eq:app:wk}
    S_{\o}(\omega,T_{\m}) =&\, 2 \Re \frac{1}{T_{\m}} \int_0^{T_{\m}} \dd{t} \int_0^t \dd{\tau}\, \e^{i \omega \tau}\, C(\tau, t).
\end{align}
where $ C(\tau, t) = \langle \O^{\dagger}(t-\tau) \O(t)\, \rangle$ is the first-order correlation function.
Furthermore, from Eq.~\eqref{eq:app:finite-time_psd_0} it is easy to prove that
\begin{align}\label{eq:app:wk_1}
    C(\tau, T_{\m}) = \bqty{ 1 + T_{\m}\, \pdv{T_{\m}} } \int_{-\infty}^{+\infty} \frac{\dd{\omega}}{2 \pi} \e^{-i \omega \tau} S_{\o}(\omega,T_{\m}),
\end{align}
which is valid for $\abs{\tau}<T_{\m}$.
Equations \eqref{eq:app:wk} and \eqref{eq:app:wk_1} constitute the nonstationary Wiener-Khinchin theorem, which is valid for both stationary and nonstationary signals.
Next, from  Eq.~\eqref{eq:app:wk} we shall prove the stationary Wiener-Khinchin theorem.
However, in order to do so, the quantum noise process represented by $\O(t)$ needs to be wide-sense stationary, i.e., the correlation function $C(\tau, t)$ must depend only on the time difference $\tau$.
This time-homogeneity condition is satisfied when a physical system described by $\O(t)$ is in a stationary steady-state, which is the situation that we will now consider.
Thus, we define the stationary first-order correlation function as
\begin{align}\label{eq:app:correlation}
     C(\tau) = \lim_{t \to \infty} C(\tau, t),
\end{align}
where the limit was included to reiterate that we are considering a system in its steady-state.
Hence, in the stationary regime, we may write Eq.~\eqref{eq:app:wk} as
\begin{align}\label{eq:app:wk-0}
    S_{\o}(\omega,T_{\m}) =&\, 2 \Re \frac{1}{T_{\m}} \int_0^{T_{\m}} \dd{t} \int_0^t \dd{\tau} \e^{i \omega \tau} C(\tau).
\end{align}
Here we can use the the integral identity
\begin{align}
    \int_0^{T_{\m}} \dd{t} \int_0^t \dd{\tau} g(\tau) = \int_0^{T_{\m}} \dd{\tau}\, g(\tau)\, (T_{\m}-\tau),
\end{align}
which can be easily demonstrated making $g(\tau)=\dd{f(\tau)}/\dd{\tau}$ (see Appendix C of Ref.~\cite{Kusse2006} for details).
Thus, Eq.~\eqref{eq:app:wk-0} takes the form
\begin{align}\label{eq:app:wk-1}
    S_{\o}(\omega,T) =&\, 2 \Re \frac{1}{T_{\m}} \int_0^{T_{\m}} \dd{\tau} \e^{i \omega \tau} C(\tau)\, (T_{\m}-\tau).
\end{align}
Now, we consider the limit of infinite measurement time ($T_{\m} \to \infty$), where the truncated PSD $S_{\o}(\omega,T_{\m})$ reduces to the the stationary PSD $S_{\o}(\omega)$,
\begin{align}\label{eq:app:infine-time-limit}
    S_{\o}(\omega) \equiv& \lim_{T_{\m} \to \infty} S_{\o}(\omega,T_{\m}).
\end{align}
Therefore, from Eq.~\eqref{eq:app:wk-1}, we have
\begin{align}\label{eq:app:wk-2}
    &S_{\o}(\omega) = 2 \Re \int_0^{\infty} \dd{\tau} \e^{i \omega \tau}  C(\tau) \nonumber \\
    & \hspace{0.8cm} = \int_{0}^{\infty} \dd{\tau} \e^{-i \omega \tau} C^*(\tau) + \int_0^{\infty} \dd{\tau} \e^{i \omega \tau} C(\tau).
\end{align}
Since the signal is wide-sense stationary, it is satisfied $C^*(-\tau) = C(\tau)$ and, then,
\begin{align}
    S_{\o}(\omega) &= \int_{-\infty}^0 \dd{\tau} \e^{i \omega \tau} C(\tau) + \int_0^{\infty} \dd{\tau} \e^{i \omega \tau} C(\tau) \nonumber \\
    &= \int_{-\infty}^{+\infty} \dd{\tau} \e^{i \omega \tau} C(\tau).
\end{align}
Finally, we got the stationary Wiener-Khinchin theorem, which relates the stationary PSD $S_{\o}(\omega)$ and the first-order correlation function $C(\tau)$ as Fourier transform pairs,
\begin{subequations}
\begin{align}
    & S_{\o}(\omega) = \mathcal{F} \big\{ C(\tau) \big\} = \int_{-\infty}^{\infty} \dd{\tau} \e^{i \omega \tau} C(\tau), \label{eq:app:wk_a}\\
    & C(\tau) = \mathcal{F}^{-1}\big\{ S_{\o}(\omega) \big\} =  \int_{-\infty}^{\infty} \frac{\dd{\omega}}{2 \pi} \e^{-i \omega \tau}  S_{\o}(\omega) \label{eq:app:wk_b};
\end{align}
\end{subequations}
where $\mathcal{F}\{ \boldsymbol{\cdot} \}$ is the Fourier transform with respect to $\tau$, while $\mathcal{F}^{-1}\{ \boldsymbol{\cdot} \}$ is its inverse transform.
From Eqs.~\eqref{eq:app:correlation} and \eqref{eq:app:wk_a}, we may write the stationary PSD $S_{\o}(\omega)$ explicitly as
\begin{align}\label{eq:app:psd_stationary}
    S_{\o}(\omega) =& \lim_{t \to \infty} \int_{-\infty}^{+\infty} \dd{\tau} \e^{i \omega\, \tau} \big\langle \O^{\dagger}(t) \O(t+\tau) \, \big\rangle.
\end{align}
Furthermore, if $\O(t)$ is Hermitian, writing $\O(t)$ as the inverse Fourier transform of $\O(\omega)$ in Eq.~\eqref{eq:app:psd_stationary} yields to
\begin{align}\label{eq:app:force_noise_spectrum_1}
    S_{\o}(\omega) =& \int_{-\infty}^{+\infty} \frac{\dd{\omega'}}{2\pi} \big\langle \O(\omega') \O(\omega)\, \big\rangle,
\end{align}
where we assumed the noise correlators $\langle \O(\omega') \O(\omega) \rangle$ to be proportional to a Dirac delta function of the form $\delta(\omega' + \omega)$, which is true in many applications including those examined in this work.
\subsection{Power spectral density: exponential window}
Now, we shall consider an alternative definition of PSD which relies on the use of an exponential window in the calculation of the involved Fourier transforms, which facilitates the analytical calculation of the PSD in the nonstationary regime.
Here we will show that although this definition is not standard, it also leads to the stationary Wiener-Khinchin theorem in the long measurement time limit.
Thus, we define the nonstationary PSD as
\begin{align}\label{eq:app:nonstationary_psd}
    S_{\o}(\omega, T_{\m}) =
    %& \big\langle | \O(\omega, T_{\m})|^2  \big\rangle \nonumber \\
    \frac{1}{T_{\m}}\, \big\langle \O^{\dagger}(-i \omega + 1/2 T_{\m}) \O(-i \omega + 1/2 T_{\m})\, \big\rangle,
\end{align}
where
\begin{align}\label{eq:app:windowed_ft}
    \O(-i \omega + 1/2 T_{\m}) = \int_{0}^{+\infty} \dd{t}\, \e^{- (-i \omega + 1/2 T_{\m})\, t} \O(t).
\end{align}
It is worth noting that Eqs.~\eqref{eq:app:nonstationary_psd} and \eqref{eq:app:windowed_ft} correspond to the classical definition of \emph{modified periodogram PSD estimator}~\cite{Prabhu2014}.
Now, following a procedure completely analogous to the one shown before, we arrive to the following expression
\begin{align}
    &S_{\o}(\omega, T_{\m}) =\, 2 \Re \frac{1}{T_{\m}} \int_0^{+\infty} \dd{t} \e^{-t/T_{\m}}\, \int_0^t \dd{\tau} \nonumber \\
    & \hspace{4cm} \times \e^{(i \omega + 1/2 T_{\m})\, \tau}\, C(\tau).
\end{align}
Here, we can use the the integral identity
\begin{align}
    \int_0^{+\infty} \dd{t} \e^{-t/T_{\m}} \int_0^t \dd{\tau} g(\tau) = T_{\m} \int_0^{+\infty} \dd{\tau}\, \e^{-\tau/T_{\m}}\, g(\tau),
\end{align}
which, as before, can be proved making $g(\tau)=\dd{f(\tau)}/\dd{\tau}$.
Therefore, we will have
\begin{align}
    S_{\o}(\omega, T_{\m}) =&\, 2 \Re \int_0^{+\infty} \dd{\tau} \e^{-\tau/T_{\m}}\, \e^{(i \omega + 1/2 T_{\m}) \tau} C(\tau),
\end{align}
which in the infinite measurement time limit yields to
\begin{align}\label{eq:app:wk-3}
    \lim_{T_{\m} \to \infty} S_{\o}(\omega, T_{\m}) =&\, 2 \Re \int_0^{+\infty} \dd{\tau} \e^{i \omega \tau} C(\tau).
\end{align}
Finally, since the right-hand side of Eq.~\eqref{eq:app:wk-3} correspond to the  right-hand side of the first part of Eq.~\eqref{eq:app:wk-2}, we arrive to the desired result
\begin{align}
    S_{\o}(\omega) = \lim_{T_{\m} \to \infty} S_{\o}(\omega, T_{\m}),
\end{align}
from which follows the stationary Wiener-Khinchin theorem described in Eqs.~\eqref{eq:app:wk_a} and \eqref{eq:app:wk_b}.
% 
%------------------------------------------------------------------------------
% Inverse filter
%
\section{Inverse filter}
\label{app:inverse_filter}
To filter the output signal of a linear measurement such as the one described in Eq.~\eqref{eq:yout_snr}, $Y_{\out}(t) = \D(t)*\widebar{F}_{\i}(t) + N(t)$, it is standard to apply an inverse filter to the complete measurement record in such a way that it is possible to recover the original signal.
In principle, the inverse filter provides an exact solution to the problem of recovering the signal of interest from a given measured output signal, however, when it is required to filter a signal in the transient nonstationary regime this solution is fraught with difficulties. 
To see this, we shall consider the particular  measurement under study, which is well described in Sec. \ref{sec:nonstationary_force_sensing}; nonetheless, the results shown here are valid for any linear measurement.
\subsection{Inverse filtering in time domain}
In time domain, the inverse filter approach corresponds to deconvolve the measurement record $Y_{\out}(t)$ using a linear filter with impulse response
\begin{align}
    h(t)=A^{-1}(t),
\end{align}
where $A^{-1}(t)$ is the convolution inverse of $A(t)$ satisfying 
\begin{align}\label{eq:app:inverse_convolution_condition}
    A^{-1}(t)*A(t) = \delta(t).
\end{align}
Taking the Fourier transform of Eq.~\eqref{eq:app:inverse_convolution_condition}, we can find that $A^{-1}(t)$ is given by
\begin{align}
    A^{-1}(t) = \mathcal{F}^{-1}\{ 1/ A(\omega) \};
\end{align}
where $A(\omega) = \mathcal{F}\{ A(t)\}$, being $\mathcal{F}\{ \boldsymbol{\cdot} \}$ the Fourier transform and $\mathcal{F}^{-1}\{ \boldsymbol{\cdot} \}$ its inverse.
Thus, the quantum estimator $F_{\mathrm{est}}(t) = h(t)*Y_{\out}(t)$ described in Eq.~\eqref{eq:f_est}, will take the form
\begin{align}\label{eq:app:stationary_spectral_measurement}
    F_{\mathrm{est}}(t) = A^{-1}(t)*Y_{\out}(t) = \widebar{F}_{\i}(t) + F_{\mathrm{noise}}(t), 
\end{align}
where 
\begin{align}
    F_{\mathrm{noise}}(t) = A^{-1}(t)*N(t).
\end{align}
From  Eq.~\eqref{eq:amplitude-t}, we can calculate $A^{-1}(t)$, which will be given by
\begin{align}
    A^{-1}(t) = -\frac{1}{\sqrt{\kappa}\, (G_- + G_+) }\, \D^{-1}(t)
\end{align}
where $\D^{-1}(t)$ is the convolution inverse of the Green's function $\D(t)$, which is given by
\begin{align}
    \D^{-1}(t) = \ddot{\delta}(t) + 2\Gamma\, \dot{\delta}(t) + \Omega^2\, \delta(t)
\end{align}
Thus, we may convolve $Y_{\out}(t)$ with $h(t)=A^{-1}(t)$ to obtain the quantum estimator,
\begin{align}
    F_{\mathrm{est}}(t) = -\frac{1}{\sqrt{\kappa}\, (G_- + G_+) }\, \D^{-1}(t)*Y_{\out}(t).
\end{align}
Now considering the output signal as given by Eq.~\eqref{eq:yout_1}, $Y_{\out}(t) =\, \sqrt{\kappa}\, \big[  Y_{\p}(t) + Y_{\h}(t) \big] - Y_{\in}(t)$, it is important to notice that
\begin{align}
    \D^{-1}(t)*Y_{\mathrm{h}}(t) = \ddot{Y_{\mathrm{h}}}(t) + 2\Gamma\, \dot{Y_{\mathrm{h}}}(t) + \Omega^2 Y_{\mathrm{h}}(t),
\end{align}
where the right-hand side is clearly zero since it corresponds to the definition of the homogeneous solution and, thus, $ \D^{-1}(t)*Y_{\mathrm{h}}(t) = 0$.
Hence,
\begin{align}
    F_{\mathrm{est}}(t) = -\frac{1}{\sqrt{\kappa}\, (G_- + G_+) }\, \D^{-1}(t)* \big[  \sqrt{\kappa}\, Y_{\p}(t) - Y_{\in}(t) \big]
\end{align}
and, therefore, all information about the initial conditions is lost, remaining in $F_{\mathrm{est}}(t)$ only the terms with information on the steady-state of the system.
\subsection{Inverse filtering in frequency domain}
In frequency domain the inverse filtering is completely equivalent to what was done in time domain.
Therefore, the outlook is not very encouraging either for the filtering of the signal in the nonstationary transient regime.
However, as a matter of completeness we shall describe the procedure. 
From Eqs.~\eqref{eq:f_est_omega} and \eqref{eq:filter_frequency_response} we have that 
\begin{align}
    F_{\mathrm{est}}(\omega) = \frac{Y_{\out}(\omega)}{A(\omega)},
\end{align}
where $Y_{\out}(\omega)$ may be calculated taking the Fourier transform of Eq.~\eqref{eq:yout_1}, such that 
\begin{align}
    Y_{\out}(\omega) =&\, \sqrt{\kappa}\, \big[  Y_{\p}(\omega) + Y_{\h}(\omega) \big] - Y_{\in}(\omega),
\end{align}
with $ Y_{\p}(\omega)$, $Y_{\mathrm{h}}(\omega)$, and $Y_{\in}(\omega)$, the Fourier transforms of $ Y_{\p}(t)$, $Y_{\mathrm{h}}(t)$, and $Y_{\in}(t)$, respectively.
Here, $Y_{\mathrm{h}}(\omega) = 0$, and using the explicit form of $A(\omega)$ in Eq.~\eqref{eq:signal_amplification}, we obtain that the frequency component of the estimated force quadrature takes the form
\begin{align}
    F_{\mathrm{est}}(\omega) = -\frac{\sqrt{\kappa}\,  Y_{\p}(\omega) - Y_{\in}(\omega) }{\sqrt{\kappa}\, (G_- + G_+)\, \D(\omega)}.
\end{align}
Therefore, the information on the initial conditions contained in $Y_{\mathrm{h}}(t)$ is as before eliminated.
% 
%------------------------------------------------------------------------------
% Stationary PSD
%
\section{Stationary force noise power spectral density}
\label{app:stationary_psd}
In this Appendix, we study the different regimes defined by the drive asymmetry $G_+/G_-$, and we consider the conditions under which the stationary force noise PSD $S_{F_{\mathrm{est}}}(\omega)$ in Eq.~\eqref{eq:force_spectrum_2} reduces to the thermal noise floor at resonance.
This conditions define the optimal configuration for the enhancement of the sensitivity of stationary force measurements in each regime.
For convenience in the analysis and presentation of the results, we shall use the added force noise PSD $S_{\mathrm{add}}(\omega)$ given by Eq.~\eqref{eq:added_noise}.
\subsection{Cavity-assisted mechanical sideband cooling \texorpdfstring{$(G_+=0,\, G_->0)$}{}}
A relevant limit to consider first is $G_+=0$, which corresponds to cavity-assisted sideband cooling (SBC) of mechanical motion \cite{Marquardt2007, Wilson-Rae2007}.
The added noise PSD in this case is
\begin{align}\label{eq:sbc}
    S_{\mathrm{add}}^{\mathrm{SBC}}(\omega)& = \frac{1}{2 \kappa} \bigg\{ \frac{1}{G_-^2} \bqty{ \pqty{\omega^2 + \frac{\gamma \kappa}{4}}^2 + \pqty{\frac{\omega}{2} }^2 (\gamma - \kappa)^2}  \nonumber \\
    & \hspace{2.2cm} +\, G_-^2 - 2 \pqty{ \omega^2 + \frac{\gamma \kappa}{4} }  \bigg\},
\end{align}
which at resonance becomes
\begin{align}\label{eq:sbc_resonance}
    S_{\mathrm{add}}^{\mathrm{SBC}}(0) = \frac{\gamma}{8} \pqty{ \frac{1}{C_-} + C_- - 2},
\end{align}
with
\begin{align}\label{eq:cooperativity_c-}
    C_- = \frac{4 G_-^2}{\gamma \kappa}
\end{align}
being the cooperativity associated with a red sideband drive.
At first glance one could think that the cooling of the harmonic oscillator to its ground state could help improve the sensitivity of the force sensor, since it eliminates the noise due to thermal fluctuations.
However, since the cooling is achieved by adding damping to the mechanical oscillator, not only the added noise is reduced but also the sensitivity to any external force \cite{Blair1991}.
Therefore, the sensitivity of the force measurement is not enhanced by the SBC protocol --- cf., Figs.~\ref{fig:force_noise_2}, \ref{fig:force_noise_3} and \ref{fig:optimal_drive_asymmetry} --- nevertheless, it will be used as a reference to compare other protocols.
Thus, minimising Eq.~\eqref{eq:sbc} with respect to $G_-^2$, we get,
\begin{subequations}
\begin{align}
    &(G_-^2)_{\mathrm{min}}^{\mathrm{SBC}} =\, \bqty{ \pqty{\omega^2 + \frac{\gamma \kappa}{4}}^2 + \pqty{\frac{\omega}{2}}^2 (\gamma - \kappa)^2 }^{1/2}, \\
    &\big[ S_{\mathrm{add}}^{\mathrm{SBC}}(\omega) \big]_\mathrm{min} =\, \frac{1}{\kappa} \Bigg\{ \bqty{ \pqty{\omega^2 + \frac{\gamma \kappa}{4}}^2 + \pqty{\frac{\omega}{2} }^2 (\gamma - \kappa)^2}^{1/2} \nonumber \\
    & \hspace{4.8cm} - \pqty{ \omega^2 + \frac{\gamma \kappa}{4} } \Bigg\}. \label{eq:psd-sbc_min}
\end{align}
\end{subequations}
At resonance, Eq.~\eqref{eq:psd-sbc_min} reduces  to
$ \big[ S_{\mathrm{add}}^{\mathrm{SBC}}(0) \big]_\mathrm{min} = 0$, which is achieved at coupling $C_-$ such that $(C_-)_{\mathrm{min}}^{\mathrm{SBC}} = 1$, as can be seen from Eq.~\eqref{eq:sbc_resonance}.
\subsection{Back-action evading measurement \texorpdfstring{$(G_- = G_+,\, G_->0)$}{}}
Early proposals for the enhancement of the sensitivity of single-quadrature force measurements relied on the idea of performing a BAE measurement of the mechanical quadrature carrying information on the force component of interest, such that the back-action due to the measurement is redirected to the unmeasured canonical conjugate quadrature \cite{Braginsky1980}. 
Using the two-tone driving scheme under consideration, we can tune the coherent drives such that $G_+=G_-$ and make a BAE measurement of $Q$ in order to sense $\widebar{F}_{\i}$, as can be followed from Eqs.~\eqref{eq:system_dynamics}.
This two-tone BAE measurement was originally due to Braginsky et al. \cite{Braginsky1980,Braginsky1992}, but was brought into the context of cavity quantum optomechanics/electromechanics in Refs. \cite{Clerk2008, Woolley2008}, and demonstrated experimentally in Refs. \cite{Hertzberg2010, Suh2014, Shomroni2019b}.
% first suggested
%
Nevertheless, despite it seems to be the most obvious approach for the ultrasensitive sensing of weak forces, there are limits for which a BAE measurement is not the best option for the enhancement of sensitivity in stationary force measurements, as we will see below.
Thus, setting $G_- = G_+$ in Eq.~\eqref{eq:force_spectrum_2}, the added force noise PSD reduces to 
\begin{align}\label{eq:bae_psd}
    S_{\mathrm{add}}^{\mathrm{BAE}}(\omega) = \frac{1}{2 \kappa} &\frac{ \pqty{\omega^2 + \gamma^2/4} \pqty{\omega^2 + \kappa^2/4} }{  4\, G_-^2},
\end{align}
where there is no contribution associated with quantum back-action noise due to radiation-pressure and the force noise PSD corresponds to imprecision noise only.
On resonance, we have
\begin{align}\label{eq:bae_1}
    S_{\mathrm{add}}^{\mathrm{BAE}}(0) =& \frac{ \gamma }{ 32\, C_- },
\end{align}
with $C_-$ as defined in Eq.~\eqref{eq:cooperativity_c-}. 
We can make the noise contribution in Eq.~\eqref{eq:bae_psd} arbitrarily small by simply increasing the driving strength.
Therefore, 
\begin{subequations}
\begin{align}
    &(G_-^2)_{\mathrm{min}}^{\mathrm{BAE}} \to \infty, \\
    &[S_{\mathrm{add}}^{\mathrm{BAE}}(\omega)]_{\mathrm{min}} = 0,
\end{align}
\end{subequations}
which clearly surpasses the sensitivity achieved using the SBC protocol given by Eq.~\eqref{eq:psd-sbc_min}.
In particular, from Eq.~\eqref{eq:bae_1} we have that the stationary force noise PSD at resonance may be written as 
\begin{align}
    S_{F_{\mathrm{est}}}^{\mathrm{BAE}}(0) =\, \gamma\, \bqty{ \frac{ 1 }{ 32\, C_- } +  ( \widebar{n}_{\mathrm{th}} + 1/2) },
\end{align}
and, therefore,
\begin{align}
    (C_-)_\mathrm{min}^\mathrm{BAE} \gg \frac{1}{32\, (\widebar{n}_{\th}+1/2)}
\end{align}
is a sufficient condition for neglecting the added force noise PSD at resonance under a BAE measurement protocol.
In fact, $(C_-)_\mathrm{min}^\mathrm{BAE} \gg 1/16$ is a valid condition for any thermal occupation $\widebar{n}_{\th}$.
\subsection{Steady-state mechanical squeezing \texorpdfstring{$(G_- \neq G_+,\, 0 < G_+/G_- < 1)$}{} }
A final approach to consider is the dissipative quantum squeezing of the variance of the mechanical position quadrature, which was initially proposed in Ref. \cite{Kronwald2013} and demonstrated recently in various experiments reported in Refs. \cite{Lecocq2015,  Wollman2015, Pirkkalainen2015, Lei2016}.
The squeezing procedure relies on the asymmetry between the input coherent drives ($G_+ \neq G_-$) adding back-action that allows cooling the mechanical oscillator to a squeezed ground state.
This procedure, as in the SBC protocol, adds additional damping to the system which reduces the sensitivity of the force sensor.
Nevertheless, the combination of both effects allows us to have in certain limits a sensitivity comparable to the BAE measurement and even better.
Thus, if $G_- \neq G_+$ and $0 < G_+/G_- < 1$, we can have steady-state mechanical squeezing (SMS) and the stationary force noise PSD in Eq.~\eqref{eq:force_spectrum_2} may be rewritten in terms of $\G$ and $r$, which are given by
\begin{subequations}
\begin{align}
    &\G^2 = G_-^2 - G_+^2, \label{eq:g2} \\
    &\tanh{r} = \frac{G_+}{G_-}.
\end{align}
\end{subequations}
Further, the condition $G_- > G_+$ guarantees the stability of the system \cite{Kronwald2013}.
%
% Thus, $G_- = \G \cosh{r}$ and $G_+ = \G \sinh{r}$.
%
Therefore, the added force noise PSD under SMS is
\begin{align}\label{eq:force_spectrum_3}
    S_{\mathrm{add}}^{\mathrm{SMS}}(\omega) =& \frac{\e^{-2r}}{2 \kappa} \Bigg[  \frac{\pqty{\omega^2 + \gamma^2/4} \pqty{\omega^2 + \kappa^2/4}}{\G^2}  + \G^2 \nonumber \\
    & \hspace{2.8cm} - 2\, \pqty{ \omega^2 + \gamma \kappa/ 4 } \Bigg].
\end{align}
At resonance ($\omega=0$), the added force noise PSD may be written in terms of the cooperativity
\begin{align}\label{eq:cooperativity_c}
    C = \frac{4 \G^2}{\gamma \kappa},
\end{align}
as
\begin{align}\label{eq:sms_resonance}
    S_{\mathrm{add}}^{\mathrm{SMS}}(0) = \frac{\gamma \e^{-2r}}{8} \pqty{ \frac{1}{C} + C - 2}.
\end{align}
We minimise the expression in Eq.~ \eqref{eq:force_spectrum_3} with respect to $\G^2$ for a fixed $\omega$, and we find
\begin{subequations}
\begin{align}
    &(\G^2)_{\mathrm{min}} =\, \Big[ \pqty{\omega^2 + \gamma^2/4} \pqty{\omega^2 + \kappa^2/4} \Big]^{1/2},   \\
    &\big[ S_{\mathrm{add}}^{\mathrm{SMS}}(\omega) \big]_\mathrm{min} =\, \frac{\e^{-2r}}{\kappa} \bigg\{ \Big[ \pqty{\omega^2 + \gamma^2/4} \pqty{\omega^2 + \kappa^2/4} \Big]^{1/2} \nonumber \\
    & \hspace{3.75cm} - \pqty{ \omega^2 + \gamma \kappa/4 } \bigg\} \geq 0. \label{eq:s_min}
\end{align}
\end{subequations}
Similarly, minimising the expression on Eq.~\eqref{eq:sms_resonance}, we have, $\big[ S_{\mathrm{add}}^{\mathrm{SMS}}(0) \big]_\mathrm{min}  = 0$ with $(C)_{\mathrm{min}}^{\mathrm{SMS}} = 1$.
\subsection{Summary of stationary force sensing under two-tone driving}
\begin{table}[ht]
    \bigskip
    \centering
    %\small
    \setlength{\tabcolsep}{7.2pt} % General space between cols (6pt standard)
    \begin{tabular}{S|SSS} 
        \toprule[1pt]\midrule[0.3pt]
        %\toprule
        \\ [-0.5em] \\ [-1.6em]
        { } & {SBC} & {BAE} & {SMS} \\
        \\ [-0.65em]
        %{ } & {$G_- \neq G_+$,} & {$G_-=G_+$,} & {$G_+=0$,} \\ 
        %{ } & {$G_->G_+$.} & {$G_->0$.} & {$G_->0$.} \\ \midrule
        { } & {$\frac{G_+}{G_-} = 0$} & {$\frac{G_+}{G_-} = 1$} & {$0 < \frac{G_+}{G_-} < 1$} \\ 
        \\ [-0.75em]
        \midrule
        %{$S_{F_{\mathrm{est}}}(\omega) =  S_{\mathrm{th}}$}    &  {---}  & {$G_- \to \infty$} & {---}  \\
        %\\ [-0.75em]
        \\ [-0.25em]
        {$S_{F_{\mathrm{est}}}(0) =  S_{\mathrm{th}}$}  & {$C_- = 1$} & {$C_- \gg \frac{1}{16} $} & {$C = 1$} \\
        \\ [-0.5em]
        \midrule[0.3pt]\bottomrule[1pt]
        %\bottomrule
    \end{tabular}
    \caption{Summary of the conditions for which the added force noise power spectral density reduces to the thermal noise floor. SMS stands for steady-state mechanical squeezing, BAE for back-action evasion measurement, and SBC for cavity-assisted side-band cooling. The cooperativities $C_-$ and $C$ are defined in Eqs.~\eqref{eq:cooperativity_c-} and \eqref{eq:cooperativity_c}, respectively.}
    \label{tab:summary}
\end{table}
We have considered three different operating conditions for the stationary sensing of a weak classical force, which where classified according to the values that the drive asymmetry $G_+/G_-$ can take.
First, we considered cavity assisted SBC, where $G_+/G_-=0$ and the system is dissipative cooled to its ground state thanks to an extra damping that also reduces the sensitivity of the system to an external force.
Second, we considered a BAE of the mechanical position quadrature, which can be achieved if $G_+/G_-= 0$.
We found that BAE measurement correspond in most of the cases to the optimal configuration for the measurement of classical forces in the steady-state.
Finally, we considered the regime where $0<G_+/G_-<1$, which can lead to dissipative cooling to a squeezed mechanical state through extra damping. 
Thus, although more damping is added to the system, the quantum fluctuations associated with one of the mechanical quadratures are also being reduced and, therefore, a trade-off of this two effects can lead to a sensitivity comparable or even better than that due to a BAE measurement.
The three studied cases are represented graphically in Fig. \ref{fig:force_noise_2}, where we show the stationary force noise PSD scaled by the thermal noise floor [$S_{F_{\mathrm{est}}}(\omega)/S_{\mathrm{th}}$] as a function of the dimensionless frequency $\omega/\kappa$ for different values of the drive asymmetry $G_+/G_-$.
We found that as a consequence of the mutual cancellation of the noise contributions due to imprecision and radiation-pressure,  the proposed two-tone driving scheme allows one to reduce the stationary force noise PSD to the thermal noise floor at resonance, i.e., $S_{F_{\mathrm{est}}}(0) = S_{\mathrm{th}}$ and $S_{\mathrm{add}}(0) = 0$. 
The conditions under which this is possible are summarised in Table \ref{tab:summary}.
%
%------------------------------------------------------------------------------

%------------------------------------------------------------------------------
% Bibliography
%
\bibliography{ms.bib}

\begin{thebibliography}{83}%
\makeatletter
\providecommand \@ifxundefined [1]{%
 \@ifx{#1\undefined}
}%
\providecommand \@ifnum [1]{%
 \ifnum #1\expandafter \@firstoftwo
 \else \expandafter \@secondoftwo
 \fi
}%
\providecommand \@ifx [1]{%
 \ifx #1\expandafter \@firstoftwo
 \else \expandafter \@secondoftwo
 \fi
}%
\providecommand \natexlab [1]{#1}%
\providecommand \enquote  [1]{``#1''}%
\providecommand \bibnamefont  [1]{#1}%
\providecommand \bibfnamefont [1]{#1}%
\providecommand \citenamefont [1]{#1}%
\providecommand \href@noop [0]{\@secondoftwo}%
\providecommand \href [0]{\begingroup \@sanitize@url \@href}%
\providecommand \@href[1]{\@@startlink{#1}\@@href}%
\providecommand \@@href[1]{\endgroup#1\@@endlink}%
\providecommand \@sanitize@url [0]{\catcode `\\12\catcode `\$12\catcode
  `\&12\catcode `\#12\catcode `\^12\catcode `\_12\catcode `\%12\relax}%
\providecommand \@@startlink[1]{}%
\providecommand \@@endlink[0]{}%
\providecommand \url  [0]{\begingroup\@sanitize@url \@url }%
\providecommand \@url [1]{\endgroup\@href {#1}{\urlprefix }}%
\providecommand \urlprefix  [0]{URL }%
\providecommand \Eprint [0]{\href }%
\providecommand \doibase [0]{https://doi.org/}%
\providecommand \selectlanguage [0]{\@gobble}%
\providecommand \bibinfo  [0]{\@secondoftwo}%
\providecommand \bibfield  [0]{\@secondoftwo}%
\providecommand \translation [1]{[#1]}%
\providecommand \BibitemOpen [0]{}%
\providecommand \bibitemStop [0]{}%
\providecommand \bibitemNoStop [0]{.\EOS\space}%
\providecommand \EOS [0]{\spacefactor3000\relax}%
\providecommand \BibitemShut  [1]{\csname bibitem#1\endcsname}%
\let\auto@bib@innerbib\@empty
%</preamble>
\bibitem [{\citenamefont {Braginsky}\ \emph {et~al.}(1980)\citenamefont
  {Braginsky}, \citenamefont {Vorontsov},\ and\ \citenamefont
  {Thorne}}]{Braginsky1980}%
  \BibitemOpen
  \bibfield  {author} {\bibinfo {author} {\bibfnamefont {V.~B.}\ \bibnamefont
  {Braginsky}}, \bibinfo {author} {\bibfnamefont {Y.~I.}\ \bibnamefont
  {Vorontsov}},\ and\ \bibinfo {author} {\bibfnamefont {K.~S.}\ \bibnamefont
  {Thorne}},\ }\bibfield  {title} {\bibinfo {title} {{Quantum Nondemolition
  Measurements}},\ }\href {https://doi.org/10.1126/science.209.4456.547}
  {\bibfield  {journal} {\bibinfo  {journal} {Science}\ }\textbf {\bibinfo
  {volume} {209}},\ \bibinfo {pages} {547} (\bibinfo {year}
  {1980})}\BibitemShut {NoStop}%
\bibitem [{\citenamefont {Caves}\ \emph {et~al.}(1980)\citenamefont {Caves},
  \citenamefont {Thorne}, \citenamefont {Drever}, \citenamefont {Sandberg},\
  and\ \citenamefont {Zimmermann}}]{Caves1980}%
  \BibitemOpen
  \bibfield  {author} {\bibinfo {author} {\bibfnamefont {C.~M.}\ \bibnamefont
  {Caves}}, \bibinfo {author} {\bibfnamefont {K.~S.}\ \bibnamefont {Thorne}},
  \bibinfo {author} {\bibfnamefont {R.~W.~P.}\ \bibnamefont {Drever}}, \bibinfo
  {author} {\bibfnamefont {V.~D.}\ \bibnamefont {Sandberg}},\ and\ \bibinfo
  {author} {\bibfnamefont {M.}~\bibnamefont {Zimmermann}},\ }\bibfield  {title}
  {\bibinfo {title} {{On the measurement of a weak classical force coupled to a
  quantum-mechanical oscillator. I. Issues of principle}},\ }\href
  {https://doi.org/10.1103/RevModPhys.52.341} {\bibfield  {journal} {\bibinfo
  {journal} {Rev. Mod. Phys.}\ }\textbf {\bibinfo {volume} {52}},\ \bibinfo
  {pages} {341} (\bibinfo {year} {1980})}\BibitemShut {NoStop}%
\bibitem [{\citenamefont {Caves}\ and\ \citenamefont
  {Milburn}(1987)}]{Caves1987}%
  \BibitemOpen
  \bibfield  {author} {\bibinfo {author} {\bibfnamefont {C.~M.}\ \bibnamefont
  {Caves}}\ and\ \bibinfo {author} {\bibfnamefont {G.~J.}\ \bibnamefont
  {Milburn}},\ }\bibfield  {title} {\bibinfo {title} {{Quantum-mechanical model
  for continuous position measurements}},\ }\href
  {https://doi.org/10.1103/PhysRevA.36.5543} {\bibfield  {journal} {\bibinfo
  {journal} {Phys. Rev. A}\ }\textbf {\bibinfo {volume} {36}},\ \bibinfo
  {pages} {5543} (\bibinfo {year} {1987})}\BibitemShut {NoStop}%
\bibitem [{\citenamefont {Braginsky}\ \emph {et~al.}(1992)\citenamefont
  {Braginsky}, \citenamefont {Khalili},\ and\ \citenamefont
  {Thorne}}]{Braginsky1992}%
  \BibitemOpen
  \bibfield  {author} {\bibinfo {author} {\bibfnamefont {V.~B.}\ \bibnamefont
  {Braginsky}}, \bibinfo {author} {\bibfnamefont {F.~Y.}\ \bibnamefont
  {Khalili}},\ and\ \bibinfo {author} {\bibfnamefont {K.~S.}\ \bibnamefont
  {Thorne}},\ }\href {https://doi.org/10.1017/CBO9780511622748} {\emph
  {\bibinfo {title} {Quantum measurement}}}\ (\bibinfo  {publisher} {Cambridge
  University Press},\ \bibinfo {address} {Cambridge},\ \bibinfo {year}
  {1992})\BibitemShut {NoStop}%
\bibitem [{\citenamefont {Bocko}\ and\ \citenamefont
  {Onofrio}(1996)}]{Bocko1996}%
  \BibitemOpen
  \bibfield  {author} {\bibinfo {author} {\bibfnamefont {M.~F.}\ \bibnamefont
  {Bocko}}\ and\ \bibinfo {author} {\bibfnamefont {R.}~\bibnamefont
  {Onofrio}},\ }\bibfield  {title} {\bibinfo {title} {On the measurement of a
  weak classical force coupled to a harmonic oscillator: experimental
  progress},\ }\href {https://doi.org/10.1103/RevModPhys.68.755} {\bibfield
  {journal} {\bibinfo  {journal} {Rev. Mod. Phys.}\ }\textbf {\bibinfo {volume}
  {68}},\ \bibinfo {pages} {755} (\bibinfo {year} {1996})}\BibitemShut
  {NoStop}%
\bibitem [{\citenamefont {Chen}(2013)}]{Chen2013}%
  \BibitemOpen
  \bibfield  {author} {\bibinfo {author} {\bibfnamefont {Y.}~\bibnamefont
  {Chen}},\ }\bibfield  {title} {\bibinfo {title} {{Macroscopic quantum
  mechanics: theory and experimental concepts of optomechanics}},\ }\href
  {https://doi.org/10.1088/0953-4075/46/10/104001} {\bibfield  {journal}
  {\bibinfo  {journal} {J. Phys. B}\ }\textbf {\bibinfo {volume} {46}},\
  \bibinfo {pages} {104001} (\bibinfo {year} {2013})}\BibitemShut {NoStop}%
\bibitem [{\citenamefont {Danilishin}\ and\ \citenamefont
  {Khalili}(2012)}]{Danilishin2012}%
  \BibitemOpen
  \bibfield  {author} {\bibinfo {author} {\bibfnamefont {S.~L.}\ \bibnamefont
  {Danilishin}}\ and\ \bibinfo {author} {\bibfnamefont {F.~Y.}\ \bibnamefont
  {Khalili}},\ }\bibfield  {title} {\bibinfo {title} {{Quantum Measurement
  Theory in Gravitational-Wave Detectors}},\ }\href
  {https://doi.org/10.12942/lrr-2012-5} {\bibfield  {journal} {\bibinfo
  {journal} {Living Rev. Relativ.}\ }\textbf {\bibinfo {volume} {15}},\
  \bibinfo {pages} {5} (\bibinfo {year} {2012})}\BibitemShut {NoStop}%
\bibitem [{\citenamefont {Zeuthen}\ \emph {et~al.}(2019)\citenamefont
  {Zeuthen}, \citenamefont {Polzik},\ and\ \citenamefont
  {Khalili}}]{Zeuthen2019}%
  \BibitemOpen
  \bibfield  {author} {\bibinfo {author} {\bibfnamefont {E.}~\bibnamefont
  {Zeuthen}}, \bibinfo {author} {\bibfnamefont {E.~S.}\ \bibnamefont
  {Polzik}},\ and\ \bibinfo {author} {\bibfnamefont {F.~Y.}\ \bibnamefont
  {Khalili}},\ }\href {http://arxiv.org/abs/1908.03416} {\bibinfo {title}
  {{Gravitational wave detection beyond the standard quantum limit using a
  negative-mass spin system and virtual rigidity}}} (\bibinfo {year} {2019}),\
  \Eprint {https://arxiv.org/abs/1908.03416} {arXiv:1908.03416 [gr-qc]}
  \BibitemShut {NoStop}%
\bibitem [{\citenamefont {Milburn}\ \emph {et~al.}(1994)\citenamefont
  {Milburn}, \citenamefont {Jacobs},\ and\ \citenamefont
  {Walls}}]{Milburn1994}%
  \BibitemOpen
  \bibfield  {author} {\bibinfo {author} {\bibfnamefont {G.~J.}\ \bibnamefont
  {Milburn}}, \bibinfo {author} {\bibfnamefont {K.}~\bibnamefont {Jacobs}},\
  and\ \bibinfo {author} {\bibfnamefont {D.~F.}\ \bibnamefont {Walls}},\
  }\bibfield  {title} {\bibinfo {title} {{Quantum-limited measurements with the
  atomic force microscope}},\ }\href {https://doi.org/10.1103/PhysRevA.50.5256}
  {\bibfield  {journal} {\bibinfo  {journal} {Phys. Rev. A}\ }\textbf {\bibinfo
  {volume} {50}},\ \bibinfo {pages} {5256} (\bibinfo {year}
  {1994})}\BibitemShut {NoStop}%
\bibitem [{\citenamefont {Butt}\ \emph {et~al.}(2005)\citenamefont {Butt},
  \citenamefont {Cappella},\ and\ \citenamefont {Kappl}}]{Butt2005}%
  \BibitemOpen
  \bibfield  {author} {\bibinfo {author} {\bibfnamefont {H.-J.}\ \bibnamefont
  {Butt}}, \bibinfo {author} {\bibfnamefont {B.}~\bibnamefont {Cappella}},\
  and\ \bibinfo {author} {\bibfnamefont {M.}~\bibnamefont {Kappl}},\ }\bibfield
   {title} {\bibinfo {title} {{Force measurements with the atomic force
  microscope: Technique, interpretation and applications}},\ }\href
  {https://doi.org/10.1016/j.surfrep.2005.08.003} {\bibfield  {journal}
  {\bibinfo  {journal} {Surf. Sci. Rep.}\ }\textbf {\bibinfo {volume} {59}},\
  \bibinfo {pages} {1} (\bibinfo {year} {2005})}\BibitemShut {NoStop}%
\bibitem [{\citenamefont {Poggio}\ and\ \citenamefont
  {Degen}(2010)}]{Poggio2010}%
  \BibitemOpen
  \bibfield  {author} {\bibinfo {author} {\bibfnamefont {M.}~\bibnamefont
  {Poggio}}\ and\ \bibinfo {author} {\bibfnamefont {C.~L.}\ \bibnamefont
  {Degen}},\ }\bibfield  {title} {\bibinfo {title} {{Force-detected nuclear
  magnetic resonance: recent advances and future challenges}},\ }\href
  {https://doi.org/10.1088/0957-4484/21/34/342001} {\bibfield  {journal}
  {\bibinfo  {journal} {Nanotechnology}\ }\textbf {\bibinfo {volume} {21}},\
  \bibinfo {pages} {342001} (\bibinfo {year} {2010})}\BibitemShut {NoStop}%
\bibitem [{\citenamefont {Davuluri}(2016)}]{Davuluri2016a}%
  \BibitemOpen
  \bibfield  {author} {\bibinfo {author} {\bibfnamefont {S.}~\bibnamefont
  {Davuluri}},\ }\bibfield  {title} {\bibinfo {title} {Optomechanics for
  absolute rotation detection},\ }\href
  {https://doi.org/10.1103/PhysRevA.94.013808} {\bibfield  {journal} {\bibinfo
  {journal} {Phys. Rev. A}\ }\textbf {\bibinfo {volume} {94}},\ \bibinfo
  {pages} {013808} (\bibinfo {year} {2016})}\BibitemShut {NoStop}%
\bibitem [{\citenamefont {Davuluri}\ and\ \citenamefont
  {Li}(2016)}]{Davuluri2016b}%
  \BibitemOpen
  \bibfield  {author} {\bibinfo {author} {\bibfnamefont {S.}~\bibnamefont
  {Davuluri}}\ and\ \bibinfo {author} {\bibfnamefont {Y.}~\bibnamefont {Li}},\
  }\bibfield  {title} {\bibinfo {title} {Absolute rotation detection by
  coriolis force measurement using optomechanics},\ }\href
  {https://doi.org/10.1088/1367-2630/18/10/103047} {\bibfield  {journal}
  {\bibinfo  {journal} {New Journal of Physics}\ }\textbf {\bibinfo {volume}
  {18}},\ \bibinfo {pages} {103047} (\bibinfo {year} {2016})}\BibitemShut
  {NoStop}%
\bibitem [{\citenamefont {Carney}\ \emph
  {et~al.}(2019{\natexlab{a}})\citenamefont {Carney}, \citenamefont {Ghosh},
  \citenamefont {Krnjaic},\ and\ \citenamefont {Taylor}}]{Carney2019b}%
  \BibitemOpen
  \bibfield  {author} {\bibinfo {author} {\bibfnamefont {D.}~\bibnamefont
  {Carney}}, \bibinfo {author} {\bibfnamefont {S.}~\bibnamefont {Ghosh}},
  \bibinfo {author} {\bibfnamefont {G.}~\bibnamefont {Krnjaic}},\ and\ \bibinfo
  {author} {\bibfnamefont {J.~M.}\ \bibnamefont {Taylor}},\ }\href
  {http://arxiv.org/abs/1903.00492} {\bibinfo {title} {{Gravitational Direct
  Detection of Dark Matter}}} (\bibinfo {year} {2019}{\natexlab{a}}),\ \Eprint
  {https://arxiv.org/abs/1903.00492} {arXiv:1903.00492 [hep-ph]} \BibitemShut
  {NoStop}%
\bibitem [{\citenamefont {Carney}\ \emph
  {et~al.}(2019{\natexlab{b}})\citenamefont {Carney}, \citenamefont {Hook},
  \citenamefont {Liu}, \citenamefont {Taylor},\ and\ \citenamefont
  {Zhao}}]{Carney2019c}%
  \BibitemOpen
  \bibfield  {author} {\bibinfo {author} {\bibfnamefont {D.}~\bibnamefont
  {Carney}}, \bibinfo {author} {\bibfnamefont {A.}~\bibnamefont {Hook}},
  \bibinfo {author} {\bibfnamefont {Z.}~\bibnamefont {Liu}}, \bibinfo {author}
  {\bibfnamefont {J.~M.}\ \bibnamefont {Taylor}},\ and\ \bibinfo {author}
  {\bibfnamefont {Y.}~\bibnamefont {Zhao}},\ }\href
  {http://arxiv.org/abs/1908.04797} {\bibinfo {title} {{Ultralight dark matter
  detection with mechanical quantum sensors}}} (\bibinfo {year}
  {2019}{\natexlab{b}}),\ \Eprint {https://arxiv.org/abs/1908.04797}
  {arXiv:1908.04797 [hep-ph]} \BibitemShut {NoStop}%
\bibitem [{\citenamefont {Bose}\ \emph {et~al.}(2017)\citenamefont {Bose},
  \citenamefont {Mazumdar}, \citenamefont {Morley}, \citenamefont {Ulbricht},
  \citenamefont {Toro\ifmmode~\check{s}\else \v{s}\fi{}}, \citenamefont
  {Paternostro}, \citenamefont {Geraci}, \citenamefont {Barker}, \citenamefont
  {Kim},\ and\ \citenamefont {Milburn}}]{Bose2017}%
  \BibitemOpen
  \bibfield  {author} {\bibinfo {author} {\bibfnamefont {S.}~\bibnamefont
  {Bose}}, \bibinfo {author} {\bibfnamefont {A.}~\bibnamefont {Mazumdar}},
  \bibinfo {author} {\bibfnamefont {G.~W.}\ \bibnamefont {Morley}}, \bibinfo
  {author} {\bibfnamefont {H.}~\bibnamefont {Ulbricht}}, \bibinfo {author}
  {\bibfnamefont {M.}~\bibnamefont {Toro\ifmmode~\check{s}\else \v{s}\fi{}}},
  \bibinfo {author} {\bibfnamefont {M.}~\bibnamefont {Paternostro}}, \bibinfo
  {author} {\bibfnamefont {A.~A.}\ \bibnamefont {Geraci}}, \bibinfo {author}
  {\bibfnamefont {P.~F.}\ \bibnamefont {Barker}}, \bibinfo {author}
  {\bibfnamefont {M.~S.}\ \bibnamefont {Kim}},\ and\ \bibinfo {author}
  {\bibfnamefont {G.}~\bibnamefont {Milburn}},\ }\bibfield  {title} {\bibinfo
  {title} {Spin entanglement witness for quantum gravity},\ }\href
  {https://doi.org/10.1103/PhysRevLett.119.240401} {\bibfield  {journal}
  {\bibinfo  {journal} {Phys. Rev. Lett.}\ }\textbf {\bibinfo {volume} {119}},\
  \bibinfo {pages} {240401} (\bibinfo {year} {2017})}\BibitemShut {NoStop}%
\bibitem [{\citenamefont {Marletto}\ and\ \citenamefont
  {Vedral}(2017)}]{Marletto2017}%
  \BibitemOpen
  \bibfield  {author} {\bibinfo {author} {\bibfnamefont {C.}~\bibnamefont
  {Marletto}}\ and\ \bibinfo {author} {\bibfnamefont {V.}~\bibnamefont
  {Vedral}},\ }\bibfield  {title} {\bibinfo {title} {{Gravitationally Induced
  Entanglement between Two Massive Particles is Sufficient Evidence of Quantum
  Effects in Gravity}},\ }\href
  {https://doi.org/10.1103/PhysRevLett.119.240402} {\bibfield  {journal}
  {\bibinfo  {journal} {Phys. Rev. Lett.}\ }\textbf {\bibinfo {volume} {119}},\
  \bibinfo {pages} {240402} (\bibinfo {year} {2017})}\BibitemShut {NoStop}%
\bibitem [{\citenamefont {Carney}\ \emph
  {et~al.}(2019{\natexlab{c}})\citenamefont {Carney}, \citenamefont {Stamp},\
  and\ \citenamefont {Taylor}}]{Carney2019a}%
  \BibitemOpen
  \bibfield  {author} {\bibinfo {author} {\bibfnamefont {D.}~\bibnamefont
  {Carney}}, \bibinfo {author} {\bibfnamefont {P.~C.~E.}\ \bibnamefont
  {Stamp}},\ and\ \bibinfo {author} {\bibfnamefont {J.~M.}\ \bibnamefont
  {Taylor}},\ }\bibfield  {title} {\bibinfo {title} {Tabletop experiments for
  quantum gravity: a user's manual},\ }\href
  {https://doi.org/10.1088/1361-6382/aaf9ca} {\bibfield  {journal} {\bibinfo
  {journal} {Classical and Quantum Gravity}\ }\textbf {\bibinfo {volume}
  {36}},\ \bibinfo {pages} {034001} (\bibinfo {year}
  {2019}{\natexlab{c}})}\BibitemShut {NoStop}%
\bibitem [{\citenamefont {Carlesso}\ \emph {et~al.}(2019)\citenamefont
  {Carlesso}, \citenamefont {Bassi}, \citenamefont {Paternostro},\ and\
  \citenamefont {Ulbricht}}]{Carlesso2019a}%
  \BibitemOpen
  \bibfield  {author} {\bibinfo {author} {\bibfnamefont {M.}~\bibnamefont
  {Carlesso}}, \bibinfo {author} {\bibfnamefont {A.}~\bibnamefont {Bassi}},
  \bibinfo {author} {\bibfnamefont {M.}~\bibnamefont {Paternostro}},\ and\
  \bibinfo {author} {\bibfnamefont {H.}~\bibnamefont {Ulbricht}},\ }\bibfield
  {title} {\bibinfo {title} {{Testing the gravitational field generated by a
  quantum superposition}},\ }\href {https://doi.org/10.1088/1367-2630/ab41c1}
  {\bibfield  {journal} {\bibinfo  {journal} {New J. Phys.}\ }\textbf {\bibinfo
  {volume} {21}},\ \bibinfo {pages} {093052} (\bibinfo {year}
  {2019})}\BibitemShut {NoStop}%
\bibitem [{\citenamefont {Carlesso}\ and\ \citenamefont
  {Paternostro}(2019)}]{Carlesso2019b}%
  \BibitemOpen
  \bibfield  {author} {\bibinfo {author} {\bibfnamefont {M.}~\bibnamefont
  {Carlesso}}\ and\ \bibinfo {author} {\bibfnamefont {M.}~\bibnamefont
  {Paternostro}},\ }\href {http://arxiv.org/abs/1906.11041} {\bibinfo {title}
  {{Opto-mechanical test of collapse models}}} (\bibinfo {year} {2019}),\
  \Eprint {https://arxiv.org/abs/1906.11041} {arXiv:1906.11041 [quant-ph]}
  \BibitemShut {NoStop}%
\bibitem [{\citenamefont {Braginsky}(1967)}]{Braginsky1967}%
  \BibitemOpen
  \bibfield  {author} {\bibinfo {author} {\bibfnamefont {V.~B.}\ \bibnamefont
  {Braginsky}},\ }\bibfield  {title} {\bibinfo {title} {Classical and quantum
  restrictions on the detection of weak disturbances of a macroscopic
  oscillator},\ }\href@noop {} {\bibfield  {journal} {\bibinfo  {journal} {Zh.
  Eksp. Teor. Fiz}\ }\textbf {\bibinfo {volume} {53}},\ \bibinfo {pages} {1434}
  (\bibinfo {year} {1967})}\BibitemShut {NoStop}%
\bibitem [{\citenamefont {Braginsky}(1968)}]{Braginsky1968}%
  \BibitemOpen
  \bibfield  {author} {\bibinfo {author} {\bibfnamefont {V.~B.}\ \bibnamefont
  {Braginsky}},\ }\bibfield  {title} {\bibinfo {title} {Classical and quantum
  restrictions on the detection of weak disturbances of a macroscopic
  oscillator},\ }\href
  {http://www.jetp.ac.ru/cgi-bin/e/index/e/26/4/p831?a=list} {\bibfield
  {journal} {\bibinfo  {journal} {J. Exp. Theor. Phys}\ }\textbf {\bibinfo
  {volume} {26}},\ \bibinfo {pages} {831} (\bibinfo {year} {1968})}\BibitemShut
  {NoStop}%
\bibitem [{\citenamefont {Caves}(1980)}]{Caves1980a}%
  \BibitemOpen
  \bibfield  {author} {\bibinfo {author} {\bibfnamefont {C.~M.}\ \bibnamefont
  {Caves}},\ }\bibfield  {title} {\bibinfo {title} {{Quantum-Mechanical
  Radiation-Pressure Fluctuations in an Interferometer}},\ }\href
  {https://doi.org/10.1103/PhysRevLett.45.75} {\bibfield  {journal} {\bibinfo
  {journal} {Phys. Rev. Lett.}\ }\textbf {\bibinfo {volume} {45}},\ \bibinfo
  {pages} {75} (\bibinfo {year} {1980})}\BibitemShut {NoStop}%
\bibitem [{\citenamefont {Clerk}\ \emph {et~al.}(2010)\citenamefont {Clerk},
  \citenamefont {Devoret}, \citenamefont {Girvin}, \citenamefont {Marquardt},\
  and\ \citenamefont {Schoelkopf}}]{Clerk2010}%
  \BibitemOpen
  \bibfield  {author} {\bibinfo {author} {\bibfnamefont {A.~A.}\ \bibnamefont
  {Clerk}}, \bibinfo {author} {\bibfnamefont {M.~H.}\ \bibnamefont {Devoret}},
  \bibinfo {author} {\bibfnamefont {S.~M.}\ \bibnamefont {Girvin}}, \bibinfo
  {author} {\bibfnamefont {F.}~\bibnamefont {Marquardt}},\ and\ \bibinfo
  {author} {\bibfnamefont {R.~J.}\ \bibnamefont {Schoelkopf}},\ }\bibfield
  {title} {\bibinfo {title} {Introduction to quantum noise, measurement, and
  amplification},\ }\href {https://doi.org/10.1103/RevModPhys.82.1155}
  {\bibfield  {journal} {\bibinfo  {journal} {Rev. Mod. Phys.}\ }\textbf
  {\bibinfo {volume} {82}},\ \bibinfo {pages} {1155} (\bibinfo {year}
  {2010})}\BibitemShut {NoStop}%
\bibitem [{\citenamefont {Braginsky}\ and\ \citenamefont
  {Vorontsov}(1974)}]{Braginsky1974}%
  \BibitemOpen
  \bibfield  {author} {\bibinfo {author} {\bibfnamefont {V.~B.}\ \bibnamefont
  {Braginsky}}\ and\ \bibinfo {author} {\bibfnamefont {Y.~I.}\ \bibnamefont
  {Vorontsov}},\ }\bibfield  {title} {\bibinfo {title} {{Quantum-mechanical
  limitations in macroscopic experiments and modern experimental technique}},\
  }\href {https://doi.org/10.3367/UFNr.0114.197409b.0041} {\bibfield  {journal}
  {\bibinfo  {journal} {Usp. Fiz. Nauk.}\ }\textbf {\bibinfo {volume} {114}},\
  \bibinfo {pages} {41} (\bibinfo {year} {1974})}\BibitemShut {NoStop}%
\bibitem [{\citenamefont {Braginsky}\ and\ \citenamefont
  {Vorontsov}(1975)}]{Braginsky1975}%
  \BibitemOpen
  \bibfield  {author} {\bibinfo {author} {\bibfnamefont {V.~B.}\ \bibnamefont
  {Braginsky}}\ and\ \bibinfo {author} {\bibfnamefont {Y.~I.}\ \bibnamefont
  {Vorontsov}},\ }\bibfield  {title} {\bibinfo {title} {Quantum-mechanical
  limitations in macroscopic experiments and modern experimental technique},\
  }\href {https://doi.org/10.1070/PU1975v017n05ABEH004362} {\bibfield
  {journal} {\bibinfo  {journal} {Sov. Phys. Usp.}\ }\textbf {\bibinfo {volume}
  {17}},\ \bibinfo {pages} {644} (\bibinfo {year} {1975})}\BibitemShut
  {NoStop}%
\bibitem [{\citenamefont {Giffard}(1976)}]{Giffard1976}%
  \BibitemOpen
  \bibfield  {author} {\bibinfo {author} {\bibfnamefont {R.~P.}\ \bibnamefont
  {Giffard}},\ }\bibfield  {title} {\bibinfo {title} {{Ultimate sensitivity
  limit of a resonant gravitational wave antenna using a linear motion
  detector}},\ }\href {https://doi.org/10.1103/PhysRevD.14.2478} {\bibfield
  {journal} {\bibinfo  {journal} {Phys. Rev. D}\ }\textbf {\bibinfo {volume}
  {14}},\ \bibinfo {pages} {2478} (\bibinfo {year} {1976})}\BibitemShut
  {NoStop}%
\bibitem [{\citenamefont {Aspelmeyer}\ \emph
  {et~al.}(2014{\natexlab{a}})\citenamefont {Aspelmeyer}, \citenamefont
  {Kippenberg},\ and\ \citenamefont {Marquardt}}]{Aspelmeyer2014}%
  \BibitemOpen
  \bibfield  {author} {\bibinfo {author} {\bibfnamefont {M.}~\bibnamefont
  {Aspelmeyer}}, \bibinfo {author} {\bibfnamefont {T.~J.}\ \bibnamefont
  {Kippenberg}},\ and\ \bibinfo {author} {\bibfnamefont {F.}~\bibnamefont
  {Marquardt}},\ }\bibfield  {title} {\bibinfo {title} {{Cavity
  optomechanics}},\ }\href {https://doi.org/10.1103/RevModPhys.86.1391}
  {\bibfield  {journal} {\bibinfo  {journal} {Rev. Mod. Phys.}\ }\textbf
  {\bibinfo {volume} {86}},\ \bibinfo {pages} {1391} (\bibinfo {year}
  {2014}{\natexlab{a}})}\BibitemShut {NoStop}%
\bibitem [{\citenamefont {Aspelmeyer}\ \emph
  {et~al.}(2014{\natexlab{b}})\citenamefont {Aspelmeyer}, \citenamefont
  {Kippenberg},\ and\ \citenamefont {Marquardt}}]{Aspelmeyer2014a}%
  \BibitemOpen
  \bibinfo {editor} {\bibfnamefont {M.}~\bibnamefont {Aspelmeyer}}, \bibinfo
  {editor} {\bibfnamefont {T.~J.}\ \bibnamefont {Kippenberg}},\ and\ \bibinfo
  {editor} {\bibfnamefont {F.}~\bibnamefont {Marquardt}},\ eds.,\ \href
  {https://doi.org/10.1007/978-3-642-55312-7} {\emph {\bibinfo {title} {Cavity
  optomechanics}}}\ (\bibinfo  {publisher} {Springer-Verlag},\ \bibinfo
  {address} {Berlin, Heidelberg},\ \bibinfo {year} {2014})\BibitemShut
  {NoStop}%
\bibitem [{\citenamefont {et~al}(2016)}]{Abbott2016}%
  \BibitemOpen
  \bibfield  {author} {\bibinfo {author} {\bibfnamefont {B.~P.~A.}\
  \bibnamefont {et~al}},\ }\bibfield  {title} {\bibinfo {title} {{Observation
  of Gravitational Waves from a Binary Black Hole Merger}},\ }\href
  {https://doi.org/10.1103/PhysRevLett.116.061102} {\bibfield  {journal}
  {\bibinfo  {journal} {Phys. Rev. Lett.}\ }\textbf {\bibinfo {volume} {116}},\
  \bibinfo {pages} {061102} (\bibinfo {year} {2016})}\BibitemShut {NoStop}%
\bibitem [{\citenamefont {Mason}\ \emph {et~al.}(2019)\citenamefont {Mason},
  \citenamefont {Chen}, \citenamefont {Rossi}, \citenamefont {Tsaturyan},\ and\
  \citenamefont {Schliesser}}]{Mason2019}%
  \BibitemOpen
  \bibfield  {author} {\bibinfo {author} {\bibfnamefont {D.}~\bibnamefont
  {Mason}}, \bibinfo {author} {\bibfnamefont {J.}~\bibnamefont {Chen}},
  \bibinfo {author} {\bibfnamefont {M.}~\bibnamefont {Rossi}}, \bibinfo
  {author} {\bibfnamefont {Y.}~\bibnamefont {Tsaturyan}},\ and\ \bibinfo
  {author} {\bibfnamefont {A.}~\bibnamefont {Schliesser}},\ }\bibfield  {title}
  {\bibinfo {title} {{Continuous force and displacement measurement below the
  standard quantum limit}},\ }\href {https://doi.org/10.1038/s41567-019-0533-5}
  {\bibfield  {journal} {\bibinfo  {journal} {Nat. Phys.}\ }\textbf {\bibinfo
  {volume} {15}},\ \bibinfo {pages} {745} (\bibinfo {year} {2019})}\BibitemShut
  {NoStop}%
\bibitem [{\citenamefont {Burd}\ \emph {et~al.}(2019)\citenamefont {Burd},
  \citenamefont {Srinivas}, \citenamefont {Bollinger}, \citenamefont {Wilson},
  \citenamefont {Wineland}, \citenamefont {Leibfried}, \citenamefont
  {Slichter},\ and\ \citenamefont {Allcock}}]{Burd2019}%
  \BibitemOpen
  \bibfield  {author} {\bibinfo {author} {\bibfnamefont {S.~C.}\ \bibnamefont
  {Burd}}, \bibinfo {author} {\bibfnamefont {R.}~\bibnamefont {Srinivas}},
  \bibinfo {author} {\bibfnamefont {J.~J.}\ \bibnamefont {Bollinger}}, \bibinfo
  {author} {\bibfnamefont {A.~C.}\ \bibnamefont {Wilson}}, \bibinfo {author}
  {\bibfnamefont {D.~J.}\ \bibnamefont {Wineland}}, \bibinfo {author}
  {\bibfnamefont {D.}~\bibnamefont {Leibfried}}, \bibinfo {author}
  {\bibfnamefont {D.~H.}\ \bibnamefont {Slichter}},\ and\ \bibinfo {author}
  {\bibfnamefont {D.~T.~C.}\ \bibnamefont {Allcock}},\ }\bibfield  {title}
  {\bibinfo {title} {{Quantum amplification of mechanical oscillator motion}},\
  }\href {https://doi.org/10.1126/science.aaw2884} {\bibfield  {journal}
  {\bibinfo  {journal} {Science}\ }\textbf {\bibinfo {volume} {364}},\ \bibinfo
  {pages} {1163} (\bibinfo {year} {2019})}\BibitemShut {NoStop}%
\bibitem [{\citenamefont {Huang}\ and\ \citenamefont
  {Agarwal}(2017)}]{Huang2017}%
  \BibitemOpen
  \bibfield  {author} {\bibinfo {author} {\bibfnamefont {S.}~\bibnamefont
  {Huang}}\ and\ \bibinfo {author} {\bibfnamefont {G.~S.}\ \bibnamefont
  {Agarwal}},\ }\bibfield  {title} {\bibinfo {title} {{Robust force sensing for
  a free particle in a dissipative optomechanical system with a parametric
  amplifier}},\ }\href {https://doi.org/10.1103/PhysRevA.95.023844} {\bibfield
  {journal} {\bibinfo  {journal} {Phys. Rev. A}\ }\textbf {\bibinfo {volume}
  {95}},\ \bibinfo {pages} {023844} (\bibinfo {year} {2017})}\BibitemShut
  {NoStop}%
\bibitem [{\citenamefont {Zhao}\ \emph {et~al.}(2019)\citenamefont {Zhao},
  \citenamefont {Zhang}, \citenamefont {Miranowicz},\ and\ \citenamefont
  {Jing}}]{Zhao2019}%
  \BibitemOpen
  \bibfield  {author} {\bibinfo {author} {\bibfnamefont {W.}~\bibnamefont
  {Zhao}}, \bibinfo {author} {\bibfnamefont {S.-D.}\ \bibnamefont {Zhang}},
  \bibinfo {author} {\bibfnamefont {A.}~\bibnamefont {Miranowicz}},\ and\
  \bibinfo {author} {\bibfnamefont {H.}~\bibnamefont {Jing}},\ }\bibfield
  {title} {\bibinfo {title} {Weak-force sensing with squeezed optomechanics},\
  }\href {https://doi.org/10.1007/s11433-019-9451-3} {\bibfield  {journal}
  {\bibinfo  {journal} {Sci. China-Phys. Mech. Astron.}\ }\textbf {\bibinfo
  {volume} {63}},\ \bibinfo {pages} {224211} (\bibinfo {year}
  {2019})}\BibitemShut {NoStop}%
\bibitem [{\citenamefont {Wang}\ \emph {et~al.}(2018)\citenamefont {Wang},
  \citenamefont {Xiong}, \citenamefont {Zhang},\ and\ \citenamefont
  {Zhou}}]{Wang2018}%
  \BibitemOpen
  \bibfield  {author} {\bibinfo {author} {\bibfnamefont {X.-Y.}\ \bibnamefont
  {Wang}}, \bibinfo {author} {\bibfnamefont {B.}~\bibnamefont {Xiong}},
  \bibinfo {author} {\bibfnamefont {W.-Z.}\ \bibnamefont {Zhang}},\ and\
  \bibinfo {author} {\bibfnamefont {L.}~\bibnamefont {Zhou}},\ }\bibfield
  {title} {\bibinfo {title} {{Improve the sensitivity of an optomechanical
  sensor with the auxiliary mechanical oscillator}},\ }\href
  {https://doi.org/10.1140/epjd/e2018-90080-4} {\bibfield  {journal} {\bibinfo
  {journal} {Eur. Phys. J. D}\ }\textbf {\bibinfo {volume} {72}},\ \bibinfo
  {pages} {117} (\bibinfo {year} {2018})}\BibitemShut {NoStop}%
\bibitem [{\citenamefont {Zhang}\ \emph {et~al.}(2019)\citenamefont {Zhang},
  \citenamefont {Chen}, \citenamefont {Cheng},\ and\ \citenamefont
  {Jiang}}]{Zhang2019}%
  \BibitemOpen
  \bibfield  {author} {\bibinfo {author} {\bibfnamefont {W.-Z.}\ \bibnamefont
  {Zhang}}, \bibinfo {author} {\bibfnamefont {L.-B.}\ \bibnamefont {Chen}},
  \bibinfo {author} {\bibfnamefont {J.}~\bibnamefont {Cheng}},\ and\ \bibinfo
  {author} {\bibfnamefont {Y.-F.}\ \bibnamefont {Jiang}},\ }\bibfield  {title}
  {\bibinfo {title} {{Quantum-correlation-enhanced weak-field detection in an
  optomechanical system}},\ }\href {https://doi.org/10.1103/PhysRevA.99.063811}
  {\bibfield  {journal} {\bibinfo  {journal} {Phys. Rev. A}\ }\textbf {\bibinfo
  {volume} {99}},\ \bibinfo {pages} {063811} (\bibinfo {year}
  {2019})}\BibitemShut {NoStop}%
\bibitem [{\citenamefont {Ivanov}\ \emph {et~al.}(2015)\citenamefont {Ivanov},
  \citenamefont {Singer}, \citenamefont {Vitanov},\ and\ \citenamefont
  {Porras}}]{Ivanov2015}%
  \BibitemOpen
  \bibfield  {author} {\bibinfo {author} {\bibfnamefont {P.~A.}\ \bibnamefont
  {Ivanov}}, \bibinfo {author} {\bibfnamefont {K.}~\bibnamefont {Singer}},
  \bibinfo {author} {\bibfnamefont {N.~V.}\ \bibnamefont {Vitanov}},\ and\
  \bibinfo {author} {\bibfnamefont {D.}~\bibnamefont {Porras}},\ }\bibfield
  {title} {\bibinfo {title} {Quantum sensors assisted by spontaneous symmetry
  breaking for detecting very small forces},\ }\href
  {https://doi.org/10.1103/PhysRevApplied.4.054007} {\bibfield  {journal}
  {\bibinfo  {journal} {Phys. Rev. Applied}\ }\textbf {\bibinfo {volume} {4}},\
  \bibinfo {pages} {054007} (\bibinfo {year} {2015})}\BibitemShut {NoStop}%
\bibitem [{\citenamefont {Motazedifard}\ \emph {et~al.}(2016)\citenamefont
  {Motazedifard}, \citenamefont {Bemani}, \citenamefont {Naderi}, \citenamefont
  {Roknizadeh},\ and\ \citenamefont {Vitali}}]{Motazedifard2016}%
  \BibitemOpen
  \bibfield  {author} {\bibinfo {author} {\bibfnamefont {A.}~\bibnamefont
  {Motazedifard}}, \bibinfo {author} {\bibfnamefont {F.}~\bibnamefont
  {Bemani}}, \bibinfo {author} {\bibfnamefont {M.~H.}\ \bibnamefont {Naderi}},
  \bibinfo {author} {\bibfnamefont {R.}~\bibnamefont {Roknizadeh}},\ and\
  \bibinfo {author} {\bibfnamefont {D.}~\bibnamefont {Vitali}},\ }\bibfield
  {title} {\bibinfo {title} {{Force sensing based on coherent quantum noise
  cancellation in a hybrid optomechanical cavity with squeezed-vacuum
  injection}},\ }\href {https://doi.org/10.1088/1367-2630/18/7/073040}
  {\bibfield  {journal} {\bibinfo  {journal} {New J. Phys.}\ }\textbf {\bibinfo
  {volume} {18}},\ \bibinfo {pages} {073040} (\bibinfo {year}
  {2016})}\BibitemShut {NoStop}%
\bibitem [{\citenamefont {Motazedifard}\ \emph {et~al.}(2019)\citenamefont
  {Motazedifard}, \citenamefont {Dalafi}, \citenamefont {Bemani},\ and\
  \citenamefont {Naderi}}]{Motazedifard2019}%
  \BibitemOpen
  \bibfield  {author} {\bibinfo {author} {\bibfnamefont {A.}~\bibnamefont
  {Motazedifard}}, \bibinfo {author} {\bibfnamefont {A.}~\bibnamefont
  {Dalafi}}, \bibinfo {author} {\bibfnamefont {F.}~\bibnamefont {Bemani}},\
  and\ \bibinfo {author} {\bibfnamefont {M.~H.}\ \bibnamefont {Naderi}},\
  }\bibfield  {title} {\bibinfo {title} {Force sensing in hybrid
  bose-einstein-condensate optomechanics based on parametric amplification},\
  }\href {https://doi.org/10.1103/PhysRevA.100.023815} {\bibfield  {journal}
  {\bibinfo  {journal} {Phys. Rev. A}\ }\textbf {\bibinfo {volume} {100}},\
  \bibinfo {pages} {023815} (\bibinfo {year} {2019})}\BibitemShut {NoStop}%
\bibitem [{\citenamefont {Davuluri}\ and\ \citenamefont
  {Li}(2018)}]{Davuluri2018}%
  \BibitemOpen
  \bibfield  {author} {\bibinfo {author} {\bibfnamefont {S.}~\bibnamefont
  {Davuluri}}\ and\ \bibinfo {author} {\bibfnamefont {Y.}~\bibnamefont {Li}},\
  }\bibfield  {title} {\bibinfo {title} {{Shot-noise-limited interferometry for
  measuring a classical force}},\ }\href
  {https://doi.org/10.1103/PhysRevA.98.043809} {\bibfield  {journal} {\bibinfo
  {journal} {Phys. Rev. A}\ }\textbf {\bibinfo {volume} {98}},\ \bibinfo
  {pages} {043809} (\bibinfo {year} {2018})}\BibitemShut {NoStop}%
\bibitem [{\citenamefont {Clerk}\ \emph {et~al.}(2008)\citenamefont {Clerk},
  \citenamefont {Marquardt},\ and\ \citenamefont {Jacobs}}]{Clerk2008}%
  \BibitemOpen
  \bibfield  {author} {\bibinfo {author} {\bibfnamefont {A.~A.}\ \bibnamefont
  {Clerk}}, \bibinfo {author} {\bibfnamefont {F.}~\bibnamefont {Marquardt}},\
  and\ \bibinfo {author} {\bibfnamefont {K.}~\bibnamefont {Jacobs}},\
  }\bibfield  {title} {\bibinfo {title} {{Back-action evasion and squeezing of
  a mechanical resonator using a cavity detector}},\ }\href
  {https://doi.org/10.1088/1367-2630/10/9/095010} {\bibfield  {journal}
  {\bibinfo  {journal} {New J. Phys.}\ }\textbf {\bibinfo {volume} {10}},\
  \bibinfo {pages} {095010} (\bibinfo {year} {2008})}\BibitemShut {NoStop}%
\bibitem [{\citenamefont {Woolley}\ \emph {et~al.}(2008)\citenamefont
  {Woolley}, \citenamefont {Doherty}, \citenamefont {Milburn},\ and\
  \citenamefont {Schwab}}]{Woolley2008}%
  \BibitemOpen
  \bibfield  {author} {\bibinfo {author} {\bibfnamefont {M.~J.}\ \bibnamefont
  {Woolley}}, \bibinfo {author} {\bibfnamefont {A.~C.}\ \bibnamefont
  {Doherty}}, \bibinfo {author} {\bibfnamefont {G.~J.}\ \bibnamefont
  {Milburn}},\ and\ \bibinfo {author} {\bibfnamefont {K.~C.}\ \bibnamefont
  {Schwab}},\ }\bibfield  {title} {\bibinfo {title} {{Nanomechanical squeezing
  with detection via a microwave cavity}},\ }\href
  {https://doi.org/10.1103/PhysRevA.78.062303} {\bibfield  {journal} {\bibinfo
  {journal} {Phys. Rev. A}\ }\textbf {\bibinfo {volume} {78}},\ \bibinfo
  {pages} {062303} (\bibinfo {year} {2008})}\BibitemShut {NoStop}%
\bibitem [{\citenamefont {Hertzberg}\ \emph {et~al.}(2010)\citenamefont
  {Hertzberg}, \citenamefont {Rocheleau}, \citenamefont {Ndukum}, \citenamefont
  {Savva}, \citenamefont {Clerk},\ and\ \citenamefont
  {Schwab}}]{Hertzberg2010}%
  \BibitemOpen
  \bibfield  {author} {\bibinfo {author} {\bibfnamefont {J.~B.}\ \bibnamefont
  {Hertzberg}}, \bibinfo {author} {\bibfnamefont {T.}~\bibnamefont
  {Rocheleau}}, \bibinfo {author} {\bibfnamefont {T.}~\bibnamefont {Ndukum}},
  \bibinfo {author} {\bibfnamefont {M.}~\bibnamefont {Savva}}, \bibinfo
  {author} {\bibfnamefont {A.~A.}\ \bibnamefont {Clerk}},\ and\ \bibinfo
  {author} {\bibfnamefont {K.~C.}\ \bibnamefont {Schwab}},\ }\bibfield  {title}
  {\bibinfo {title} {{Back-action-evading measurements of nanomechanical
  motion}},\ }\href {https://doi.org/10.1038/nphys1479} {\bibfield  {journal}
  {\bibinfo  {journal} {Nat. Phys.}\ }\textbf {\bibinfo {volume} {6}},\
  \bibinfo {pages} {213} (\bibinfo {year} {2010})}\BibitemShut {NoStop}%
\bibitem [{\citenamefont {Suh}\ \emph {et~al.}(2014)\citenamefont {Suh},
  \citenamefont {Weinstein}, \citenamefont {Lei}, \citenamefont {Wollman},
  \citenamefont {Steinke}, \citenamefont {Meystre}, \citenamefont {Clerk},\
  and\ \citenamefont {Schwab}}]{Suh2014}%
  \BibitemOpen
  \bibfield  {author} {\bibinfo {author} {\bibfnamefont {J.}~\bibnamefont
  {Suh}}, \bibinfo {author} {\bibfnamefont {A.~J.}\ \bibnamefont {Weinstein}},
  \bibinfo {author} {\bibfnamefont {C.~U.}\ \bibnamefont {Lei}}, \bibinfo
  {author} {\bibfnamefont {E.~E.}\ \bibnamefont {Wollman}}, \bibinfo {author}
  {\bibfnamefont {S.~K.}\ \bibnamefont {Steinke}}, \bibinfo {author}
  {\bibfnamefont {P.}~\bibnamefont {Meystre}}, \bibinfo {author} {\bibfnamefont
  {A.~A.}\ \bibnamefont {Clerk}},\ and\ \bibinfo {author} {\bibfnamefont
  {K.~C.}\ \bibnamefont {Schwab}},\ }\bibfield  {title} {\bibinfo {title}
  {{Mechanically detecting and avoiding the quantum fluctuations of a microwave
  field}},\ }\href {https://doi.org/10.1126/science.1253258} {\bibfield
  {journal} {\bibinfo  {journal} {Science}\ }\textbf {\bibinfo {volume}
  {344}},\ \bibinfo {pages} {1262} (\bibinfo {year} {2014})}\BibitemShut
  {NoStop}%
\bibitem [{\citenamefont {Shomroni}\ \emph
  {et~al.}(2019{\natexlab{a}})\citenamefont {Shomroni}, \citenamefont {Qiu},
  \citenamefont {Malz}, \citenamefont {Nunnenkamp},\ and\ \citenamefont
  {Kippenberg}}]{Shomroni2019b}%
  \BibitemOpen
  \bibfield  {author} {\bibinfo {author} {\bibfnamefont {I.}~\bibnamefont
  {Shomroni}}, \bibinfo {author} {\bibfnamefont {L.}~\bibnamefont {Qiu}},
  \bibinfo {author} {\bibfnamefont {D.}~\bibnamefont {Malz}}, \bibinfo {author}
  {\bibfnamefont {A.}~\bibnamefont {Nunnenkamp}},\ and\ \bibinfo {author}
  {\bibfnamefont {T.~J.}\ \bibnamefont {Kippenberg}},\ }\bibfield  {title}
  {\bibinfo {title} {{Optical backaction-evading measurement of a mechanical
  oscillator}},\ }\href {https://doi.org/10.1038/s41467-019-10024-3} {\bibfield
   {journal} {\bibinfo  {journal} {Nat. Commun.}\ }\textbf {\bibinfo {volume}
  {10}},\ \bibinfo {pages} {2086} (\bibinfo {year}
  {2019}{\natexlab{a}})}\BibitemShut {NoStop}%
\bibitem [{\citenamefont {Kronwald}\ \emph {et~al.}(2013)\citenamefont
  {Kronwald}, \citenamefont {Marquardt},\ and\ \citenamefont
  {Clerk}}]{Kronwald2013}%
  \BibitemOpen
  \bibfield  {author} {\bibinfo {author} {\bibfnamefont {A.}~\bibnamefont
  {Kronwald}}, \bibinfo {author} {\bibfnamefont {F.}~\bibnamefont
  {Marquardt}},\ and\ \bibinfo {author} {\bibfnamefont {A.~A.}\ \bibnamefont
  {Clerk}},\ }\bibfield  {title} {\bibinfo {title} {{Arbitrarily large
  steady-state bosonic squeezing via dissipation}},\ }\href
  {https://doi.org/10.1103/PhysRevA.88.063833} {\bibfield  {journal} {\bibinfo
  {journal} {Phys. Rev. A}\ }\textbf {\bibinfo {volume} {88}},\ \bibinfo
  {pages} {063833} (\bibinfo {year} {2013})}\BibitemShut {NoStop}%
\bibitem [{\citenamefont {Kronwald}\ \emph {et~al.}(2014)\citenamefont
  {Kronwald}, \citenamefont {Marquardt},\ and\ \citenamefont
  {Clerk}}]{Kronwald2014}%
  \BibitemOpen
  \bibfield  {author} {\bibinfo {author} {\bibfnamefont {A.}~\bibnamefont
  {Kronwald}}, \bibinfo {author} {\bibfnamefont {F.}~\bibnamefont
  {Marquardt}},\ and\ \bibinfo {author} {\bibfnamefont {A.~A.}\ \bibnamefont
  {Clerk}},\ }\bibfield  {title} {\bibinfo {title} {Dissipative optomechanical
  squeezing of light},\ }\href {https://doi.org/10.1088/1367-2630/16/6/063058}
  {\bibfield  {journal} {\bibinfo  {journal} {New J. Phys.}\ }\textbf {\bibinfo
  {volume} {16}},\ \bibinfo {pages} {063058} (\bibinfo {year}
  {2014})}\BibitemShut {NoStop}%
\bibitem [{\citenamefont {Lecocq}\ \emph {et~al.}(2015)\citenamefont {Lecocq},
  \citenamefont {Clark}, \citenamefont {Simmonds}, \citenamefont {Aumentado},\
  and\ \citenamefont {Teufel}}]{Lecocq2015}%
  \BibitemOpen
  \bibfield  {author} {\bibinfo {author} {\bibfnamefont {F.}~\bibnamefont
  {Lecocq}}, \bibinfo {author} {\bibfnamefont {J.~B.}\ \bibnamefont {Clark}},
  \bibinfo {author} {\bibfnamefont {R.~W.}\ \bibnamefont {Simmonds}}, \bibinfo
  {author} {\bibfnamefont {J.}~\bibnamefont {Aumentado}},\ and\ \bibinfo
  {author} {\bibfnamefont {J.~D.}\ \bibnamefont {Teufel}},\ }\bibfield  {title}
  {\bibinfo {title} {{Quantum Nondemolition Measurement of a Nonclassical State
  of a Massive Object}},\ }\href {https://doi.org/10.1103/PhysRevX.5.041037}
  {\bibfield  {journal} {\bibinfo  {journal} {Phys. Rev. X}\ }\textbf {\bibinfo
  {volume} {5}},\ \bibinfo {pages} {041037} (\bibinfo {year}
  {2015})}\BibitemShut {NoStop}%
\bibitem [{\citenamefont {Wollman}\ \emph {et~al.}(2015)\citenamefont
  {Wollman}, \citenamefont {Lei}, \citenamefont {Weinstein}, \citenamefont
  {Suh}, \citenamefont {Kronwald}, \citenamefont {Marquardt}, \citenamefont
  {Clerk},\ and\ \citenamefont {Schwab}}]{Wollman2015}%
  \BibitemOpen
  \bibfield  {author} {\bibinfo {author} {\bibfnamefont {E.~E.}\ \bibnamefont
  {Wollman}}, \bibinfo {author} {\bibfnamefont {C.~U.}\ \bibnamefont {Lei}},
  \bibinfo {author} {\bibfnamefont {A.~J.}\ \bibnamefont {Weinstein}}, \bibinfo
  {author} {\bibfnamefont {J.}~\bibnamefont {Suh}}, \bibinfo {author}
  {\bibfnamefont {A.}~\bibnamefont {Kronwald}}, \bibinfo {author}
  {\bibfnamefont {F.}~\bibnamefont {Marquardt}}, \bibinfo {author}
  {\bibfnamefont {A.~A.}\ \bibnamefont {Clerk}},\ and\ \bibinfo {author}
  {\bibfnamefont {K.~C.}\ \bibnamefont {Schwab}},\ }\bibfield  {title}
  {\bibinfo {title} {{Quantum squeezing of motion in a mechanical resonator}},\
  }\href {https://doi.org/10.1126/science.aac5138} {\bibfield  {journal}
  {\bibinfo  {journal} {Science}\ }\textbf {\bibinfo {volume} {349}},\ \bibinfo
  {pages} {952} (\bibinfo {year} {2015})}\BibitemShut {NoStop}%
\bibitem [{\citenamefont {Pirkkalainen}\ \emph {et~al.}(2015)\citenamefont
  {Pirkkalainen}, \citenamefont {Damsk{\"{a}}gg}, \citenamefont {Brandt},
  \citenamefont {Massel},\ and\ \citenamefont
  {Sillanp{\"{a}}{\"{a}}}}]{Pirkkalainen2015}%
  \BibitemOpen
  \bibfield  {author} {\bibinfo {author} {\bibfnamefont {J.-M.}\ \bibnamefont
  {Pirkkalainen}}, \bibinfo {author} {\bibfnamefont {E.}~\bibnamefont
  {Damsk{\"{a}}gg}}, \bibinfo {author} {\bibfnamefont {M.}~\bibnamefont
  {Brandt}}, \bibinfo {author} {\bibfnamefont {F.}~\bibnamefont {Massel}},\
  and\ \bibinfo {author} {\bibfnamefont {M.~A.}\ \bibnamefont
  {Sillanp{\"{a}}{\"{a}}}},\ }\bibfield  {title} {\bibinfo {title} {{Squeezing
  of Quantum Noise of Motion in a Micromechanical Resonator}},\ }\href
  {https://doi.org/10.1103/PhysRevLett.115.243601} {\bibfield  {journal}
  {\bibinfo  {journal} {Phys. Rev. Lett.}\ }\textbf {\bibinfo {volume} {115}},\
  \bibinfo {pages} {243601} (\bibinfo {year} {2015})}\BibitemShut {NoStop}%
\bibitem [{\citenamefont {Lei}\ \emph {et~al.}(2016)\citenamefont {Lei},
  \citenamefont {Weinstein}, \citenamefont {Suh}, \citenamefont {Wollman},
  \citenamefont {Kronwald}, \citenamefont {Marquardt}, \citenamefont {Clerk},\
  and\ \citenamefont {Schwab}}]{Lei2016}%
  \BibitemOpen
  \bibfield  {author} {\bibinfo {author} {\bibfnamefont {C.~U.}\ \bibnamefont
  {Lei}}, \bibinfo {author} {\bibfnamefont {A.~J.}\ \bibnamefont {Weinstein}},
  \bibinfo {author} {\bibfnamefont {J.}~\bibnamefont {Suh}}, \bibinfo {author}
  {\bibfnamefont {E.~E.}\ \bibnamefont {Wollman}}, \bibinfo {author}
  {\bibfnamefont {A.}~\bibnamefont {Kronwald}}, \bibinfo {author}
  {\bibfnamefont {F.}~\bibnamefont {Marquardt}}, \bibinfo {author}
  {\bibfnamefont {A.~A.}\ \bibnamefont {Clerk}},\ and\ \bibinfo {author}
  {\bibfnamefont {K.~C.}\ \bibnamefont {Schwab}},\ }\bibfield  {title}
  {\bibinfo {title} {{Quantum Nondemolition Measurement of a Quantum Squeezed
  State beyond the 3 dB Limit}},\ }\href
  {https://doi.org/10.1103/PhysRevLett.117.100801} {\bibfield  {journal}
  {\bibinfo  {journal} {Phys. Rev. Lett.}\ }\textbf {\bibinfo {volume} {117}},\
  \bibinfo {pages} {100801} (\bibinfo {year} {2016})}\BibitemShut {NoStop}%
\bibitem [{\citenamefont {Shomroni}\ \emph
  {et~al.}(2019{\natexlab{b}})\citenamefont {Shomroni}, \citenamefont
  {Youssefi}, \citenamefont {Sauerwein}, \citenamefont {Qiu}, \citenamefont
  {Seidler}, \citenamefont {Malz}, \citenamefont {Nunnenkamp},\ and\
  \citenamefont {Kippenberg}}]{Shomroni2019a}%
  \BibitemOpen
  \bibfield  {author} {\bibinfo {author} {\bibfnamefont {I.}~\bibnamefont
  {Shomroni}}, \bibinfo {author} {\bibfnamefont {A.}~\bibnamefont {Youssefi}},
  \bibinfo {author} {\bibfnamefont {N.}~\bibnamefont {Sauerwein}}, \bibinfo
  {author} {\bibfnamefont {L.}~\bibnamefont {Qiu}}, \bibinfo {author}
  {\bibfnamefont {P.}~\bibnamefont {Seidler}}, \bibinfo {author} {\bibfnamefont
  {D.}~\bibnamefont {Malz}}, \bibinfo {author} {\bibfnamefont {A.}~\bibnamefont
  {Nunnenkamp}},\ and\ \bibinfo {author} {\bibfnamefont {T.~J.}\ \bibnamefont
  {Kippenberg}},\ }\bibfield  {title} {\bibinfo {title} {{Two-Tone
  Optomechanical Instability and Its Fundamental Implications for
  Backaction-Evading Measurements}},\ }\href
  {https://doi.org/10.1103/PhysRevX.9.041022} {\bibfield  {journal} {\bibinfo
  {journal} {Phys. Rev. X}\ }\textbf {\bibinfo {volume} {9}},\ \bibinfo {pages}
  {41022} (\bibinfo {year} {2019}{\natexlab{b}})}\BibitemShut {NoStop}%
\bibitem [{\citenamefont {Woolley}\ and\ \citenamefont
  {Clerk}(2013)}]{Woolley2013}%
  \BibitemOpen
  \bibfield  {author} {\bibinfo {author} {\bibfnamefont {M.~J.}\ \bibnamefont
  {Woolley}}\ and\ \bibinfo {author} {\bibfnamefont {A.~A.}\ \bibnamefont
  {Clerk}},\ }\bibfield  {title} {\bibinfo {title} {{Two-mode
  back-action-evading measurements in cavity optomechanics}},\ }\href
  {https://doi.org/10.1103/PhysRevA.87.063846} {\bibfield  {journal} {\bibinfo
  {journal} {Phys. Rev. A}\ }\textbf {\bibinfo {volume} {87}},\ \bibinfo
  {pages} {063846} (\bibinfo {year} {2013})}\BibitemShut {NoStop}%
\bibitem [{\citenamefont {Ockeloen-Korppi}\ \emph {et~al.}(2016)\citenamefont
  {Ockeloen-Korppi}, \citenamefont {Damsk{\"{a}}gg}, \citenamefont
  {Pirkkalainen}, \citenamefont {Clerk}, \citenamefont {Woolley},\ and\
  \citenamefont {Sillanp{\"{a}}{\"{a}}}}]{Ockeloen-Korppi2016}%
  \BibitemOpen
  \bibfield  {author} {\bibinfo {author} {\bibfnamefont {C.~F.}\ \bibnamefont
  {Ockeloen-Korppi}}, \bibinfo {author} {\bibfnamefont {E.}~\bibnamefont
  {Damsk{\"{a}}gg}}, \bibinfo {author} {\bibfnamefont {J.~M.}\ \bibnamefont
  {Pirkkalainen}}, \bibinfo {author} {\bibfnamefont {A.~A.}\ \bibnamefont
  {Clerk}}, \bibinfo {author} {\bibfnamefont {M.~J.}\ \bibnamefont {Woolley}},\
  and\ \bibinfo {author} {\bibfnamefont {M.~A.}\ \bibnamefont
  {Sillanp{\"{a}}{\"{a}}}},\ }\bibfield  {title} {\bibinfo {title} {{Quantum
  Backaction Evading Measurement of Collective Mechanical Modes}},\ }\href
  {https://doi.org/10.1103/PhysRevLett.117.140401} {\bibfield  {journal}
  {\bibinfo  {journal} {Phys. Rev. Lett.}\ }\textbf {\bibinfo {volume} {117}},\
  \bibinfo {pages} {140401} (\bibinfo {year} {2016})}\BibitemShut {NoStop}%
\bibitem [{\citenamefont {Woolley}\ and\ \citenamefont
  {Clerk}(2014)}]{Woolley2014}%
  \BibitemOpen
  \bibfield  {author} {\bibinfo {author} {\bibfnamefont {M.~J.}\ \bibnamefont
  {Woolley}}\ and\ \bibinfo {author} {\bibfnamefont {A.~A.}\ \bibnamefont
  {Clerk}},\ }\bibfield  {title} {\bibinfo {title} {{Two-mode squeezed states
  in cavity optomechanics via engineering of a single reservoir}},\ }\href
  {https://doi.org/10.1103/PhysRevA.89.063805} {\bibfield  {journal} {\bibinfo
  {journal} {Phys. Rev. A}\ }\textbf {\bibinfo {volume} {89}},\ \bibinfo
  {pages} {063805} (\bibinfo {year} {2014})}\BibitemShut {NoStop}%
\bibitem [{\citenamefont {Ockeloen-Korppi}\ \emph {et~al.}(2018)\citenamefont
  {Ockeloen-Korppi}, \citenamefont {Damsk{\"{a}}gg}, \citenamefont
  {Pirkkalainen}, \citenamefont {Asjad}, \citenamefont {Clerk}, \citenamefont
  {Massel}, \citenamefont {Woolley},\ and\ \citenamefont
  {Sillanp{\"{a}}{\"{a}}}}]{Ockeloen-Korppi2018}%
  \BibitemOpen
  \bibfield  {author} {\bibinfo {author} {\bibfnamefont {C.~F.}\ \bibnamefont
  {Ockeloen-Korppi}}, \bibinfo {author} {\bibfnamefont {E.}~\bibnamefont
  {Damsk{\"{a}}gg}}, \bibinfo {author} {\bibfnamefont {J.-M.}\ \bibnamefont
  {Pirkkalainen}}, \bibinfo {author} {\bibfnamefont {M.}~\bibnamefont {Asjad}},
  \bibinfo {author} {\bibfnamefont {A.~A.}\ \bibnamefont {Clerk}}, \bibinfo
  {author} {\bibfnamefont {F.}~\bibnamefont {Massel}}, \bibinfo {author}
  {\bibfnamefont {M.~J.}\ \bibnamefont {Woolley}},\ and\ \bibinfo {author}
  {\bibfnamefont {M.~A.}\ \bibnamefont {Sillanp{\"{a}}{\"{a}}}},\ }\bibfield
  {title} {\bibinfo {title} {{Stabilized entanglement of massive mechanical
  oscillators}},\ }\href {https://doi.org/10.1038/s41586-018-0038-x} {\bibfield
   {journal} {\bibinfo  {journal} {Nature}\ }\textbf {\bibinfo {volume}
  {556}},\ \bibinfo {pages} {478} (\bibinfo {year} {2018})}\BibitemShut
  {NoStop}%
\bibitem [{\citenamefont {Massel}(2019)}]{Massel2019}%
  \BibitemOpen
  \bibfield  {author} {\bibinfo {author} {\bibfnamefont {F.}~\bibnamefont
  {Massel}},\ }\bibfield  {title} {\bibinfo {title} {{Backaction-evading
  measurement of entanglement in optomechanics}},\ }\href
  {https://doi.org/10.1103/PhysRevA.100.023824} {\bibfield  {journal} {\bibinfo
   {journal} {Phys. Rev. A}\ }\textbf {\bibinfo {volume} {100}},\ \bibinfo
  {pages} {023824} (\bibinfo {year} {2019})}\BibitemShut {NoStop}%
\bibitem [{\citenamefont {Vitali}\ \emph {et~al.}(2001)\citenamefont {Vitali},
  \citenamefont {Mancini},\ and\ \citenamefont {Tombesi}}]{Vitali2001}%
  \BibitemOpen
  \bibfield  {author} {\bibinfo {author} {\bibfnamefont {D.}~\bibnamefont
  {Vitali}}, \bibinfo {author} {\bibfnamefont {S.}~\bibnamefont {Mancini}},\
  and\ \bibinfo {author} {\bibfnamefont {P.}~\bibnamefont {Tombesi}},\
  }\bibfield  {title} {\bibinfo {title} {Optomechanical scheme for the
  detection of weak impulsive forces},\ }\href
  {https://doi.org/10.1103/PhysRevA.64.051401} {\bibfield  {journal} {\bibinfo
  {journal} {Phys. Rev. A}\ }\textbf {\bibinfo {volume} {64}},\ \bibinfo
  {pages} {051401(R)} (\bibinfo {year} {2001})}\BibitemShut {NoStop}%
\bibitem [{\citenamefont {Vitali}\ \emph
  {et~al.}(2004{\natexlab{a}})\citenamefont {Vitali}, \citenamefont {Mancini},\
  and\ \citenamefont {Tombesi}}]{Vitali2001Erratum}%
  \BibitemOpen
  \bibfield  {author} {\bibinfo {author} {\bibfnamefont {D.}~\bibnamefont
  {Vitali}}, \bibinfo {author} {\bibfnamefont {S.}~\bibnamefont {Mancini}},\
  and\ \bibinfo {author} {\bibfnamefont {P.}~\bibnamefont {Tombesi}},\
  }\bibfield  {title} {\bibinfo {title} {Erratum: Optomechanical scheme for the
  detection of weak impulsive forces {[Phys. Rev. A 64, 051401(R) (2001)]}},\
  }\href {https://doi.org/10.1103/PhysRevA.69.049904} {\bibfield  {journal}
  {\bibinfo  {journal} {Phys. Rev. A}\ }\textbf {\bibinfo {volume} {69}},\
  \bibinfo {pages} {049904(E)} (\bibinfo {year}
  {2004}{\natexlab{a}})}\BibitemShut {NoStop}%
\bibitem [{\citenamefont {Vitali}\ \emph {et~al.}(2002)\citenamefont {Vitali},
  \citenamefont {Mancini}, \citenamefont {Ribichini},\ and\ \citenamefont
  {Tombesi}}]{Vitali2002}%
  \BibitemOpen
  \bibfield  {author} {\bibinfo {author} {\bibfnamefont {D.}~\bibnamefont
  {Vitali}}, \bibinfo {author} {\bibfnamefont {S.}~\bibnamefont {Mancini}},
  \bibinfo {author} {\bibfnamefont {L.}~\bibnamefont {Ribichini}},\ and\
  \bibinfo {author} {\bibfnamefont {P.}~\bibnamefont {Tombesi}},\ }\bibfield
  {title} {\bibinfo {title} {Mirror quiescence and high-sensitivity position
  measurements with feedback},\ }\href
  {https://doi.org/10.1103/PhysRevA.65.063803} {\bibfield  {journal} {\bibinfo
  {journal} {Phys. Rev. A}\ }\textbf {\bibinfo {volume} {65}},\ \bibinfo
  {pages} {063803} (\bibinfo {year} {2002})}\BibitemShut {NoStop}%
\bibitem [{\citenamefont {Vitali}\ \emph
  {et~al.}(2004{\natexlab{b}})\citenamefont {Vitali}, \citenamefont {Mancini},
  \citenamefont {Ribichini},\ and\ \citenamefont
  {Tombesi}}]{Vitali2002Erratum}%
  \BibitemOpen
  \bibfield  {author} {\bibinfo {author} {\bibfnamefont {D.}~\bibnamefont
  {Vitali}}, \bibinfo {author} {\bibfnamefont {S.}~\bibnamefont {Mancini}},
  \bibinfo {author} {\bibfnamefont {L.}~\bibnamefont {Ribichini}},\ and\
  \bibinfo {author} {\bibfnamefont {P.}~\bibnamefont {Tombesi}},\ }\bibfield
  {title} {\bibinfo {title} {Erratum: Mirror quiescence and high-sensitivity
  position measurements with feedback {[Phys. Rev. A 65, 063803 (2002)]}},\
  }\href {https://doi.org/10.1103/PhysRevA.69.029901} {\bibfield  {journal}
  {\bibinfo  {journal} {Phys. Rev. A}\ }\textbf {\bibinfo {volume} {69}},\
  \bibinfo {pages} {029901(E)} (\bibinfo {year}
  {2004}{\natexlab{b}})}\BibitemShut {NoStop}%
\bibitem [{\citenamefont {Ford}\ \emph {et~al.}(1988)\citenamefont {Ford},
  \citenamefont {Lewis},\ and\ \citenamefont {O'Connell}}]{Ford1988}%
  \BibitemOpen
  \bibfield  {author} {\bibinfo {author} {\bibfnamefont {G.~W.}\ \bibnamefont
  {Ford}}, \bibinfo {author} {\bibfnamefont {J.~T.}\ \bibnamefont {Lewis}},\
  and\ \bibinfo {author} {\bibfnamefont {R.~F.}\ \bibnamefont {O'Connell}},\
  }\bibfield  {title} {\bibinfo {title} {{Quantum Langevin equation}},\ }\href
  {https://doi.org/10.1103/PhysRevA.37.4419} {\bibfield  {journal} {\bibinfo
  {journal} {Phys. Rev. A}\ }\textbf {\bibinfo {volume} {37}},\ \bibinfo
  {pages} {4419} (\bibinfo {year} {1988})}\BibitemShut {NoStop}%
\bibitem [{\citenamefont {Giovannetti}\ and\ \citenamefont
  {Vitali}(2001)}]{Giovannetti2001}%
  \BibitemOpen
  \bibfield  {author} {\bibinfo {author} {\bibfnamefont {V.}~\bibnamefont
  {Giovannetti}}\ and\ \bibinfo {author} {\bibfnamefont {D.}~\bibnamefont
  {Vitali}},\ }\bibfield  {title} {\bibinfo {title} {{Phase-noise measurement
  in a cavity with a movable mirror undergoing quantum Brownian motion}},\
  }\href {https://doi.org/10.1103/PhysRevA.63.023812} {\bibfield  {journal}
  {\bibinfo  {journal} {Phys. Rev. A}\ }\textbf {\bibinfo {volume} {63}},\
  \bibinfo {pages} {023812} (\bibinfo {year} {2001})}\BibitemShut {NoStop}%
\bibitem [{\citenamefont {Bowen}\ and\ \citenamefont
  {Milburn}(2015)}]{Bowen2015}%
  \BibitemOpen
  \bibfield  {author} {\bibinfo {author} {\bibfnamefont {W.}~\bibnamefont
  {Bowen}}\ and\ \bibinfo {author} {\bibfnamefont {G.}~\bibnamefont
  {Milburn}},\ }\href {https://doi.org/10.1201/b19379} {\emph {\bibinfo {title}
  {Quantum optomechanics}}}\ (\bibinfo  {publisher} {CRC Press},\ \bibinfo
  {address} {Boca Raton},\ \bibinfo {year} {2015})\BibitemShut {NoStop}%
\bibitem [{\citenamefont {Papoulis}\ and\ \citenamefont
  {Pillai}(2002)}]{Papoulis2002}%
  \BibitemOpen
  \bibfield  {author} {\bibinfo {author} {\bibfnamefont {A.}~\bibnamefont
  {Papoulis}}\ and\ \bibinfo {author} {\bibfnamefont {S.~U.}\ \bibnamefont
  {Pillai}},\ }\href {http://www.mhhe.com/engcs/electrical/papoulis/} {\emph
  {\bibinfo {title} {Probability, random variables and stochastic
  processes}}},\ \bibinfo {edition} {4th}\ ed.\ (\bibinfo  {publisher}
  {McGraw-Hill},\ \bibinfo {address} {New York},\ \bibinfo {year}
  {2002})\BibitemShut {NoStop}%
\bibitem [{\citenamefont {Anandan}(1990)}]{Anandan1990}%
  \BibitemOpen
  \bibfield  {author} {\bibinfo {author} {\bibfnamefont {J.}~\bibnamefont
  {Anandan}},\ }\bibfield  {title} {\bibinfo {title} {Geometric phase for
  cyclic motions and the quantum state space metric},\ }\href
  {https://doi.org/10.1016/0375-9601(90)90003-7} {\bibfield  {journal}
  {\bibinfo  {journal} {Phys. Lett. A}\ }\textbf {\bibinfo {volume} {147}},\
  \bibinfo {pages} {3} (\bibinfo {year} {1990})}\BibitemShut {NoStop}%
\bibitem [{\citenamefont {Pati}\ \emph {et~al.}(2015)\citenamefont {Pati},
  \citenamefont {Singh},\ and\ \citenamefont {Sinha}}]{Pati2015}%
  \BibitemOpen
  \bibfield  {author} {\bibinfo {author} {\bibfnamefont {A.~K.}\ \bibnamefont
  {Pati}}, \bibinfo {author} {\bibfnamefont {U.}~\bibnamefont {Singh}},\ and\
  \bibinfo {author} {\bibfnamefont {U.}~\bibnamefont {Sinha}},\ }\bibfield
  {title} {\bibinfo {title} {Measuring non-hermitian operators via weak
  values},\ }\href {https://doi.org/10.1103/PhysRevA.92.052120} {\bibfield
  {journal} {\bibinfo  {journal} {Phys. Rev. A}\ }\textbf {\bibinfo {volume}
  {92}},\ \bibinfo {pages} {052120} (\bibinfo {year} {2015})}\BibitemShut
  {NoStop}%
\bibitem [{\citenamefont {Cooper}\ and\ \citenamefont
  {McGillem}(1998)}]{Cooper1998}%
  \BibitemOpen
  \bibfield  {author} {\bibinfo {author} {\bibfnamefont {G.~R.}\ \bibnamefont
  {Cooper}}\ and\ \bibinfo {author} {\bibfnamefont {C.~D.}\ \bibnamefont
  {McGillem}},\ }\href
  {https://global.oup.com/academic/product/probabilistic-methods-of-signal-and-system-analysis-9780195123548?q=9780195123548{\&}cc=au{\&}lang=en}
  {\emph {\bibinfo {title} {Probabilistic methods of signal and system
  Analysis}}},\ \bibinfo {edition} {3rd}\ ed.\ (\bibinfo  {publisher} {Oxford
  University Press},\ \bibinfo {address} {Oxford, UK},\ \bibinfo {year}
  {1998})\BibitemShut {NoStop}%
\bibitem [{\citenamefont {{Grover Brown}}\ and\ \citenamefont
  {Hwang}(2012)}]{GroverBrown2012}%
  \BibitemOpen
  \bibfield  {author} {\bibinfo {author} {\bibfnamefont {R.}~\bibnamefont
  {{Grover Brown}}}\ and\ \bibinfo {author} {\bibfnamefont {P.~Y.~C.}\
  \bibnamefont {Hwang}},\ }\href
  {https://www.wiley.com/en-us/exportProduct/pdf/9780470609699} {\emph
  {\bibinfo {title} {Introduction to random signals and applied {Kalman}
  filtering}}},\ \bibinfo {edition} {4th}\ ed.\ (\bibinfo  {publisher} {John
  Wiley {\&} Sons},\ \bibinfo {year} {2012})\BibitemShut {NoStop}%
\bibitem [{\citenamefont {Prabhu}(2014)}]{Prabhu2014}%
  \BibitemOpen
  \bibfield  {author} {\bibinfo {author} {\bibfnamefont {K.~M.~M.}\
  \bibnamefont {Prabhu}},\ }\href {https://doi.org/10.1201/9781315216386}
  {\emph {\bibinfo {title} {Window functions and their applications in signal
  processing}}}\ (\bibinfo  {publisher} {CRC Press},\ \bibinfo {address} {Boca
  Raton},\ \bibinfo {year} {2014})\BibitemShut {NoStop}%
\bibitem [{\citenamefont {Lucamarini}\ \emph {et~al.}(2006)\citenamefont
  {Lucamarini}, \citenamefont {Vitali},\ and\ \citenamefont
  {Tombesi}}]{Lucamarini2006}%
  \BibitemOpen
  \bibfield  {author} {\bibinfo {author} {\bibfnamefont {M.}~\bibnamefont
  {Lucamarini}}, \bibinfo {author} {\bibfnamefont {D.}~\bibnamefont {Vitali}},\
  and\ \bibinfo {author} {\bibfnamefont {P.}~\bibnamefont {Tombesi}},\
  }\bibfield  {title} {\bibinfo {title} {{Scheme for a quantum-limited force
  measurement with an optomechanical device}},\ }\href
  {https://doi.org/10.1103/PhysRevA.74.063816} {\bibfield  {journal} {\bibinfo
  {journal} {Phys. Rev. A}\ }\textbf {\bibinfo {volume} {74}},\ \bibinfo
  {pages} {063816} (\bibinfo {year} {2006})}\BibitemShut {NoStop}%
\bibitem [{\citenamefont {Wimmer}\ \emph {et~al.}(2014)\citenamefont {Wimmer},
  \citenamefont {Steinmeyer}, \citenamefont {Hammerer},\ and\ \citenamefont
  {Heurs}}]{Wimmer2014}%
  \BibitemOpen
  \bibfield  {author} {\bibinfo {author} {\bibfnamefont {M.~H.}\ \bibnamefont
  {Wimmer}}, \bibinfo {author} {\bibfnamefont {D.}~\bibnamefont {Steinmeyer}},
  \bibinfo {author} {\bibfnamefont {K.}~\bibnamefont {Hammerer}},\ and\
  \bibinfo {author} {\bibfnamefont {M.}~\bibnamefont {Heurs}},\ }\bibfield
  {title} {\bibinfo {title} {Coherent cancellation of backaction noise in
  optomechanical force measurements},\ }\href
  {https://doi.org/10.1103/PhysRevA.89.053836} {\bibfield  {journal} {\bibinfo
  {journal} {Phys. Rev. A}\ }\textbf {\bibinfo {volume} {89}},\ \bibinfo
  {pages} {053836} (\bibinfo {year} {2014})}\BibitemShut {NoStop}%
\bibitem [{\citenamefont {Harris}\ \emph {et~al.}(2013)\citenamefont {Harris},
  \citenamefont {McAuslan}, \citenamefont {Stace}, \citenamefont {Doherty},\
  and\ \citenamefont {Bowen}}]{Harris2013}%
  \BibitemOpen
  \bibfield  {author} {\bibinfo {author} {\bibfnamefont {G.~I.}\ \bibnamefont
  {Harris}}, \bibinfo {author} {\bibfnamefont {D.~L.}\ \bibnamefont
  {McAuslan}}, \bibinfo {author} {\bibfnamefont {T.~M.}\ \bibnamefont {Stace}},
  \bibinfo {author} {\bibfnamefont {A.~C.}\ \bibnamefont {Doherty}},\ and\
  \bibinfo {author} {\bibfnamefont {W.~P.}\ \bibnamefont {Bowen}},\ }\bibfield
  {title} {\bibinfo {title} {{Minimum Requirements for Feedback Enhanced Force
  Sensing}},\ }\href {https://doi.org/10.1103/PhysRevLett.111.103603}
  {\bibfield  {journal} {\bibinfo  {journal} {Phys. Rev. Lett.}\ }\textbf
  {\bibinfo {volume} {111}},\ \bibinfo {pages} {103603} (\bibinfo {year}
  {2013})}\BibitemShut {NoStop}%
\bibitem [{\citenamefont {Hosseini}\ \emph {et~al.}(2014)\citenamefont
  {Hosseini}, \citenamefont {Guccione}, \citenamefont {Slatyer}, \citenamefont
  {Buchler},\ and\ \citenamefont {Lam}}]{Hosseini2014}%
  \BibitemOpen
  \bibfield  {author} {\bibinfo {author} {\bibfnamefont {M.}~\bibnamefont
  {Hosseini}}, \bibinfo {author} {\bibfnamefont {G.}~\bibnamefont {Guccione}},
  \bibinfo {author} {\bibfnamefont {H.~J.}\ \bibnamefont {Slatyer}}, \bibinfo
  {author} {\bibfnamefont {B.~C.}\ \bibnamefont {Buchler}},\ and\ \bibinfo
  {author} {\bibfnamefont {P.~K.}\ \bibnamefont {Lam}},\ }\bibfield  {title}
  {\bibinfo {title} {{Multimode laser cooling and ultra-high sensitivity force
  sensing with nanowires}},\ }\href {https://doi.org/10.1038/ncomms5663}
  {\bibfield  {journal} {\bibinfo  {journal} {Nat. Commun.}\ }\textbf {\bibinfo
  {volume} {5}},\ \bibinfo {pages} {4663} (\bibinfo {year} {2014})}\BibitemShut
  {NoStop}%
\bibitem [{\citenamefont {Gardiner}\ and\ \citenamefont
  {Collett}(1985)}]{Gardiner1985}%
  \BibitemOpen
  \bibfield  {author} {\bibinfo {author} {\bibfnamefont {C.~W.}\ \bibnamefont
  {Gardiner}}\ and\ \bibinfo {author} {\bibfnamefont {M.~J.}\ \bibnamefont
  {Collett}},\ }\bibfield  {title} {\bibinfo {title} {{Input and output in
  damped quantum systems: Quantum stochastic differential equations and the
  master equation}},\ }\href {https://doi.org/10.1103/PhysRevA.31.3761}
  {\bibfield  {journal} {\bibinfo  {journal} {Phys. Rev. A}\ }\textbf {\bibinfo
  {volume} {31}},\ \bibinfo {pages} {3761} (\bibinfo {year}
  {1985})}\BibitemShut {NoStop}%
\bibitem [{\citenamefont {Gardiner}\ and\ \citenamefont
  {Zoller}(2004)}]{Gardiner2004}%
  \BibitemOpen
  \bibfield  {author} {\bibinfo {author} {\bibfnamefont {C.}~\bibnamefont
  {Gardiner}}\ and\ \bibinfo {author} {\bibfnamefont {P.}~\bibnamefont
  {Zoller}},\ }\href {https://www.springer.com/gp/book/9783540223016} {\emph
  {\bibinfo {title} {Quantum noise}}}\ (\bibinfo  {publisher}
  {Springer-Verlag},\ \bibinfo {address} {Berlin, Heidelberg},\ \bibinfo {year}
  {2004})\BibitemShut {NoStop}%
\bibitem [{\citenamefont {Walls}\ and\ \citenamefont
  {Milburn}(2007)}]{Walls2007}%
  \BibitemOpen
  \bibfield  {author} {\bibinfo {author} {\bibfnamefont {D.~F.}\ \bibnamefont
  {Walls}}\ and\ \bibinfo {author} {\bibfnamefont {G.~J.}\ \bibnamefont
  {Milburn}},\ }\href {https://doi.org/10.1007/978-3-540-28574-8} {\emph
  {\bibinfo {title} {Quantum optics}}},\ \bibinfo {edition} {2nd}\ ed.\
  (\bibinfo  {publisher} {Springer-Verlag},\ \bibinfo {address} {Berlin,
  Heidelberg},\ \bibinfo {year} {2007})\BibitemShut {NoStop}%
\bibitem [{\citenamefont {Gea-Banacloche}\ \emph {et~al.}(1990)\citenamefont
  {Gea-Banacloche}, \citenamefont {Lu}, \citenamefont {Pedrotti}, \citenamefont
  {Prasad}, \citenamefont {Scully},\ and\ \citenamefont
  {W{\'{o}}dkiewicz}}]{Gea-Banacloche1990}%
  \BibitemOpen
  \bibfield  {author} {\bibinfo {author} {\bibfnamefont {J.}~\bibnamefont
  {Gea-Banacloche}}, \bibinfo {author} {\bibfnamefont {N.}~\bibnamefont {Lu}},
  \bibinfo {author} {\bibfnamefont {L.~M.}\ \bibnamefont {Pedrotti}}, \bibinfo
  {author} {\bibfnamefont {S.}~\bibnamefont {Prasad}}, \bibinfo {author}
  {\bibfnamefont {M.~O.}\ \bibnamefont {Scully}},\ and\ \bibinfo {author}
  {\bibfnamefont {K.}~\bibnamefont {W{\'{o}}dkiewicz}},\ }\bibfield  {title}
  {\bibinfo {title} {{Treatment of the spectrum of squeezing based on the modes
  of the universe. I. Theory and a physical picture}},\ }\href
  {https://doi.org/10.1103/PhysRevA.41.369} {\bibfield  {journal} {\bibinfo
  {journal} {Phys. Rev. A}\ }\textbf {\bibinfo {volume} {41}},\ \bibinfo
  {pages} {369} (\bibinfo {year} {1990})}\BibitemShut {NoStop}%
\bibitem [{\citenamefont {Dechant}\ and\ \citenamefont
  {Lutz}(2015)}]{Dechant2015}%
  \BibitemOpen
  \bibfield  {author} {\bibinfo {author} {\bibfnamefont {A.}~\bibnamefont
  {Dechant}}\ and\ \bibinfo {author} {\bibfnamefont {E.}~\bibnamefont {Lutz}},\
  }\bibfield  {title} {\bibinfo {title} {{Wiener-Khinchin Theorem for
  Nonstationary Scale-Invariant Processes}},\ }\href
  {https://doi.org/10.1103/PhysRevLett.115.080603} {\bibfield  {journal}
  {\bibinfo  {journal} {Phys. Rev. Lett.}\ }\textbf {\bibinfo {volume} {115}},\
  \bibinfo {pages} {080603} (\bibinfo {year} {2015})}\BibitemShut {NoStop}%
\bibitem [{\citenamefont {Kusse}\ and\ \citenamefont
  {Westwig}(2006)}]{Kusse2006}%
  \BibitemOpen
  \bibfield  {author} {\bibinfo {author} {\bibfnamefont {B.~R.}\ \bibnamefont
  {Kusse}}\ and\ \bibinfo {author} {\bibfnamefont {E.~A.}\ \bibnamefont
  {Westwig}},\ }\href {https://doi.org/10.1002/9783527618132} {\emph {\bibinfo
  {title} {Mathematical physics}}},\ \bibinfo {edition} {2nd}\ ed.\ (\bibinfo
  {publisher} {Wiley-VCH Verlag GmbH},\ \bibinfo {address} {Weinheim,
  Germany},\ \bibinfo {year} {2006})\BibitemShut {NoStop}%
\bibitem [{\citenamefont {Marquardt}\ \emph {et~al.}(2007)\citenamefont
  {Marquardt}, \citenamefont {Chen}, \citenamefont {Clerk},\ and\ \citenamefont
  {Girvin}}]{Marquardt2007}%
  \BibitemOpen
  \bibfield  {author} {\bibinfo {author} {\bibfnamefont {F.}~\bibnamefont
  {Marquardt}}, \bibinfo {author} {\bibfnamefont {J.~P.}\ \bibnamefont {Chen}},
  \bibinfo {author} {\bibfnamefont {A.~A.}\ \bibnamefont {Clerk}},\ and\
  \bibinfo {author} {\bibfnamefont {S.~M.}\ \bibnamefont {Girvin}},\ }\bibfield
   {title} {\bibinfo {title} {Quantum theory of cavity-assisted sideband
  cooling of mechanical motion},\ }\href
  {https://doi.org/10.1103/PhysRevLett.99.093902} {\bibfield  {journal}
  {\bibinfo  {journal} {Phys. Rev. Lett.}\ }\textbf {\bibinfo {volume} {99}},\
  \bibinfo {pages} {093902} (\bibinfo {year} {2007})}\BibitemShut {NoStop}%
\bibitem [{\citenamefont {Wilson-Rae}\ \emph {et~al.}(2007)\citenamefont
  {Wilson-Rae}, \citenamefont {Nooshi}, \citenamefont {Zwerger},\ and\
  \citenamefont {Kippenberg}}]{Wilson-Rae2007}%
  \BibitemOpen
  \bibfield  {author} {\bibinfo {author} {\bibfnamefont {I.}~\bibnamefont
  {Wilson-Rae}}, \bibinfo {author} {\bibfnamefont {N.}~\bibnamefont {Nooshi}},
  \bibinfo {author} {\bibfnamefont {W.}~\bibnamefont {Zwerger}},\ and\ \bibinfo
  {author} {\bibfnamefont {T.~J.}\ \bibnamefont {Kippenberg}},\ }\bibfield
  {title} {\bibinfo {title} {Theory of ground state cooling of a mechanical
  oscillator using dynamical backaction},\ }\href
  {https://doi.org/10.1103/PhysRevLett.99.093901} {\bibfield  {journal}
  {\bibinfo  {journal} {Phys. Rev. Lett.}\ }\textbf {\bibinfo {volume} {99}},\
  \bibinfo {pages} {093901} (\bibinfo {year} {2007})}\BibitemShut {NoStop}%
\bibitem [{\citenamefont {Blair}(1991)}]{Blair1991}%
  \BibitemOpen
  \bibinfo {editor} {\bibfnamefont {D.~G.}\ \bibnamefont {Blair}},\ ed.,\ \href
  {https://doi.org/10.1017/CBO9780511600104} {\emph {\bibinfo {title} {The
  detection of gravitational waves}}}\ (\bibinfo  {publisher} {Cambridge
  University Press},\ \bibinfo {address} {Cambridge},\ \bibinfo {year}
  {1991})\BibitemShut {NoStop}%
\end{thebibliography}%
% 
%------------------------------------------------------------------------------

\end{document}